\documentclass[letterpaper, oneside, 12pt]{article}
\usepackage[round]{natbib}
\usepackage{amsmath}
\usepackage{amssymb}
\usepackage{graphicx}
\usepackage{amsfonts}
\usepackage{latexsym}
\usepackage{subfigure}
\usepackage{url}
\usepackage{algorithmic}\label{•}
\usepackage{longtable}
\usepackage{multirow}
\usepackage{fancyvrb}
\usepackage{float}
\usepackage{setspace}
\usepackage{slashbox}
\usepackage[hmargin=1.0in,vmargin=1.12in]{geometry}
\usepackage{setspace}
\usepackage{rotating}
\usepackage{enumerate}
\usepackage{comment}
\bibpunct{(}{)}{;}{a}{}{,}

\newcommand{\tp}{\tilde p}

\DefineVerbatimEnvironment{Code}{Verbatim}{fontsize=\small}
\newcommand{\ignore}[1]{}

\newcommand{\bet}{ \mbox{\boldmath $ \eta $} }

\newcommand{\bbeta}{ \mbox{\boldmath $ \beta $} }

\newcommand{\bphi}{ \mbox{\boldmath $\phi$}}

\newcommand{\beps}{ \mbox{\boldmath $\epsilon$}}
\newcommand{\bepsilon}{ \mbox{\boldmath $\epsilon$}}

\newcommand{\btheta}{ \mbox{\boldmath $ \theta $} }

\newcommand{\bmu}{ \mbox{\boldmath $\mu$} }

\newcommand{\bGam}{ \mbox{\boldmath $\Gamma$} }

\newcommand{\bSigma}{ \mbox{\boldmath $\Sigma$} }

\newcommand{\bLambda}{ \mbox{\boldmath $\Lambda$} }

\newcommand{\boeta}{ \mbox{\boldmath $\eta$} }

\newcommand{\bzero}{\textbf{0}}

\newcommand{\bZ}{\textbf{Z}}

\newcommand{\ba}{\textbf{a}}
\newcommand{\bA}{\textbf{A}}

\newcommand{\bB}{\textbf{B}}

\newcommand{\bC}{\textbf{C}}

\newcommand{\bD}{\textbf{D}}
\newcommand{\be}{\textbf{e}}

\newcommand{\bF}{\textbf{F}}

\newcommand{\bh}{\textbf{h}}

\newcommand{\bI}{\textbf{I}}

\newcommand{\bL}{\textbf{L}}
\newcommand{\bm}{\textbf{m}}

\newcommand{\bO}{\textbf{O}}

\newcommand{\bs}{\textbf{s}}

\newcommand{\bt}{\textbf{t}}

\newcommand{\bu}{\textbf{u}}

\newcommand{\bv}{\textbf{v}}
\newcommand{\bV}{\textbf{V}}
\newcommand{\bw}{\textbf{w}}
\newcommand{\bW}{\textbf{W}}
\newcommand{\bx}{\textbf{x}}
\newcommand{\bX}{\textbf{X}}
\newcommand{\by}{\textbf{y}}
\newcommand{\bY}{\textbf{Y}}

\newcommand{\bR}{\textbf{R}}

\newcommand{\tildebC}{\tilde{\bC}}

\newcommand{\given}{\,|\,}

\newcommand{\taus}{\tau^2}
\newcommand{\sigs}{\sigma^2}

\newcommand{\calS}{{\cal S}}
\newcommand{\calK}{{\cal K}}

\newcommand{\calD}{{\cal D}}
\newcommand{\calG}{{\cal G}}
\newcommand{\calT}{{\cal T}}
\newcommand{\calU}{{\cal U}}
\newcommand{\calV}{{\cal V}}
\newcommand{\calZ}{{\cal Z}}

\usepackage{xr}
\begin{document}

\title{Hierarchical Nearest-Neighbor Gaussian Process Models for Large Geostatistical Datasets}
\author{Abhirup Datta, Sudipto Banerjee, Andrew O. Finley and Alan E. Gelfand 
}
\date{}
\maketitle
\vspace{-0.25in}
\begin{abstract}
Spatial process models for analyzing geostatistical data entail computations that become prohibitive as the number of spatial locations become large. This manuscript develops a class of highly scalable Nearest Neighbor Gaussian Process (NNGP) models to provide fully model-based inference for large geostatistical datasets. We establish that the NNGP is a well-defined spatial process providing legitimate finite-dimensional Gaussian densities with sparse precision matrices. We embed the NNGP as a sparsity-inducing prior within a rich hierarchical modeling framework and outline how computationally efficient Markov chain Monte Carlo (MCMC) algorithms can be executed without storing or decomposing large matrices. The floating point operations (flops) per iteration of this algorithm is linear in the number of spatial locations, thereby rendering substantial scalability. We illustrate the computational and inferential benefits of the NNGP over competing methods using simulation studies and also analyze forest biomass from a massive United States Forest Inventory dataset at a scale that precludes alternative dimension-reducing methods. 
\end{abstract}


\section{Introduction}\label{Sec: Intro}

\noindent With the growing capabilities of Geographical Information Systems (GIS) and user-friendly software, statisticians today routinely encounter geographically referenced datasets containing a large number of irregularly located observations on multiple variables. This has, in turn, fueled considerable interest in statistical modeling for location-referenced spatial data; see, for example, the books by \citet{stein99}, \citet{moll03}, \citet{ban14}, \citet{scha04}, and \citet{creswikle11} for a variety of methods and applications. Spatial process models introduce spatial dependence between observations using an underlying random field, $\{w(\bs): \bs\in \calD\}$, over a region of interest $\calD$, which is endowed with a probability law that specifies the joint distribution for any finite set of random variables. For example, a zero-centered Gaussian process ensures that $\bw = (w(\bs_1), w(\bs_2)\ldots, w(\bs_n))' \sim N(\bzero, \bC(\btheta))$, where $\bC(\btheta)$ is a family of covariance matrices, indexed by an unknown set of parameters $\btheta$. Such processes offer a rich modeling framework and are being widely deployed to help researchers comprehend complex spatial phenomena in the sciences. However, model fitting usually involves the inverse and determinant of $\bC(\btheta)$, which typically require $\sim n^3$ floating point operations (flops) and storage of the order of $n^2$. These become prohibitive when $n$ is large and $\bC(\btheta)$ has no exploitable structure. 

Broadly speaking, modeling large spatial datasets proceeds from either exploiting ``low-rank'' models or using sparsity. The former
attempts to construct spatial processes on a lower-dimensional subspace \citep[see, e.g.,][]{hig01, kam03, stein07,
stein08, ban08,  cres08, cra08, rasm08, fin09} by regressing the original (\emph{parent}) process on its realizations over a
smaller set of $r << n$ locations (``knots'' or ``centers''). 
The algorithmic cost for model fitting typically decreases from $O(n^3)$ to $O(nr^2+r^3)\approx O(nr^2)$ flops since $n >> r$. However, when $n$ is large, empirical investigations suggest that $r$ must be fairly large to adequately approximate the parent process and the $nr^2$ flops becomes exorbitant (see Section~\ref{Sec: Simulation_Experiments}).
Furthermore, low rank models perform poorly when neighboring observations are strongly correlated and the spatial signal dominates the noise \citep{stein13}. Although bias-adjusted low-rank models tend to perform better \citep{fin09,ban10,sang12}, they increase the computational burden.

Sparse methods include covariance tapering \citep[see, e.g.,][]{fur06,kauf08,du09,shabytaper}, which introduces sparsity in $\bC(\btheta)$ using compactly supported covariance functions. This is effective for parameter estimation and  interpolation of the response (``kriging''), but it has not been fully developed or explored for more general inference on residual or latent processes. Introducing sparsity in $\bC(\btheta)^{-1}$ is prevalent in approximating Gaussian process likelihoods using Markov random fields \citep[e.g.,][]{rueheld04}, products of lower dimensional conditional distributions \citep{ve88,ve92,stein04}, or composite likelihoods \citep[e.g.,][]{bevilacqua14,eidsvik14}. However, unlike low rank processes, these do not, necessarily, extend to new random variables at arbitrary locations. There may not be a corresponding process, which restricts inference to the estimation of spatial covariance parameters. Spatial prediction (``kriging'') at arbitrary locations proceeds by imputing estimates into an interpolator derived from a different process model. This may not reflect accurate estimates of predictive uncertainty and is undesirable.


Our intended inferential contribution is to offer substantial scalability for fully process-based inference on underlying, perhaps completely unobserved, spatial processes. Moving from finite-dimensional sparse likelihoods to sparsity-inducing spatial processes can be complicated. We first introduce sparsity in finite-dimensional probability models using specified neighbor sets constructed from directed acyclic graphs. We use these sets to extend these finite-dimensional models to a valid spatial process over uncountable sets. We call this process a \emph{Nearest-Neighbor Gaussian Process} (NNGP). Its finite-dimensional realizations have sparse precision matrices available in closed form. While sparsity has been effectively exploited by \cite{ve88,stein04,emory09,gram14,gram14p} and \cite{stroud14} for approximating expensive likelihoods cheaply, a fully process-based modeling and inferential framework has, hitherto, proven elusive. The NNGP fills this gap and enriches the inferential capabilities of existing methods by subsuming estimation of model parameters, prediction of outcomes and interpolation of underlying processes into one highly scalable unifying framework.

To demonstrate its full inferential capabilities, we deploy the NNGP as a sparsity-inducing prior for spatial processes in a Bayesian framework. Unlike low rank processes, the NNGP always specifies non-degenerate finite dimensional distributions making it a legitimate proper prior for random fields and is applicable to any class of distributions that support a spatial stochastic process. It can, therefore, model an underlying process that is never actually observed. The modeling provides structured dependence for random effects, e.g. intercepts or coefficients, at a second stage of specification where the first stage need not be Gaussian. We cast a multivariate NNGP within a versatile spatially-varying regression framework \citep{gel03, ban08} and conveniently obtain entire posteriors for all model parameters as well as for the spatial processes at both observed and unobserved locations. 
Using a forestry example, we show how the NNGP delivers process-based inference for spatially-varying regression models at a scale where even low-rank processes, let alone full Gaussian processes, are unimplementable even in high-performance computing environments. 
   
Here is a brief outline. Section~\ref{Sec: NNGP} formulates the NNGP using multivariate Gaussian processes. Section~\ref{Sec: Implementation} outlines Bayesian estimation and prediction within a very flexible hierarchical modeling setup. Section~{\ref{Sec: alt} discusses alternative NNGP models and algorithms. Section~\ref{Sec: Illustrations} presents simulation studies to highlight the inferential benefits of the NNGP and also analyzes forest biomass from a massive USDA dataset. Finally, Section~\ref{Sec: Conclusions} concludes the manuscript with a brief summary and pointers toward future work.

\section{Nearest-Neighbor Gaussian Process}\label{Sec: NNGP}
\subsection{Gaussian density on sparse directed acyclic graphs}\label{Sec: NNGD}
\noindent We will consider a $q$-variate spatial process over $\Re^d$. Let $\bw(\bs) \sim GP(\bzero, \bC(\cdot,\cdot \given \btheta))$ denote a zero-centered $q$-variate Gaussian process, where $\bw(\bs)\in \Re^{q}$ for all $\bs\in\calD\subseteq \Re^d$. The process is completely specified by a valid cross-covariance function $\bC(\cdot,\cdot \given \btheta)$, which maps a pair of locations $\bs$ and $\bt$ in $\calD\times \calD$ into a $q\times q$ real valued matrix $\bC(\bs,\bt)$ with entries $\mbox{cov}\{w_i(\bs),w_j(\bt)\}$. Here, $\btheta$ denotes the parameters associated with the cross-covariance function. Let ${\cal S} = \{\bs_1,\bs_2,\ldots,\bs_k\}$ be a fixed collection of distinct locations in $\calD$, which we call the \emph{reference set}.
So, $\bw_ \calS \sim N(\bzero, \bC_ \calS (\btheta))$, where $\bw_\calS = (\bw(\bs_1)',\bw(\bs_2)',\ldots,\bw(\bs_k)')'$ and
$\bC_ \calS (\btheta)$ is a positive definite $qk\times qk$ block matrix with $\bC(\bs_i,\bs_j)$ as its blocks. Henceforth, we write $\bC_ \calS (\btheta)$ as $\bC_ \calS$, the dependence on $\btheta$ being implicit, with similar notation for all spatial covariance matrices. 

The reference set $\calS$ need not coincide with or be a part of the observed locations, so $k$ need not equal $n$, although we later show that the observed locations are a convenient practical choice for $\calS$. When $k$ is large, parameter estimation becomes computationally cumbersome, perhaps even unfeasible, because it entails the inverse and determinant of $\tildebC_\calS$. Here, we benefit from expressing the joint density of $\bw_\calS$ as the product of conditional densities, i.e.,
\begin{equation}\label{eq:tel}
p(\bw_\calS)=p(\bw(\bs_1))\; p(\bw(\bs_2) \given \bw(\bs_1))\; \ldots \, p(\bw(\bs_k) \given \bw(\bs_{k-1}),\ldots,\bw(\bs_1))\; ,
\end{equation}
and replacing the larger conditioning sets on the right hand side of (\ref{eq:tel}) with smaller, carefully chosen, conditioning sets of size at most $m$, where $m \ll k$ \citep[see, e.g.,][]{ve88,stein04,gram14,gram14p}. 
So, for every $\bs_i \in \calS$, a smaller conditioning set $N(\bs_i) \subset \calS \setminus \{ \bs_i \}$ is used to construct
\begin{equation}\label{Eq: nngp_gen}
\tp(\bw_ \calS) = \prod_{i=1}^k p(\bw(\bs_i)\given \bw_{N(\bs_i)})\; ,
\end{equation}
where $\bw_{N(\bs_i)}$ is the vector formed by stacking the realizations of $\bw(\bs)$ over $N(\bs_i)$.

Let $N_{\calS}=\{N(\bs_i); i=1,2,\ldots,k\}$ be the collection of all conditioning sets over $\calS$. We can view the pair $\{\calS,N_ \calS\}$ as a directed graph $\calG$ with ${\cal S} = \{\bs_1,\bs_2,\ldots,\bs_k\}$ being the set of nodes and $N_{\calS}$ the set of directed edges. For every two nodes $\bs_i$ and $\bs_j$, we say $\bs_j$ is a directed neighbor of $\bs_i$ if there is a directed edge from $\bs_i$ to $\bs_j$. So, $N(\bs_i)$ denotes the set of directed neighbors of $\bs_i$ and is, henceforth, referred to as the ``neighbor set'' for $\bs_i$. A ``directed cycle'' in a directed graph is a chain of nodes $\bs_{i_1},\bs_{i_2},\ldots,\bs_{i_b}$ such that $\bs_{i_1}=\bs_{i_b}$ and there is a directed edge between $\bs_{i_j}$ and $\bs_{i_{j+1}}$ for every $j=1,2,\ldots,b-1$. A directed graph with no directed cycles is known as a `directed acyclic graph'.

If $\calG$ is a directed acyclic graph, then $\tilde p(\bw_\calS)$, as defined above, is a proper multivariate joint density (see Appendix~\ref{sec:dag} or \cite{lauritzen96} for a similar result). Starting from a joint multivariate density $p(\bw_\calS)$, we derive a new density $\tilde p(\bw_\calS)$ using a directed acyclic graph $\calG$. While this holds for any original density $p(\bw_\calS)$, it is especially useful in our context, where $p(\bw_\calS)$ is a multivariate Gaussian density and $\calG$ is sufficiently sparse. To be precise, let $\bC_{N(\bs_i)}$ be the 
covariance matrix of $\bw_{N(\bs_i)}$ and let $\bC_{\bs_i,N(\bs_i)}$ be the $q\times mq$ cross-covariance matrix between the random vectors $\bw(\bs_i)$ and $\bw_{N(\bs_i)}$. Standard distribution theory reveals 
\begin{equation}\label{eq:cl}
\tilde p(\bw_\calS) = \prod_{i=1}^k  N(\bw(\bs_i)\given \bB_{\bs_i}\bw_{N(\bs_i)}, \bF_{\bs_i})\;,
\end{equation}
where $\bB_{\bs_i} = \bC_{\bs_i,N(\bs_i)}\bC_{N(\bs_i)}^{-1}$ and $\bF_{\bs_i} = \bC(\bs_i,\bs_i ) - \bC_{\bs_i,N(\bs_i)}\bC_{N(\bs_i)}^{-1}\bC_{N(\bs_i),\bs_i}$.
Appendix~\ref{sec:sparse} shows that $\tp(\bw_\calS)$ in (\ref{eq:cl}) is a multivariate Gaussian density with covariance matrix $\tildebC_\calS$, which, obviously, is different from $\bC_\calS$. Furthermore, if $N(\bs_i)$ has at most $m$ members for each $\bs_i$ in $\calS$, where $m \ll k$, then $\tildebC_\calS^{-1}$  is sparse with at most $km(m+1)q^2/2$ non-zero entries. 
Thus, for a very general class of neighboring sets, $\tilde p(\bw_\calS)$ defined in (\ref{Eq: nngp_gen}) is the joint density of a multivariate Gaussian distribution with a sparse precision matrix. 

Turning to the neighbor sets, choosing $N(\bs_i)$ to be any subset of $\{\bs_1, \bs_2, \ldots, \bs_{i-1}\}$ ensures an acyclic $\calG$ and, hence, a valid probability density in (\ref{eq:cl}). Several special cases exist in likelihood approximation contexts. For example, \cite{ve88} and \cite{stroud14} specified $N(\bs_i)$ to be the $m$ nearest neighbors of $\bs_i$ among $\bs_1, \bs_2, \ldots, \bs_{i-1}$ with respect to Euclidean distance. 
\cite{stein04} considered nearest as well as farthest neighbors from $\{\bs_1, \bs_2, \ldots, \bs_{i-1}\}$. \cite{gram14} offer greater flexibility in choosing $N(\bs_i)$, but may require several approximations to be efficient. 

All of the above choices depend upon an ordering of the locations. Spatial locations are not ordered naturally, so one imposes order by, for example, ordering on one of the coordinates. Of course, any other function of the coordinates can be used to impose order. However, the aforementioned authors have cogently demonstrated that the choice of the ordering has no discernible impact on the approximation of (\ref{eq:tel}) by (\ref{eq:cl}). Our own simulation experiments (see Appendix \ref{sec:order}) concur with these findings; inference based upon $\tp(\bw_\calS)$ is extremely robust to the ordering of the locations. This is not entirely surprising. Clearly, whatever order we choose in (\ref{eq:tel}), $p(\bw_\calS)$ produces the full joint density. Note that we reduce (\ref{eq:tel}) to (\ref{Eq: nngp_gen}) based upon neighbor sets constructed with respect to the \emph{specific} ordering in (\ref{eq:tel}). A different ordering in (\ref{eq:tel}) will produce a different set of neighbors for (\ref{Eq: nngp_gen}). Since $\tp(\bw_\calS)$ ultimately relies upon the information borrowed from the neighbors, its effectiveness is often determined by the number of neighbors we specify and \emph{not} the specific ordering.          


In the following section, we will extend the density $\tp(\bw_\calS)$ to a legitimate spatial process. We remark that our subsequent development holds true for any choice of $N(\bs_i)$ that ensures an acyclic $\calG$. In general, identifying a ``best subset'' of $m$ locations for obtaining optimal predictions for $\bs_i$ is a non-convex optimization problem, which is difficult to implement and defeats our purpose of using smaller conditioning sets to ease computations. Nevertheless, we have found Vecchia's choice of $m$-nearest neighbors from $\{\bs_1, \bs_2, \ldots, \bs_{i-1}\}$ to be simple and to perform extremely well for a wide range of simulation experiments. In what ensues, this will be our choice for $N(\bs_i)$ and the corresponding density $\tp(\bw_\calS)$ will be referred to as the `nearest neighbor' density of $\bw_\calS$.

\subsection{Extension to a Gaussian Process}\label{Sec: NNSP}
\noindent 

Let $\bu$ be any location in $\calD$ outside $\calS$. Consistent with the definition of $N(\bs_i)$, let $N(\bu)$ be the set of
$m$-nearest neighbors of $\bu$ in $\calS$. Hence, for any finite set $\calU=\{\bu_1,\bu_2,\ldots,\bu_r\}$ such that $\calS\cap
\calU$ is empty, we define the nearest neighbor density of $\bw_\calU$ conditional on $\bw_\calS$ as
\begin{equation}\label{Eq: nngp_t}
\tp(\bw_ \calU \given \bw_ \calS) = \prod_{i=1}^r p(\bw(\bu_i) \given \bw_{N(\bu_i)})\; .
\end{equation}
This conditional density is akin to (\ref{Eq: nngp_gen}) except that all the neighbor sets are subsets of $\calS$. This ensures a proper conditional density. Indeed (\ref{Eq: nngp_gen}) and (\ref{Eq: nngp_t}) are sufficient to describe the joint density of \emph{any} finite set over the domain $\calD$. More precisely, if $\calV=\{\bv_1,\bv_2,\ldots,\bv_n\}$ is \emph{any} finite subset in $\calD$, then, using (\ref{Eq: nngp_t}) we obtain the density of $\bw_\calV$ as,
\begin{equation}\label{Eq: nngp_finite}
\tp(\bw_ \calV) = \int  \tp(\bw_\calU \given \bw_ \calS ) \; \tp(\bw_ \calS) \prod_{\{\bs_i \in \calS \setminus \calV \}} d(\bw(\bs_i))\; \mbox{ where } \calU=\calV \setminus \calS\; .
\end{equation}
If $\calU$ is empty, then (\ref{Eq: nngp_t}) implies that $\tp(\bw_ \calU \given \bw_ \calS) =1$ in (\ref{Eq: nngp_finite}). If
$\calS \setminus \calV$ is empty, then the integration in (\ref{Eq: nngp_finite}) is not needed.

These probability densities, defined on finite topologies, conform to Kolmogorov's consistency criteria and, hence, correspond to a valid spatial process over $\calD$ (Appendix~\ref{sec:kolcon}). So, given any original (parent) spatial process and any \emph{fixed} reference set $\calS$, we can construct a new process over the domain $\calD$ using a collection of neighbor sets in $\calS$. We refer to this process as the `nearest neighbor process' derived from the original parent process. If the parent process is $GP(\bzero,\bC(\cdot,\cdot \given \btheta))$, then  
\begin{equation}\label{Eq: wt}
\tp(\bw_\calU \given \bw_ \calS) = \prod_{i=1}^r N(\bw(\bu_i) \given \bB_{\bu_i}\bw_ {N(\bu_i)} , \bF_{\bu_i}) = N(\bB_\calU  \bw_ \calS, \bF_ \calU)
\end{equation}
for any finite set $\calU=\{\bu_1,\bu_2,\ldots,\bu_r\}$ in $\calD$ outside $\calS$, where $\bB_{\bu_i}$ and $\bF_{\bu_i}$ are defined analogous to (\ref{eq:cl}) based on the neighbor sets $N(\bu_i)$, $\bF_ \calU=diag(\bF_{\bu_1},\bF_{\bu_2},\ldots,\bF_{\bu_r})$ and $\bB_ \calU$ is a sparse $nq \times kq$ matrix with each row having at most $mq$ non-zero entries (see Appendix~\ref{sec:nngp}). 

For any finite set $\calV$ in $\calD$, $\tp(\bw_\calV)$ is the density of the realizations of a Gaussian Process over $\calV$ with cross covariance function
\begin{equation}\label{Eq: tildefunc}
 \tildebC(\bv_1,\bv_2; \btheta) = \left\{\begin{array}{l}
                                        \tildebC_{\bs_i,\bs_j}\;,\quad \mbox{if $\bv_1=\bs_i$ and $\bv_2=\bs_j$ are both in $\calS$,}\\
					\bB_{\bv_1}\tildebC_{N(\bv_2),\bs_j}\;\quad \mbox{if $\bv_1 \notin \calS$ and $\bv_2=\bs_j\in \calS$,}\\
					\bB_{\bv_1}\tildebC _{N(\bv_1),N(\bv_2)}\bB_{\bv_2}' + \delta_{(\bv_1=\bv_2)}\bF_{\bv_1} \; \quad \mbox{if $\bv_1$ and $\bv _2$ are not in $\calS$ }
                                         \end{array} \right.
\end{equation}
where $\bv_1$ and $\bv_2$ are any two locations in ${\cal D}$,  $\tildebC_{A,B}$ denotes submatrices of $\tildebC_\calS$ indexed by the locations in the sets $A$ and $B$, and $\delta_{(\bv_1=\bv_2)}$ is the Kronecker delta. Appendix~\ref{sec:nngp} also shows that $\tildebC(\bv_1, \bv_2 \given \btheta)$ is continuous for all pairs $(\bv_1,\bv_2)$ outside a set of Lebesgue measure zero.
This completes the construction of a well-defined \emph{Nearest Neighbor Gaussian Process},
$NNGP(\bzero,\tildebC(\cdot,\cdot \given \btheta))$, derived from a \emph{parent Gaussian process}, $GP(\bzero,\bC(\cdot,\cdot
\given \btheta))$. In the NNGP, the size of $\calS$, i.e., $k$, can be as large, or even larger than the size of the dataset. The reduction in computational complexity is achieved through sparsity of the NNGP precision matrices. Unlike low-rank processes, the NNGP is \emph{not} a degenerate process. It is a proper, sparsity-inducing Gaussian process, immediately available as a prior in hierarchical modeling, and, as we show in the next section, delivers massive computational benefits.


\section{Bayesian estimation and implementation}\label{Sec: Implementation}

\subsection{A hierarchical model}\label{Sec: model}
\noindent Consider a vector of $l$ dependent variables, say $\by(\bt)$, at location $\bt \in
{\cal D} \subseteq \Re^d$ in a spatially-varying regression model,
\begin{equation}\label{Eq: Multi_Spatial_Regression}
\by(\bt) = \bX(\bt)'\bbeta + \bZ(\bt)'\bw(\bt) + \bepsilon(\bt)\; ,
\end{equation}
where $\bX(\bt)'$ is the $l\times p$ matrix of fixed spatially-referenced predictors, $\bw(\bt)$ is a $q\times 1$ spatial
process forming the coefficients of the $l\times q$ fixed design matrix $\bZ(\bt)'$, and $\bepsilon(\bt)\stackrel{iid}{\sim}
N(\bzero,\bD)$ is an $l\times 1$ white noise process capturing measurement error or micro-scale variability with dispersion
matrix $\bD$, which we assume is diagonal with entries $\tau_j^2$, $j=1,2,\ldots,l$. The matrix $\bX(\bt)'$ is block diagonal
with $p=\sum_{i=1}^{l}p_i$, where the $1\times p_i$ vector $\bx_i(\bt)'$, including perhaps an intercept, is the $i$-th block
for each $i=1,2,\ldots,l$. The model in (\ref{Eq: Multi_Spatial_Regression}) subsumes several specific spatial models. For
instance, letting $q=l$ and $\bZ(\bt)'= \bI_{l\times l}$ leads to a multivariate spatial regression model where $\bw(\bt)$ acts
as a \emph{spatially-varying intercept}. On the other hand, we could envision all coefficients to be spatially-varying and set
$q=p$ with $\bZ(\bt)' = \bX(\bt)'$. 

For scalability, instead of a customary Gaussian process prior for $\bw(t)$ in (\ref{Eq: Multi_Spatial_Regression}), we
assume $\bw(\bt) \sim NNGP(\bzero, \tildebC(\cdot,\cdot\given \btheta))$ derived from the parent $GP(\bzero,\bC(\cdot,\cdot
\given\btheta))$. Any valid isotropic cross covariance function \citep[see, e.g., ][]{gelban10} can be used to construct
$\bC(\cdot,\cdot \given \btheta)$. To elucidate, let $\calT = \{\bt_1,\bt_2,\ldots,\bt_n\}$ be the set of locations where the outcomes
and predictors have been observed. This set may, but need not, intersect with the reference set $\calS=\{\bs_1,\bs_2,\ldots,\bs_k\}$ for the NNGP. Without loss of generality, we split up $\calT$ into $\calS^*$ and $\calU$, where $\calS^* = \calS \cap \calT = \{\bs_{i_1},\bs_{i_2},\ldots,\bs_{i_r}\} $ with $\bs_{i_j} = \bt_j$ for $j=1,2,\ldots,r$ and $\calU = \calT \setminus \calS = \{\bt_{r+1},\bt_{r+2},\ldots,\bt_{n}\}$. Since $\calS \cup \calT = \calS \cup \calU$, we can completely specify the realizations of the NNGP in terms of the realizations of the parent process over $\calS$ and $\calU$, hierarchically, as $\bw_ {\calU} \given \bw_\calS \sim N(\bB_ {\calU}\bw_ \calS, \bF_ {\calU})$ and $\bw_ \calS \sim N(\bzero,\tildebC_ \calS )$. For a full Bayesian specification, we further specify prior distributions on $\bbeta$, $\btheta$ and the $\tau_j^2$'s. For example, with customary prior specifications, we obtain the joint distribution
\begin{align}\label{Eq: Bayesian_Hierarchical_NNGP}
& p(\btheta) \times \prod_{j=1}^q IG(\tau_j^2\given a_{\tau_j},b_{\tau_j})\times N(\bbeta\given \bmu_{\beta}, \bV_{\beta}) \times N(\bw_ {\calU} \given \bB_ {\calU}\bw_ \calS, \bF_ {\calU})  \nonumber \\
 &\qquad \qquad \times N(\bw_ \calS \given \bzero, \tildebC_ \calS) \times \prod_{i=1}^n N(\by(\bt_i)\given \bX(\bt_i)'\bbeta + \bZ(\bt_i)'\bw(\bt_i), \bD)\; ,
\end{align}
where $p(\btheta)$ is the prior on $\btheta$ and $IG(\tau_j^2\given a_{\tau_j}, b_{\tau_j})$ denotes the Inverse-Gamma density.

\subsection{Estimation and prediction}\label{Sec: estpred}
\noindent To describe a Gibbs sampler for estimating (\ref{Eq: Bayesian_Hierarchical_NNGP}), we define $\by = (\by(\bt_1)',\by(\bt_2)',\ldots,\by(\bt_n)')'$, and $\bw$ and $\beps$ similarly. Also, we introduce $\bX = [\bX(\bt_1) : \bX(\bt_2) : \ldots : \bX(\bt_n)]'$, $\bZ = \mbox{diag}(\bZ(\bt_1)',\ldots,\bZ(\bt_n)')$, and $\bD_n=\mbox{Cov}(\beps)=\mbox{diag}(\bD,\ldots,\bD)$. The full conditional distribution for $\bbeta$ is $N(\bV^*_{\beta}\bmu^*_{\beta},\bV^*_{\bbeta})$, where $\bV^*_{\beta} = (\bV_{\beta}^{-1}+\bX'\bD_n^{-1}\bX)^{-1}$, $\bmu^*_ \beta = (\bV_{\beta}^{-1}\bmu_{\beta}+\bX'\bD_n^{-1}(\by-\bZ\bw))$. Inverse-Gamma priors for the $\taus_j$'s leads to conjugate full conditional distribution $IG(a_{\tau_j}+\frac {n}2,b_{\tau_j}+\frac {1}{2}(\by_{*j}-\bX_{*j}\bbeta-\bZ_{*j}\bw)'(\by_{*j}-\bX_{*j}\bbeta-\bZ_{*j}\bw)$ where $\by_{*j}$ refers to the $n\times 1$ vector containing the $j^{th}$ co-ordinates of the $\by(\bt_i)$'s, $\bX_{*j}$ and $\bZ_{*j}$ are the corresponding fixed and spatial effect covariate matrices respectively. For updating $\btheta$, we use a random walk Metropolis step with target density $p(\btheta)\times N(\bw_ \calS|\bzero,\tildebC_ \calS) \times N(\bw_ {\calU} \given \bB_ {\calU}\bw_ \calS, \bF_ {\calU})$, where
\begin{equation}\label{Eq: densityw}
\begin{array}{c}
 N(\bw_ \calS \given \bzero,\tildebC_ \calS ) = \prod_{i=1}^kN(\bw(\bs_i) \given \bB_{\bs_i}\bw_ {N(\bs_i)}, \bF_ {\bs_i}) \mbox{ and } \\
 N(\bw_ {\calU} \given \bB_ {\calU}\bw_ \calS, \bF_ {\calU}) = \prod_{i=r+1}^n N(\bw(\bt_i) \given \bB_{\bt_i}\bw_ {N(\bt_i)}, \bF_ {\bt_i})
\end{array}
\end{equation}
Each of the component densities under the product sign on the right hand side of (\ref{Eq: densityw}) can be evaluated without any $n$-dimensional matrix operations rendering the NNGP suitable for efficient Metropolis (Hastings) block updates for $\btheta$.

Since the components of $\bw_ {\calU} \given \bw_ \calS$ are independent, we can update $\bw(\bt_i)$ from its full conditional $N(\bV_{\bt_i}\bmu_{\bt_i} , \bV_{\bt_i} )$ for $i=r+1,r+2,\ldots,n$ where $\bV_{\bt_i}= \left( \bZ(\bt_i) \bD^{-1} \bZ(\bt_i)'+\bF_{\bt_i}^{-1} \right)^{-1}$ and $\bmu_{\bt_i} = \bZ(\bt_i)\bD^{-1}\left(\by(\bt_i)-\bX(\bt_i)'\bbeta\right) +\bF_{\bt_i}^{-1}\bB_{\bt_i}\bw_{N(\bt_i)} $. Finally, we update the components of $\bw_ \calS$ individually. For any two locations $\bs$ and $\bt$ in $\calD$, if $\bs \in N(\bt)$ and is the $l$-th component of $N(\bt)$, i.e., say $\bs = N(\bt)(l)$, then define $\bB_{\bt,\bs}$ as the $l\times l$ submatrix formed by  columns $(l-1)q + 1,(l-1)q + 2,\ldots,lq $ of $\bB_{\bt}$. Let $U(\bs_i)=\{\bt \in \calS \cup \calT \given \bs_i \in N(\bt) \}$ and for every $\bt \in U(\bs_i)$ define, $\ba_{\bt,\bs_i}=\bw(\bt)-\sum_{\bs \in N(\bt), \bs \neq \bs_i} \bB_{\bt,\bs} \bw(\bs)$. Then, for $i=1,2,\ldots,k$, we have the full conditional $\bw_ {\bs_i} \given \cdot \sim N(\bV_{\bs_i} \bmu_{\bs_i},\bV_{\bs_i})$ where $\bV_{\bs_i} = (In(\bs_i \in \calS^*) \bZ(\bs_i)\bD^{-1}\bZ(\bs_i)' + \bF_{\bs_i}^{-1} + \sum_{\bt \in U(\bs_i)} \bB_{\bt,\bs_i}' \bF_{\bt}^{-1}\bB_{\bt,\bs_i} )^{-1}$, $\bmu_{\bs_i} = In(\bs_i \in \calS^*)\bZ(\bs_i)\bD^{-1}(\by(\bs_i)-\bX(\bs_i)'\bbeta) + \bF_{\bs_i}^{-1}\bB_{\bs_i}\bw_{N(\bs_i)} + \sum_{\bt \in U(\bs_i)} \bB_{\bt,\bs_i}' \bF_{\bt}^{-1}\ba_{\bt,\bs_i}$ and $In(\cdot)$ denotes the indicator function. Hence, the $\bw$'s can also be updated without requiring storage or factorization of any $n\times n$ matrices.

Turning to predictions, let $\bt$ be a new location where we intend to predict $\by(\bt)$ given $\bX(\bt)$ and $\bZ(\bt)$. The
Gibbs sampler for estimation also generates the posterior samples $\bw_\calS \given \by$. So, if $\bt \in \calS$, then we simply get samples of $\by(\bt) \given \by$ from $N(\bX(\bt)'\bbeta+\bZ(\bt)'\bw(\bt),\bD)$. If $\bt$ is outside $\calS$, then we generate
samples of $\bw(\bt)$ from its full conditional, $N(\bV_{\bt}\bmu_{\bt},\bV_{\bt})$,  where $\bV_{\bt}= \left( \bZ(\bt) \bD^{-1} \bZ(\bt)'+\bF_{\bt}^{-1} \right)^{-1}$ and $\bmu_{\bt} = \bZ(\bt)\bD^{-1}\left(\by(\bt)-\bX(\bt)'\bbeta\right) +\bF_{\bt}^{-1}\bB_{\bt}\bw_{N(\bt)}$, and subsequently generate posterior samples of $\by(\bt) \given \by$ similar to the earlier case.

\subsection{Computational complexity}\label{Sec: compcost}
\noindent Implementing the NNGP model in Section~\ref{Sec: estpred} reveals that one entire pass of the Gibbs sampler can be
completed without any large matrix operations. The only difference between (\ref{Eq: Bayesian_Hierarchical_NNGP})  and a full
geostatistical hierarchical model is that the spatial process is modeled as an NNGP prior as opposed to a standard GP.
For comparisons, we offer rough estimates of the flop counts to generate $\btheta$ and $\bw$ per iteration of the sampler. We express the computational complexity only in terms of the sample size $n$, size of the reference set $k$ and the size of the neighbor sets $m$ as other dimensions are assumed to be small. For all locations, $\bt \in \calS \cup \calT$, $\bB_ \bt$ and $\bF_\bt$ can be calculated using $O(m^3)$ flops. So, from (\ref{Eq: densityw}) it is easy to see that $p(\btheta \given \cdot)$ can be calculated using $O((n+k)m^3)$ flops. All subsequent calculations to generate a set of posterior samples for $\bw$ and $\btheta$ require around $O((n+k) m^2)$ flops. 

So, the total flop counts is of the order $(n+k) m^3$ and is ,therefore, linear in the total number of locations in $\calS \cup \calT$. This ensures scalability of the NNGP
to large datasets. Compare this with a full GP model with a dense correlation matrix, which requires $O(n^3)$ flops for
updating $\bw$ in each iteration. Simulation results in Section~\ref{Sec: Simulation_Experiments} and Appendix~\ref{sec:ci} indicate that NNGP models with
usually very small values of $m$ ($\approx 10$) provides inference almost indistinguishable to full geostatistical models.
Therefore, for large $n$, this linear flop count is drastically less.
Also, linearity with respect to $k$ ensures a feasible implementation even for $k \approx n$.
This offers substantial improvement over low rank models where the computational cost is quadratic in the number of ``knots,'' limiting the size of the set of knots. Also, both the full geostatistical and the predictive process models
require storage of the $n \times n$ distance matrix, which can potentially exhaust storage resources for large datasets. An NNGP
model only requires the distance matrix between neighbors for every location, thereby storing $n+k$ small matrices, each of
order $m \times m$.
Hence, NNGP accrues
substantial computational benefits over existing methods for very large spatial datasets and may be the only feasible option for
fully model-based inference in certain cases, as seen in the forestry data example (Section \ref{sec:biomass}).

\subsection{Model comparison and choice of $\calS$ and $m$} \label{Sec: Choice}
\noindent As elaborated in Section~\ref{Sec: NNGP}, given any parent Gaussian process and \emph{any} fixed reference set of locations $\calS$, we can construct a valid NNGP. The resulting finite dimensional likelihoods of the NNGP depend upon the choice of the reference set $\calS$ and the size of each $N(\bs_i)$, i.e., $m$.
Choosing the reference set is similar to selecting the knots for a predictive process. Unlike the number of ``knots'' in low rank models, the number of points in $\calS$ do not thwart computational scalability. From Section~\ref{Sec: compcost}, we observe that the flop count in an NNGP model only increases linearly with the size of $\calS$.
Hence, the number of locations in $\calS$ can, in theory, be large and this provides a lot of flexibility in choosing $\calS$.

Points over a grid across the entire domain seem to be a plausible choice for $\calS$. For example, we can construct a large
$\calS$ using a dense grid to improve performance without adversely affecting computational costs. Another, perhaps even
simpler, option for large datasets is to simply fix $\calS=\calT$, the set of observed locations. Since the NNGP is a legitimate
process for any fixed $\calS$, this choice is legitimate and it reduces computational costs even further by avoiding additional
sampling of $\bw_{\calU}$ in the Gibbs sampler. Our empirical investigations (see Section \ref{Sec: Simulation_Experiments})
reveal that choosing $\calS=\calT$ deliver inference almost indistinguishable from choosing $\calS$ to be a grid over the domain for large datasets.

\cite{stein04} and \cite{eidsvik14} proposed using a sandwich variance estimator for evaluating the inferential abilities of neighbor-based pseudo-likelihoods. \cite{shabyofs} developed a post sampling sandwich variance adjustment for posterior credible intervals of the parameters for quasi-Bayesian approaches using pseudo-likelihoods. However, all these adjustments concede accrual of additional computational costs. Also, the asymptotic results used to obtain the sandwich variance estimators are based on assumptions which are hard to verify in spatial settings with irregularly placed data points. Moreover, we view the NNGP as an independent model for fitting the data and not as an approximation to the original GP. Hence, we refrain from such sandwich variance adjustments. Instead, we can simply use any standard model comparison metrics such as DIC \citep{spieg02}, GPD \citep{gelf98} or RMSPE(RMSECV) \citep{rmspe02} to compare the performance of NNGP and any other candidate model. The same model comparison metrics are also used  for selecting $m$. However, as we illustrate later in Section \ref{Sec: Simulation_Experiments}, usually a small value of $m$ between $10$ to $15$ produces performance at par with the full geostatistical model. While larger $m$ may be beneficial for massive datasets, perhaps under a different design scheme, it is still going to be much smaller than the number of knots required in low rank models (see Section~\ref{Sec: Simulation_Experiments}).

\section{Alternate NNGP models and algorithms}\label{Sec: alt}
\subsection{Block update of $\bw_\calS$ using sparse Cholesky}
\noindent
The Gibbs' sampling algorithm detailed in Section \ref{Sec: estpred} is extremely efficient for large datasets with linear flop counts per iteration. However, it can sometimes experience slow convergence issues due to sequential updating of the elements in $\bw_\calS$. 
An alternative to sequential updating is to perform block updates of $\bw_\calS$. We choose $\calS=\calT$ so that $\bs_i=\bt_i$ for all $i=1,2,\ldots,n$ and we denote $\bw_\calS=\bw_\calT$ by $\bw$. Then,
\begin{equation}\label{Eq: Varw}
\bw|\cdot \sim N(\bV_\calS\bZ'\bD_n^{-1}(\by-\bX\bbeta),\bV_\calS)\;,\; \mbox{ where }\; \bV_\calS=(\bZ'\bD_n^{-1}\bZ+\tilde\bC_\calS^{-1})^{-1}\; .
\end{equation}
Recall that $\tildebC_\calS^{-1}$ is sparse. Since $\bZ$ and $\bD_n$ are block diagonal, $\bV_\calS^{-1}$ retains the sparsity of $\tildebC_\calS^{-1}$. So, a sparse Cholesky factorization of $\bV_\calS^{-1}$ will efficiently produce the Cholesky factors of $\bV_\calS$. This will facilitate block updating of $\bw$ in the Gibbs sampler.

\subsection{NNGP models for the response}\label{Sec: NNGP_Response}
\noindent Another possible approach involves NNGP models for the response $\by(\bs)$. If $\bw(\bs)$ is a Gaussian Process, then so is
$\by(\bs)= \bZ(\bs)'\bw(\bs)+\beps$ (without loss of generality we assume $\beta=\bzero$). One can directly use the NNGP
specification for $\by(\bs)$ instead of $\bw(\bs)$. That is, we derive $\by(\bs) \sim NNGP(\bzero,\tilde \bSigma (\cdot,\cdot))$
from the parent Gaussian process $GP(\bzero,\bSigma(\cdot,\cdot \given \btheta))$. The Gibbs sampler analogous to Section
\ref{Sec: Implementation} now enjoys the additional advantage of avoiding full conditionals for $\bw$. This results in a
Bayesian analogue for \cite{ve88} and \cite{stein04} 
but precludes inference on the spatial residual surface $\bw(\bs)$. Modeling $\bw(\bs)$ provides additional insight into residual spatial contours and is often important in identifying lurking covariates or eliciting unexplained spatial patterns. \cite{ve92} used the nearest neighbor approximation on a spatial model for observations ($\by$) with independent measurement error (nuggets) in addition to the usual spatial component ($\bw$). However, it may not be possible to recover $\bw$ using this approach. For example, a univariate stationary process $\by(\bs)$ with a nugget effect can be decomposed as $\by(\bs) = \bw(\bs)+\beps(\bs)$ (letting $\bbeta =\bzero)$ for some $\bw(\bs) \sim GP(\bzero,\bC(\cdot,\cdot \given \btheta)) $ and white noise process $\beps(\bs)$. If $\by=\bw+\beps$, where $\bw \sim N(\bzero,\bC)$, $\beps \sim N(\bzero,\taus\bI_n)$, then $\mbox{Cov}(\by) =\bC+\taus\bI = \bSigma$, all eigenvalues of $\bSigma$ are greater than $\taus$ and $\mbox{Cov}(\bw\given\by) = \taus\bI_n-\tau^4\bSigma^{-1}$.
For $\by(\bs) \sim NNGP(\bzero,\tilde \bSigma (\cdot,\cdot))$, however, the eigenvalues of $\tilde{\bSigma}$ may be less than $\tau^2$, so $\taus\bI_n-\tau^4\tilde \bSigma^{-1}$ need not be positive definite for every $\tau^2>0$ and $p(\bw \given \by)$ is no longer well-defined.

A different model is obtained by using an NNGP prior for $\bw$, as in (\ref{Eq: Bayesian_Hierarchical_NNGP}), and then
integrating out $\bw$. The resulting likelihood is $N(\by\given \bX\bbeta,\bSigma_y)$, where $\bSigma_y = \bZ\tildebC_\calS\bZ' + \bD_n$ and the Bayesian specification is completed using priors on $\bbeta$, $\taus_j$'s and $\btheta$ as in (\ref{Eq: Bayesian_Hierarchical_NNGP}).
This model drastically reduces the number of variables in the Gibbs sampler, while preserving the nugget effect in the parent
model. We can generate the full conditionals for the
parameters in the marginalized model as follows: $\bbeta |\;\by,\bphi \sim N( (\bV_ \beta ^ {-1}+\bX'\bSigma_y^{-1}\bX)^{-1}(\bV
_\beta ^{-1} \bmu_ \beta + \bX'\bSigma_y^{-1}\by) \, , \,  (\bV_ \beta ^ {-1}+\bX'\bSigma_y^{-1}\bX)^{-1})$. It is difficult to
factor out $\taus_j$'s from $\bSigma_y^{-1}$, so conjugacy is lost with respect to any standard prior. Metropolis block updates for $\btheta$ are feasible for any tractable prior $p(\btheta)$. This involves computing $\bX'\bSigma_y^{-1}\bX$, $\bX'\bSigma_y^{-1}\by$ and $(\by-\bX\bbeta)'\bSigma_y^{-1}(\by-\bX\bbeta)$. Since $\bSigma_y^{-1}=\bD_n^{-1}-\bD_n^{-1}\bZ(\tilde \bC_\calS^{-1} +\bZ'\bD_n^{-1}\bZ)^{-1}\bZ'\bD_n^{-1} = \bD_n^{-1} - \bD_n^{-1}\bZ\bV_\calS\bZ'\bD_n^{-1}$, where $\bV_\calS$ is given by (\ref{Eq: Varw}), a sparse Cholesky factorization of $\bV_\calS^{-1}$ will be beneficial. We draw posterior samples for $\bw$ from $p(\bw\given \by) = \int p(\bw\given \btheta,\bbeta,\{\taus_j\},\by)p(\btheta,\bbeta,\{\taus_j\}\given\by)$ using composition sampling---we draw $\bw^{(g)}$ from $p(\bw\given \btheta^{(g)},\bbeta^{(g)},\{{\taus_j}^{(g)}\},\by)$ one-for-one for each sampled parameter.

Using block updates for $\bw_\calS$ in (\ref{Eq: Bayesian_Hierarchical_NNGP}) and fitting the marginalized version of (\ref{Eq:
Bayesian_Hierarchical_NNGP}) both require an efficient sparse Cholesky solver for $\bV_\calS^{-1}$. Note that computational expenses for most sparse Cholesky algorithms depend on the precise nature of the sparse structure (mostly on the bandwidth) of $\tildebC_\calS^{-1}$ \citep[see, e.g.][]{davis06}. The number of flops required for Gibbs sampling and prediction in this marginalized model depends upon the sparse structure of $\tildebC_\calS^{-1}$ and may, sometimes, heavily exceed the linear usage achieved by the unmarginalized model with individual updates for $\bw_i$. Therefore, a prudent choice of the precise fitting algorithms should be based on the sparsity structure of $\tildebC_\calS^{-1}$ for the given dataset.

\subsection{Spatiotemporal and GLM versions}\label{Sec: Spatiotemporal_GLM_Versions}
In spatiotemporal settings where we seek spatial interpolation at discrete time-points (e.g., weekly, monthly or yearly data), we write the response (possibly vector-valued) as $\by_t(\bs)$ and the random effects as $\bw_t(\bs)$. One could, for example, envision that the data arise as a time series of spatial processes, i.e., there is a time series at each location. An alternative scenario is cross-sectional data being collected at a set of locations associated with each time point and these locations can differ from time point to time point. Desired inference includes spatial interpolation for each time point. Spatial dynamic models incorporating the NNGP are easily formulated as below:
\begin{equation}\label{eq:model}
\begin{array}{c}
\by_t(\bs) = \bX_t(\bs)'\bbeta_t + \bu_t(s) + \beps_t(\bs), \; \beps_t(\bs) \overset{iid}{\sim} N(0,D) \\
\bbeta_t = \bbeta_{t-1}+\boeta_t, \; \boeta_t \overset{iid}{\sim} N(0,\bSigma_\eta), \; \bbeta_0 \sim N(\bm_0, \bSigma_0) \\
\bu_t(\bs)=\bu_{t-1}(\bs)+\bw_t(\bs), \; \bw_t(\bs) \overset{ind}{\sim} NNGP(\bzero,\tildebC(\cdot,\cdot \given \btheta_t))\; . 
\end{array}
\end{equation}
Thus, one retains exactly the same structure of process-based spatial dynamic models, e.g., as in \cite{gel05}, and simply replaces the independent Gaussian process priors for $\bw_t(\bs)$ with independent NNGP's to achieve computational tractability.

The above is illustrative of how attractive and extremely convenient the NNGP is for model building. One simply writes down the parent model and subsequently replaces the full GP with an NNGP. Being a well-defined process, the NNGP ensures a valid spatial dynamic model. Similarly NNGP versions of dynamic spatiotemporal Kalman-filtering \citep[as, e.g., in]{wiklecres1999} can be constructed.   

Handling non-Gaussian (e.g., binary or count) data is also straightforward using spatial generalized linear models (GLM's)  \citep{dig98,lin00,kam03,ban14}. Here, the NNGP provides structured dependence for random effects at the second stage. First, we replace $\mbox{E}[\by(\bt)]$ in (\ref{Eq: Multi_Spatial_Regression}) with $g(E(\by(\bt)))$ where $g(\cdot)$ is a suitable link function such that $\boeta(\bt)=g(E(\by(\bt)))=\bX(\bt)'\bbeta+\bZ(\bt)'\bw(\bt)$. In the second stage, we model the $\bw(\bt)$ as an NNGP. The benefits of the algorithms in Sections~\ref{Sec: estpred}~and~\ref{Sec: compcost} still hold, but some of the alternative algorithms in Section~\ref{Sec: alt} may not apply. For example, we do obtain tractable marginalized likelihoods by integrating out the spatial effects.    

\section{Illustrations}\label{Sec: Illustrations}
\noindent We conduct simulation experiments and analyze a large forestry dataset. Additional simulation experiments are detailed in Appendices~\ref{sec:order} through \ref{sec:wave}. Posterior inference for subsequent analysis were based upon three chains of 25000 iterations (with a burn-in of 5000 iterations). All the samplers were programmed in \texttt{C++} and leveraged Intel’s Math Kernel Library's (MKL) threaded \texttt{BLAS} and \texttt{LAPACK} routines for matrix computations on a Linux workstation with 384 GB of RAM and two Intel Nehalem quad-Xeon processors. 

\subsection{Simulation experiment}\label{Sec: Simulation_Experiments}
\noindent We generated observations using $2500$ locations within a unit square domain from the model (\ref{Eq: Multi_Spatial_Regression})
with $q=l=1$ (univariate outcome), $p=2$, $\bZ(\bt)'=1$ (scalar), the spatial covariance matrix $\bC(\btheta) =
\sigma^2\bR(\bphi)$, where $\bR(\bphi)$ is a $n\times n$ correlation matrix, and $\bD=\tau^2$ (scalar). The model included an
intercept and a covariate $\bx_1$ drawn from $N(0,1)$. The $(i,j)$th element of $\bR(\bphi)$ was calculated using the Mat\'ern
function
\begin{equation}\label{matern}
\rho(\bt_i, \bt_j; \bphi) = \frac{1}{2^{\nu - 1}\Gamma(\nu)}(||\bt_i-\bt_j||\phi)^{\nu}\calK_{\nu}(||\bt_i-\bt_j||\phi);\; \phi > 0, \nu >0,
\end{equation}
where $||\bt_i-\bt_j||$ is the Euclidean distance between locations $\bt_i$ and $\bt_j$, $\bphi=(\phi, \nu)$ with $\phi$
controlling the decay in spatial correlation and $\nu$ controlling the process smoothness, $\Gamma$ is the usual Gamma function
while $\calK_{\nu}$ is a modified Bessel function of the second kind with order $\nu$ \citep{stein99}
Evaluating the Gamma function for each matrix element within each iteration requires
substantial computing time and can obscure differences in sampler run times; hence, we fixed $\nu$ at 0.5 which reduces
(\ref{matern}) to the exponential correlation function.  The first column in Table~\ref{tab:uni-params} gives the \emph{true}
values used to generate the responses. Figure~\ref{uni-w-obs} illustrates the $w(\bt)$ surface interpolated over the domain.

\begin{table}[h!]
\centering
\caption{Univariate synthetic data analysis parameter estimates and computing time in minutes for NNGP and full GP models. Parameter posterior summary 50 (2.5, 97.5) percentiles. }

{\tiny
\begin{tabular}{cccccc}
  \hline
           &      &\multicolumn{2}{c}{NNGP ($\calS\neq \calT$)} &\multicolumn{2}{c}{NNGP ($\calS=\calT$)}\\
           & True & $m=10, k=2000$ & $m=20, k=2000$           &$m=10$ & $m=20$           \\
  \hline
  $\beta_{0}$ &1&0.99 (0.71, 1.48)&1.02 (0.73, 1.49)&1.00 (0.62, 1.31)&1.03 (0.65, 1.34)\\
  $\beta_{1}$  &5&5.00 (4.98, 5.03)&5.01 (4.98, 5.03)&5.01 (4.99, 5.03)&5.01 (4.99, 5.03)\\
  $\sigma^2$  &1&1.09 (0.89, 1.49)&1.04 (0.85, 1.40)&0.96 (0.78, 1.23)&0.94 (0.77, 1.20)\\
  $\tau^2$  &0.1&0.07 (0.04, 0.10)&0.07 (0.04, 0.10)&0.10 (0.08, 0.13)&0.10 (0.08, 0.13)\\
  $\phi$  &12&11.81 (8.18, 15.02)&12.21 (8.83, 15.62)&12.93 (9.70, 16.77)&13.36 (9.99, 17.15)\\
  \hline
  p$_D$  &--&1491.08&1478.61&1243.32&1249.57\\
  DIC &--&1856.85&1901.57&2390.65&2377.51\\
  G &--&33.67&35.68&77.84&76.40\\
 P &--&253.03&259.13&340.40&337.88\\
 D &--&286.70&294.82&418.24&414.28\\
  \hline
  RMSPE  &--&1.22&1.22&1.2&1.2\\
  95\% CI cover \% &--&97.2&97.2&97.6&97.6\\
  95\% CI width &--&2.19&2.18&2.13&2.12\\
  \hline
  Time &--&14.2&47.08&9.98&33.5\\
  \hline
  &&&&&\\
   \cline{1-4}
 
           &      &Predictive Process &Full\\
           & True &64 knots          & Gaussian Process \\
    \cline{1-4}
  $\beta_{0}$ &1&1.30 (0.54, 2.03)&1.03 (0.69, 1.34)\\
  $\beta_{1}$  &5&5.03 (4.99, 5.06)&5.01 (4.99, 5.03)\\
  $\sigma^2$  &1&1.29 (0.96, 2.00)&0.94 (0.76, 1.23)\\
  $\tau^2$  &0.1&0.08 (0.04, 0.13)&0.10 (0.08, 0.12)\\
  $\phi$  &12&\textbf{5.61 (3.48, 8.09)}&13.52 (9.92, 17.50)\\
    \cline{1-4}
  p$_D$  &--&1258.27&1260.68\\
  DIC &--&13677.97&2364.80\\
  G &--&1075.63&74.80\\
 P &--&200.39&333.27\\
 D &--&1276.03&408.08\\
    \cline{1-4}
  RMSPE  &--&1.68&1.2\\
  95\% CI cover \% &--&95.6&97.6\\
  95\% CI width &--&2.97&2.12\\
    \cline{1-4}
  Time &--&43.36&560.31\\
    \cline{1-4}

\end{tabular}
}
\label{tab:uni-params}
\end{table}

We then estimated the following models from the full data: $i$) the full Gaussian Process (\emph{Full GP}); $ii$) the NNGP with
$m=\{1,2,\ldots,25\}$ for $\calS\neq \calT$ and $\calS=\calT$, and; $iii$) a Gaussian Predictive Process (GPP) model
\citep{ban08} with 64 knots placed on a grid over the domain. For the NNGP with $\calS\neq \calT$ we considered 2000 randomly
placed reference locations within the domain. The 64 knot GPP was chosen because its computing time was comparable to that of
NNGP models. We used an efficient marginalized sampling algorithm for the Full GP and GPP models as implemented in the
\texttt{spBayes} package in \texttt{R} \citep{fin13}. All the models were trained using 2000 of the 2500 observed locations, while the
remaining 500 observations were withheld to assess predictive performance.

For all models, the intercept and slope regression parameters, $\beta_0$ and $\beta_1$, were given \emph{flat} prior
distributions. The variance components $\sigma^2$ and $\tau^2$ were assigned inverse-Gamma $IG(2,1)$ and $IG(2,0.1)$ priors,
respectively, and the spatial decay $\phi$ received a uniform prior $U(3, 30)$, which corresponds to a spatial range between approximately 0.1 and 1 units.

\begin{figure}[t!]
\begin{center}
\includegraphics[width=15cm]{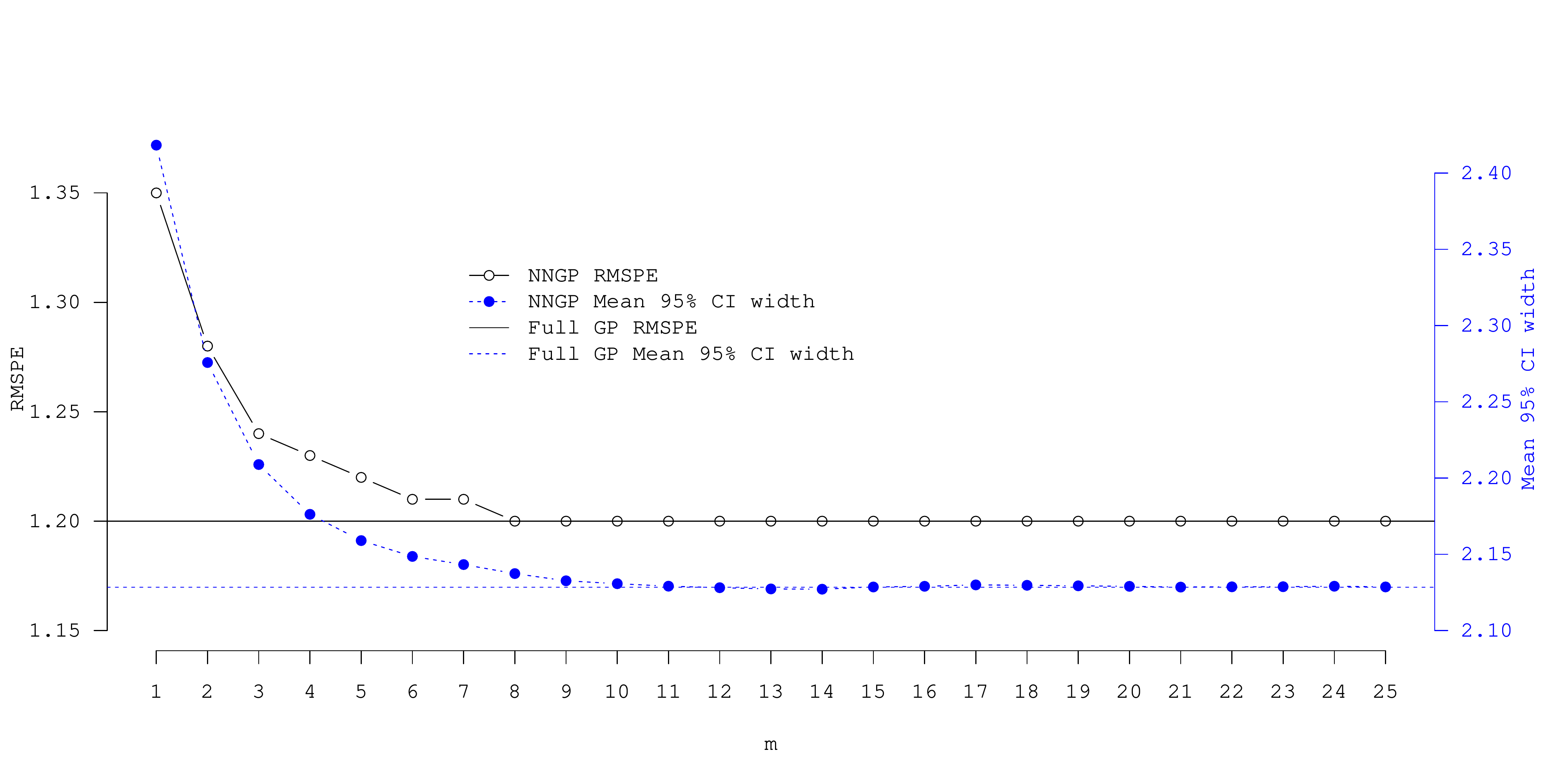}\label{uni-nn-pred}
\end{center}
\vspace{-1cm}
\caption{Choice of $m$ in NNGP models: Out-of-sample Root Mean Squared Prediction Error (RMSPE) and mean width between the upper and lower 95\% posterior predictive credible intervals for a range of $m$ for the univariate synthetic data analysis}\label{Fig: uni-nn-pred}
\end{figure}

Parameter estimates and performance metrics for the NNGP (with $m=10$ and $m=20$), GPP, and the Full GP models are provided in
Table~\ref{tab:uni-params}. All model specifications produce similar posterior median and 95\% credible intervals estimates,
with the exception of $\phi$ in the 64 knot GPP model. Larger values of DIC and D suggest that the GPP model does not fit the
data as well as the NNGP and Full GP models. The NNGP $\calS=\calT$ models provide DIC, GPD scores that are comparable to 
those of the Full GP model. These fit metrics suggest the NNGP $\calS\neq \calT$ models provide better fit to the data than that
achieved by the full GP model which is probably due to overfitting caused by a very large reference set $\calS$. The last row in
Table~\ref{tab:uni-params} shows computing times in minutes for one chain of 25000 iterations reflecting on the enormous
computational gains of NNGP models over full GP model.  

Turning to out-of-sample predictions, the Full model's RMSPE and mean width between the upper and lower $95\%$
posterior predictive credible interval is 1.2 and 2.12, respectively. As seen in Figure~\ref{Fig: uni-nn-pred}, comparable
RMSPE and mean interval width for the NNGP $\calS=\calT$ model is achieved within $m\approx 10$. There are negligible difference
between the predictive performances of the NNGP $\calS\neq \calT$ and $\calS = \calT$ models. Both the NNGP and Full GP model
have better predictive performance than the Predictive Process models when the number of knots is small, e.g., 64. All models
showed appropriate 95\% credible interval coverage rates.

\begin{figure}[t]
\begin{center}
\subfigure[True $\bw$]{\includegraphics[width=5cm]{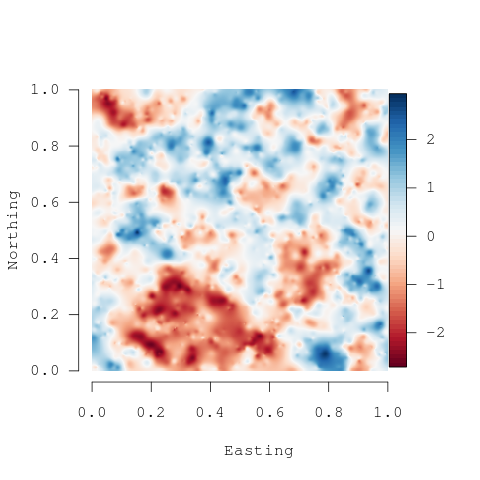}\label{uni-w-obs}}
\subfigure[Full GP]{\includegraphics[width=5cm]{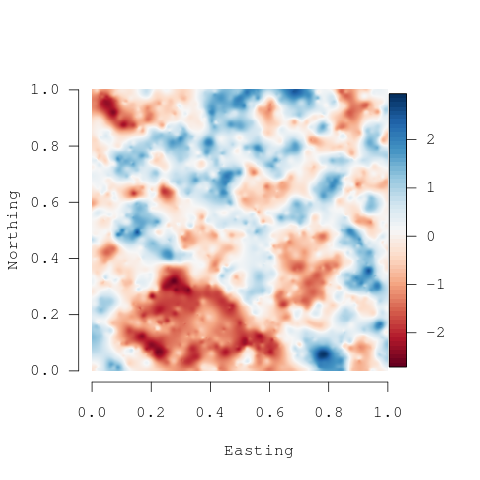}\label{uni-w-gs}}
\subfigure[GPP 64 knots]{\includegraphics[width=5cm]{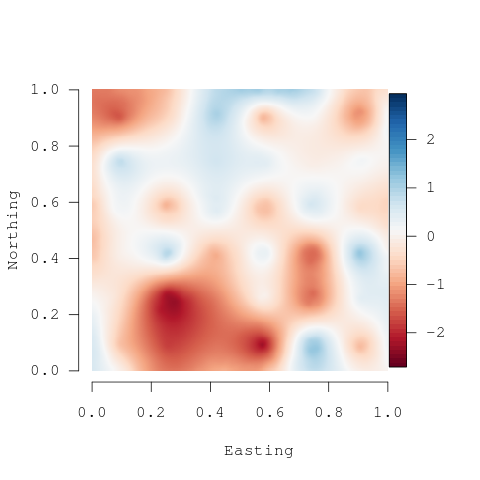}\label{uni-pp64-gs}}\\
\subfigure[NNGP ($\calS=\calT$) $m=10$]{\includegraphics[width=5cm]{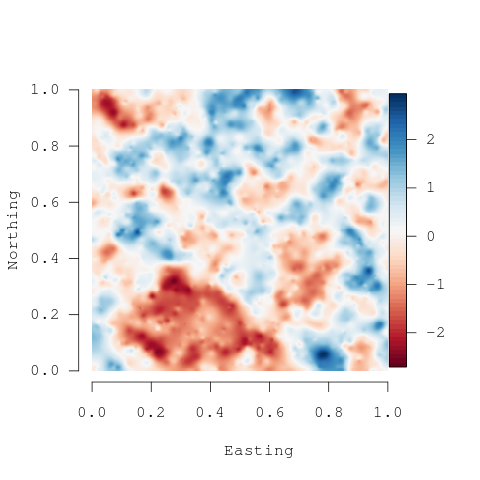}\label{uni-k10-gs}}
\subfigure[NNGP ($\calS=\calT$) $m=20$]{\includegraphics[width=5cm]{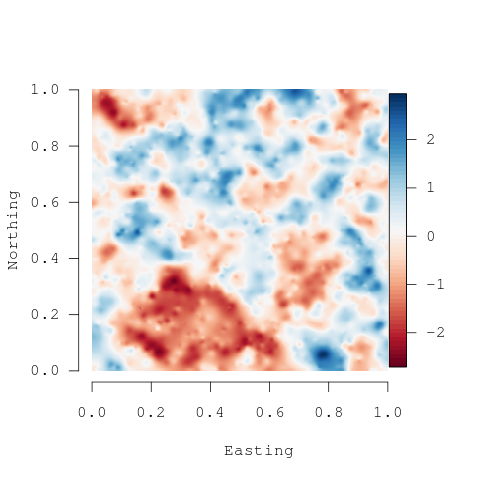}\label{uni-k20-gs}}
\subfigure[NNGP ($\calS \neq \calT$) $m=10$]{\includegraphics[width=5cm]{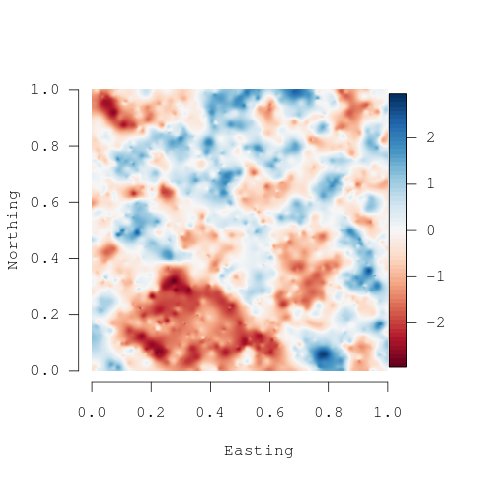}\label{uni-k10knots-gs}}
\end{center}
\caption{Univariate synthetic data analysis: Interpolated surfaces of the true spatial random effects and posterior median estimates for different models}
\label{fig:uni-w}
\end{figure}

Figures~\ref{fig:uni-w}(b-f) illustrate the posterior median estimates of the spatial random effects from the Full GP, NNGP
($\calS = \calT$) with $m=10$ and $m=20$, NNGP ($\calS \neq \calT$) with $m=10$ and GPP models. These surfaces can be compared
to the \emph{true} surface depicted in Figure~\ref{uni-w-obs}. This comparison shows: $i$) the NNGP models closely approximates
the true surface and that estimated by the Full GP model, and; $ii$) the reduced rank predictive process model based on 64 knots
greatly smooths over small-scale patterns. This last observation highlights one of the major criticisms of reduced rank models
\citet{stein13} and illustrates why these models often provide compromised predictive performance when the true surface has fine
spatial resolution details. Overall, we see the clear computational advantage of the NNGP over the Full GP model, and both
inferential and computational advantage over the GPP model.


\subsection{Forest biomass data analysis}\label{sec:biomass}
\noindent Information about the spatial distribution of forest biomass is needed to support global, regional, and local scale
decisions, including assessment of current carbon stock and flux, bio-feedstock for emerging bio-economies, and impact of
deforestation. In the United States, the Forest Inventory and Analysis (FIA) program of the USDA Forest Service collects the
data needed to support these assessments. The program has established field plot centers in permanent locations using a sampling
design that produces an equal probability sample \citep{patt05}. Field crews recorded stem measurements for all trees with
diameter at breast height (DBH); 1.37 m above the forest floor) of 12.7 cm or greater. Given these data, established allometric
equations were used to estimate each plot's forest biomass. For the subsequent analysis, plot biomass was scaled to metric tons
per ha then square root transformed. The transformation ensures that back transformation of subsequent predicted values have
support greater than zero and helps to meet basic regression models assumptions.

Figure~\ref{bio-data-fia} illustrates the georeferenced forest inventory data consisting of $114,371$ forested FIA plots
measured between 1999 and 2006 across the conterminous United States. The two blocks of missing observations in the Western and
Southwestern United States correspond to Wyoming and New Mexico, which have not yet released FIA data.
Figure~\ref{bio-data-bio-interp} shows a deterministic interpolation of forest biomass observed on the FIA plots. Dark blue
indicates high forest biomass, which is primarily seen in the Pacific Northwest, Western Coastal ranges, Eastern Appalachian
Mountains, and in portions of New England. In contrast, dark red indicates regions where climate or land use limit vegetation
growth.

A July 2006 Normalized Difference Vegetation Index (NDVI) image from the MODerate-resolution Imaging Spectroradiometer (MODIS);
\url{http://glcf.umd.edu/data/ndvi}) sensor was used as a single predictor. NDVI is calculated from the visible and
near-infrared light reflected by vegetation, and can be viewed as a measure of greenness. In this image,
Figure~\ref{bio-data-ndvi}, dark green corresponds to dense vegetation whereas brown identifies regions of sparse or no
vegetation, e.g., in the Southwest. NDVI is commonly used as a covariate in forest biomass regression models, see, for e.g.,
\citet{allo06}. Results from these and similar studies show a positive linear relationship between forest biomass and NDVI. The
strength of this relationship, however, varies by forest tree species composition, age, canopy structure, and level of
reflectance. We expect a space-varying relationship between biomass and NDVI, given tree species composition and disturbance
regimes generally exhibit strong spatial dependence across forested landscapes.

The $\sim$38 gigabytes of memory in our workstation was insufficient for storage of distance matrices required to fit a Full GP
or GPP model. Subsequently, we explore the relationship between forest biomass and NDVI using a non-spatial model,
a NNGP space-varying intercept (SVI) model (i.e., $q=l=1$ and $\bZ(\bt)=1$) in (\ref{Eq: Multi_Spatial_Regression}), and a NNGP
spatially-varying coefficients (SVC) regression model with $l=1$, $q=p=2$ and $\bZ(\bt)=\bX(\bt)$ in  (\ref{Eq:
Multi_Spatial_Regression}).
The reference sets for the NNGP models
were again the observed locations and $m$ was chosen to be $5$ or $10$. The parent process $\bw(\bt)$ is a bivariate Gaussian process with a isotropic
cross-covariance specification $\bC(\bt_i,\bt_j \given \btheta) = \bA\bGam(\bphi)\bA'$, where $\bA$ is $2\times 2$ lower-triangular
with positive diagonal elements, $\bGam$ is $2\times 2$ diagonal with $\rho(\bt_i,\bt_j; \bphi_b)$ (defined in (\ref{matern}))
as the $b^{th}$ diagonal entry, $b=1,2$ and $\bphi_b=(\phi_b,\nu_b)'$ \citep[see, e.g.,][]{gelban10}.

For all models, the intercept and slope regression parameters were given \emph{flat} prior distributions. The variance
components $\tau^2$ and $\sigma^2$ were assigned inverse-Gamma $IG(2,1)$ priors, the SVC model cross-covariance matrix $\bA\bA'$
followed an inverse-Wishart $IW(3,0.1)$, and the Mat\'ern spatial decay and smoothness parameters received uniform prior
supports $U(0.01, 3)$ and $U(0.1, 2)$, respectively. These prior distributions on $\phi$ and $\nu$ correspond to support between
approximately 0.5 and 537 km. Candidate models are assessed using the metrics described in Section~\ref{Sec: Choice}, inference drawn from mapped estimates of the regression coefficients, and out-of-sample prediction.

Parameter estimates and performance metrics for NNGP with $m=5$ are shown in Table~\ref{tab:bio-params}. The corresponding numbers for $m=10$ were similar. Relative to the spatial models, the non-spatial model has higher values of DIC and D which suggests
NDVI alone does not adequately capture the spatial structure of forest biomass. This observation is corroborated using a
variogram fit to the non-spatial model's residuals, Figure~\ref{bio-data-vario}. The variogram shows a nugget of $\sim$0.42,
partial sill of $\sim$0.05, and range of $\sim$150 km. This residual spatial dependence is apparent when we map the SVI model
spatial random effects as shown in Figure~\ref{bio-svi-w}. This map, and the estimate of a non-negligible spatial variance
$\sigma^2$ in Table~\ref{tab:bio-params}, suggests the addition of a spatial random effect was warranted and helps satisfy the
model assumption of uncorrelated residuals.

\begin{figure}[]
\begin{center}
\subfigure[Observed locations]{\includegraphics[width=8cm]{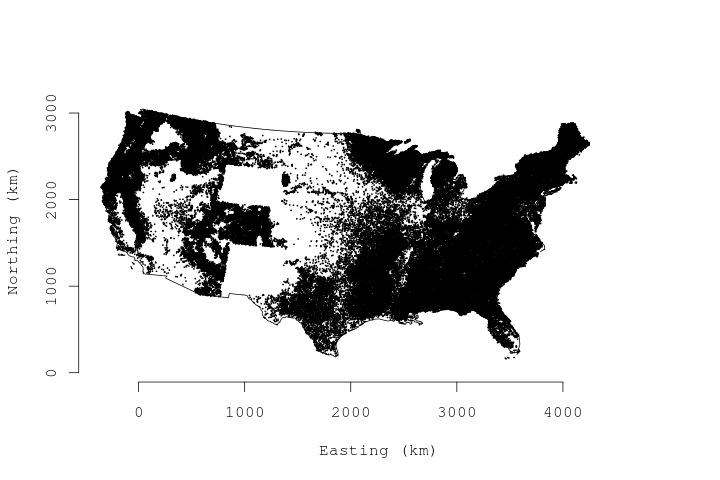}\label{bio-data-fia}}
\subfigure[Observed biomass]{\includegraphics[width=8cm]{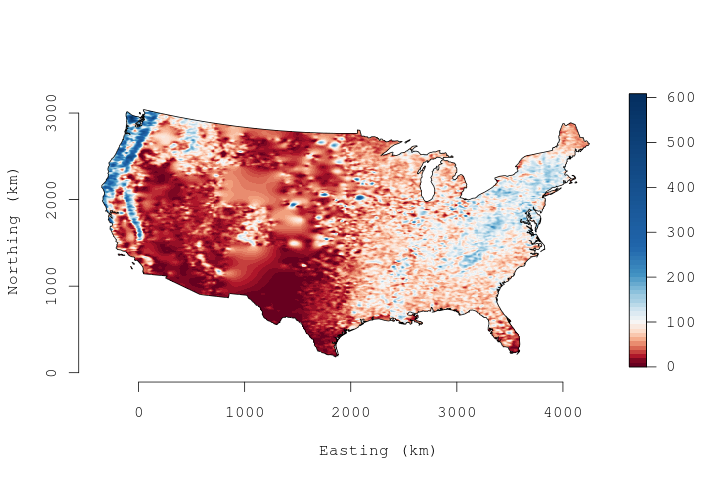}\label{bio-data-bio-interp}}\\
\subfigure[NDVI]{\includegraphics[width=8cm]{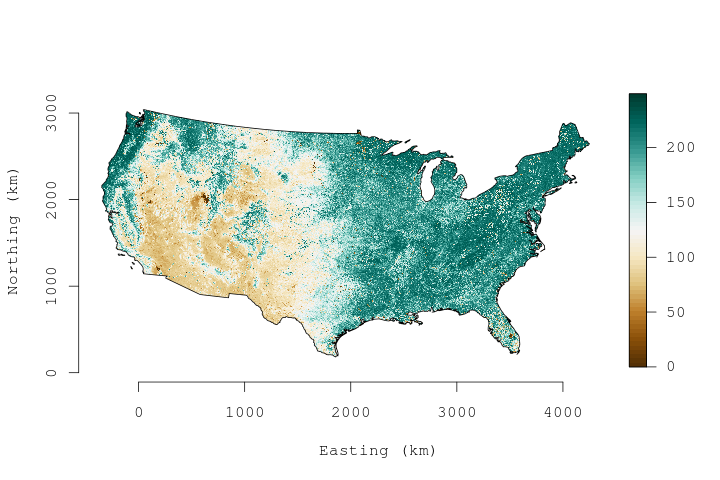}\label{bio-data-ndvi}}
\subfigure[Non-spatial model residuals]{\includegraphics[width=5cm]{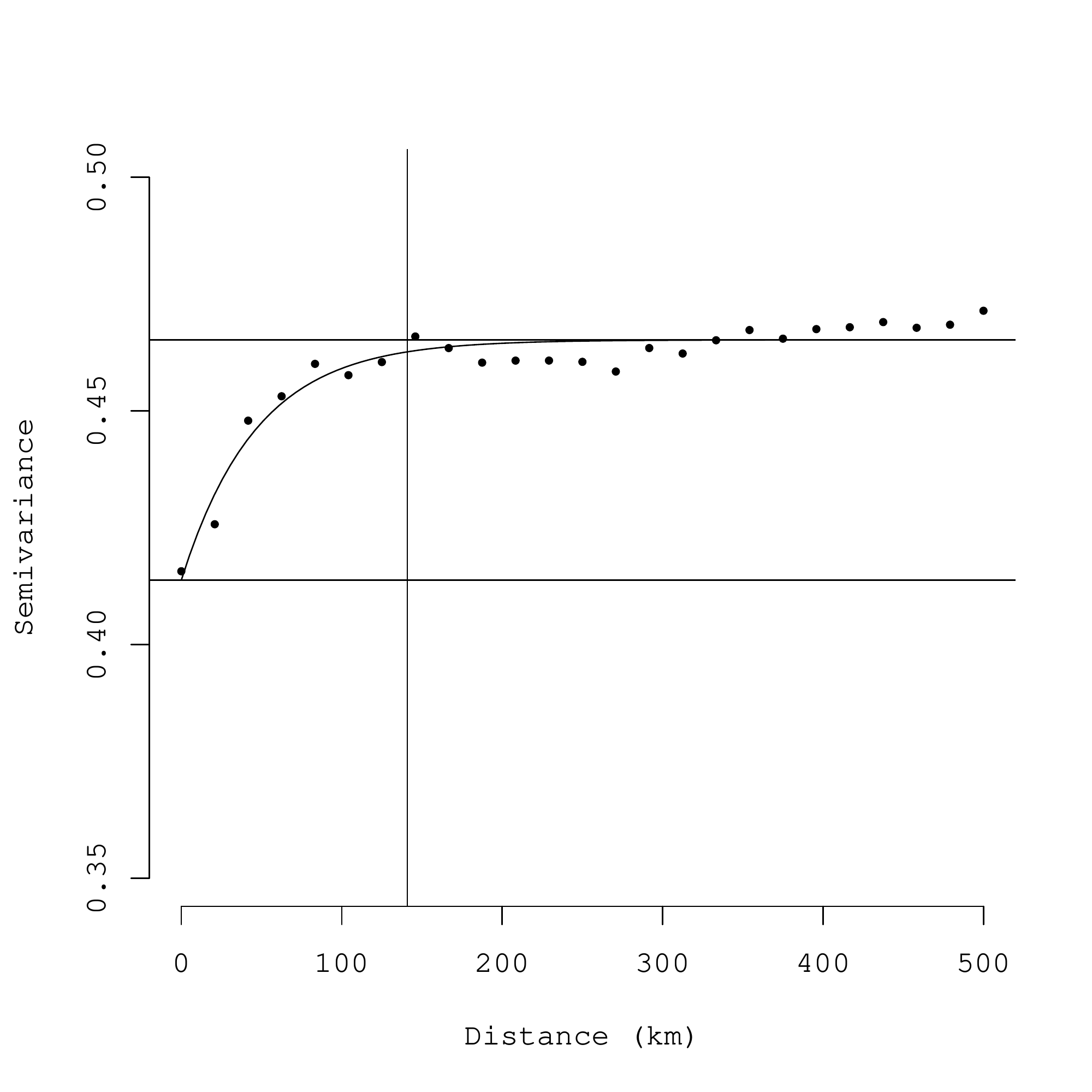}\label{bio-data-vario}}\\
\subfigure[SVI $\beta_0(\bt)$]{\includegraphics[width=8cm]{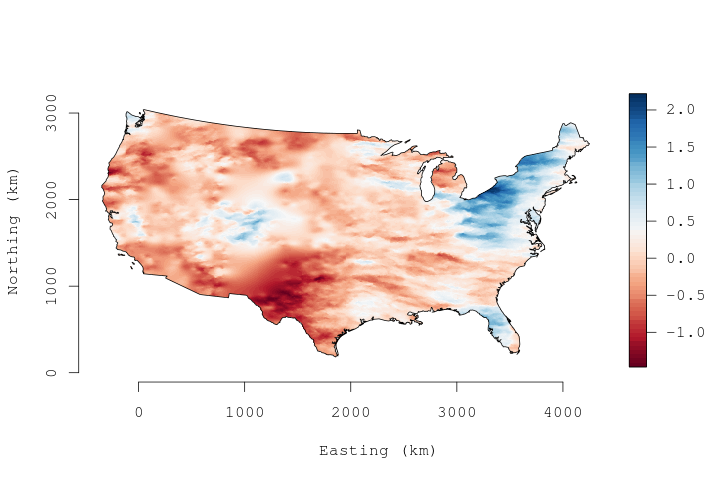}\label{bio-svi-w}}
\end{center}
\caption{Forest biomass data analysis: \subref{bio-data-fia} locations of observed biomass, \subref{bio-data-bio-interp} interpolated biomass response variable, \subref{bio-data-ndvi} NDVI regression covariate, \subref{bio-data-vario} variogram of non-spatial model residuals, and \subref{bio-svi-w} surface of the SVI model random spatial effects posterior medians. Following our FIA data sharing agreement, plot locations depicted in \subref{bio-data-fia} have been ``fuzzed'' to hide the true coordinates. }
\label{fig:bio-data}
\end{figure}

\begin{table}[t]
\caption{Forest biomass data analysis parameter estimates and computing time in hours for candidate models. Parameter posterior summary 50 (2.5, 97.5) percentiles.}\label{tab:bio-params}
\centering
\begin{tabular}{cccc}
  \hline
   &                  &NNGP&NNGP\\
   &Non-spatial       & Space-varying intercept & Space-varying coefficients\\
  \hline
  $\beta_{0}$ &1.043 (1.02, 1.065)&1.44 (1.39, 1.48)&1.23 (1.20, 1.26)\\
  $\beta_{NDVI}$ &0.0093 (0.009, 0.0095)&0.0061 (0.0059, 0.0062)& 0.0072 (0.0071, 0.0074)\\
$\sigma^2$ &--&0.16 (0.15, 0.17)&--\\
  $\bA\bA'_{1,1}$ &-- &--&0.24 (0.23, 0.24)\\
  $\bA\bA'_{2,1}$ &-- &--&-0.00088 (-0.00093, -0.00083)\\
  $\bA\bA'_{2,2}$ &-- &--&0.0000052 (0.0000047, 0.0000056)\\
  $\tau^2$ & 0.52 (0.51, 0.52)&0.39 (0.39, 0.40)&0.39 (0.38, 0.40)\\
  $\phi_1$ &-- &0.016 (0.015, 0.016)&0.022 (0.021, 0.023)\\
  $\phi_2$ &-- &--&0.030 (0.029, 0.031)\\
  $\nu_1$ &-- &0.66 (0.64, 0.67)&0.92 (0.90, 0.93)\\
  $\nu_2$ &-- &--&0.92 (0.89, 0.93)\\
  \hline
  p$_D$ & 2.94&6526.95&4976.13\\
  DIC & 250137&224484.2&222845.1\\
  G & 59765.30&42551.08&43117.37\\
 P & 59667.15&47603.47&46946.49\\
 D & 119432.45&90154.55&90063.86\\
  \hline
  Time  &--&14.53 & 41.35\\
  \hline
\end{tabular}
\end{table}

The values of the SVC model's goodness of fit metrics suggest that allowing the NDVI regression coefficient to vary spatially
improves model fit over that achieved by the SVI model. Figures~\ref{bio-coefs-b0} and \ref{bio-coefs-b1} show maps of posterior estimates for the spatially varying intercept and NDVI, respectively. The clear regional patterns seen in
Figure~\ref{bio-coefs-b1} suggest the relationship between NDVI and biomass does vary spatially---with stronger positive
regression coefficients in the Pacific Northwest and northern California areas. Forest in the Pacific Northwest and northern
California is dominated by conifers and support the greatest range in biomass per unit area within the entire conterminous
United States. The other strong regional pattern seen in Figure~\ref{bio-coefs-b1} is across western New England, where near
zero regression coefficients suggest that NDVI is not as effective at discerning differences in forest biomass. This result is
not surprising. For deciduous forests, NDVI can explain variability in low to moderate vegetation density. However, in high
biomass deciduous forests, like those found across western New England, NDVI \emph{saturates} and is no longer sensitive to
changes in vegetation structure \citep{ndvi05}. Hence, we see a higher intercept in this region but lower slope coefficient
on NDVI.
\begin{figure}[h!]
\begin{center}
\subfigure[$\beta_0(\bt)$]{\includegraphics[width=8cm]{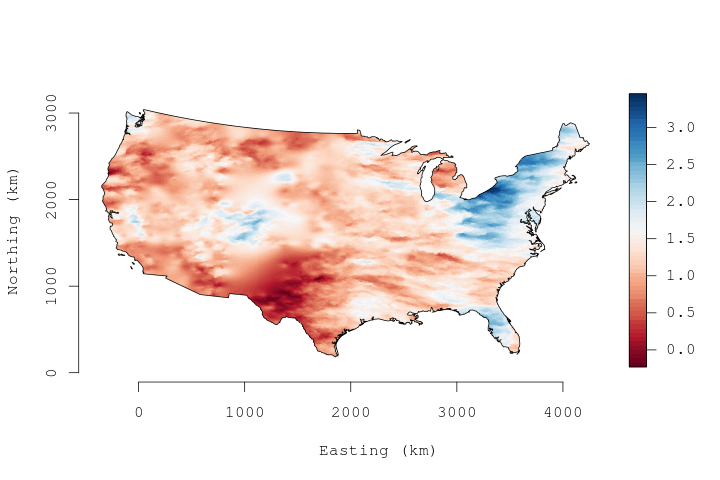}\label{bio-coefs-b0}}
\subfigure[$\beta_{NDVI}(\bt)$]{\includegraphics[width=8cm]{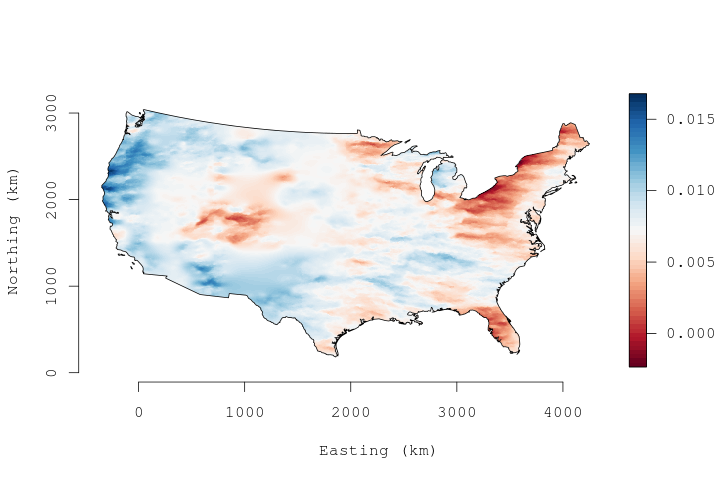}\label{bio-coefs-b1}}\\
\subfigure[Fitted biomass]{\includegraphics[width=8cm]{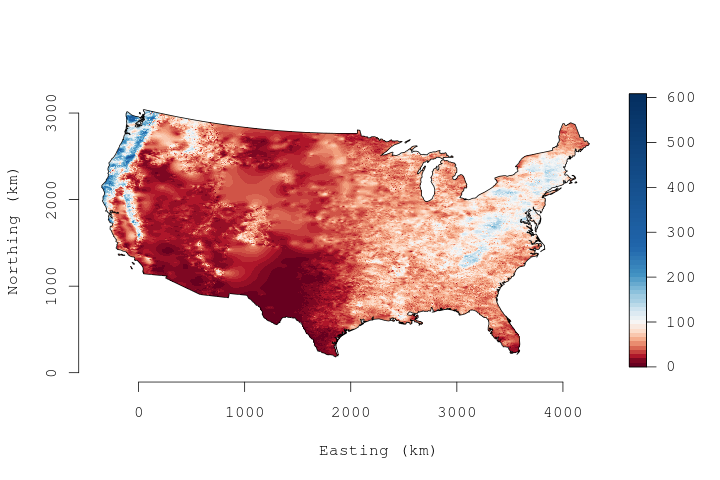}\label{bio-fitted}}
\subfigure[95\% CI width]{\includegraphics[width=8cm]{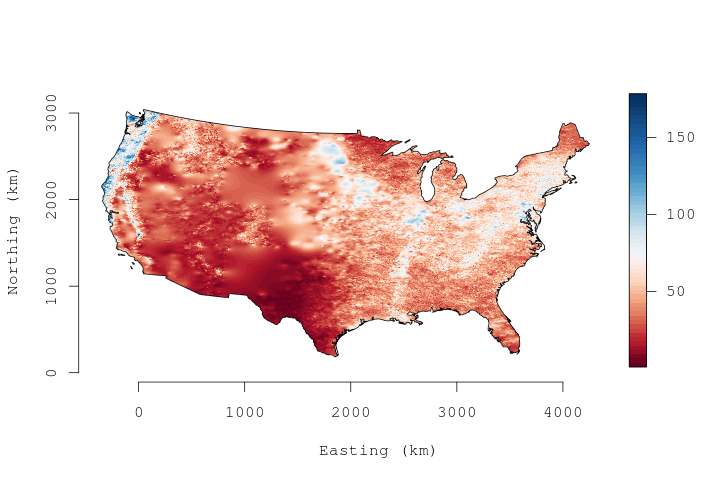}\label{bio-fitted-95-range}}
\end{center}
\caption{Forest biomass data analysis using SVC model: \subref{bio-coefs-b0} Posterior medians of the intercept, \subref{bio-coefs-b1} NDVI regression coefficients,
\subref{bio-fitted} median of biomass posterior predictive distribution, and \subref{bio-fitted-95-range} range between the upper and lower 95\% percentiles of the posterior predictive distribution.}
\label{fig:bio-coefs}
\end{figure}
Figures~\ref{bio-fitted} and \ref{bio-fitted-95-range} map each location's posterior predictive median and the range between the upper and lower 95\% credible interval, respectively, from the SVC model. Figure~\ref{bio-fitted} shows strong correspondence with the deterministic interpolation of biomass in Figure~\ref{bio-data-bio-interp}. The prediction uncertainty in Figure~\ref{bio-fitted-95-range} provides a realistic depiction of the model's ability to quantify forest biomass across the United States.

We also used prediction mean squared error (PMSE) to assess predictive performance. We fit the candidate models using $100,000$ observations and withheld $14,371$ for validation. PMSE for the non-spatial, SVI, and SVC models was $0.52$, $0.41$, and $0.42$ respectively. Lower PMSE for the spatial models, versus the non-spatial model, corroborates the results from the model fit metrics and further supports the need for spatial random effects in the analysis.

\section{Summary and conclusions}\label{Sec: Conclusions}
\noindent 
We regard the NNGP as a highly scalable model, rather than a likelihood approximation, for large geostatistical datasets. It significantly outperforms competing low-rank processes such as the GPP, in terms of inferential capabilities as well as scalability. A reference set ${\calS}$ and the resulting neighbor sets (of size $m$) define the NNGP. Larger $m$'s would increase costs, but there is no apparent benefit to increasing $m$ for larger datasets (see Appendix~\ref{sec:ci}). Selecting $\calS$ is akin to choosing the ``knots'' or ``centers'' in low-rank methods. While some sensitivity to $m$ and the choice of points in $\calS$ is expected, our results indicate that inference is very robust with respect to $\calS$ and very modest values of $m$ ($\ll 20$) typically suffice. Larger reference sets may be needed for larger datasets, but its size does not thwart computations. In fact, we observed that a very convenient choice for the reference set is the observed locations. 

A potential concern with this choice is that if the observed locations have large gaps, then the resulting NNGP may be a poor approximation of the full Gaussian Process. This arises from the fact that observations at locations outside the reference set are correlated via their respective neighbor sets and large gaps may imply two very near points have very different neighbor sets leading to low correlation. Our simulations in Appendix~\ref{sec:gaps} indeed reveal that in such a situation, the NNGP covariance field is very flat at points in the gap. However, even with this choice of $\calS$ the NNGP model performs at par with the full GP model as the latter also fails to provide strong information about observations located in large gaps. Of course, one can always choose a grid over the entire domain as $\calS$ to construct a NNGP with covariance function similar to the full GP (see Figure~\ref{fig:gapsgrid}). Another choice for $\calS$ could be based upon configurations for treed Gaussian processes \citep{gram08}. .  

Our simulation experiments revealed that estimation and kriging based on NNGP models closely emulate those from the true Mat\'ern GP models, even for slow decaying covariances (see Appendix~\ref{sec:nn}). The Mat\'ern covariance function is monotonically decreasing with distance and satisfies theoretical {\em screening} conditions, i.e. the ability to predict accurately based on a few neighbors \citep{stein02}. This, perhaps, explains the excellent performance of NNGP models with Mat\'ern covariances. We also investigated the performance of NNGP models using a wave covariance function, which does not satisfy the screening conditions, in a setting where a significant proportion of nearest neighbors had negative correlation with the corresponding locations. The NNGP estimates were still close to the true model parameters and the kriged surface closely resembled the true surface (see Appendix~\ref{sec:wave}). 

Most wave covariance functions (like the damped cosine or the cardinal sine function) produce covariance matrices with several small eigenvalues. The full GP model cannot be implemented for such models because the matrix inversion is numerically unstable. The NNGP model involves much smaller matrix inversions and can be implemented in some cases (e.g. for the damped cosine model). However, for the cardinal sine covariance, the NNGP also faces numerical issues as even the small $m \times m$ covariance matrices are numerically unstable. Bias-adjusted low-rank GPs \citep{fin09} possess a certain advantage in this aspect as the covariance matrix is guaranteed to have eigen values bounded away from zero. However, computations involving low-rank processes with numerically unstable covariance functions cannot be carried out with the efficient Sherman-Woodbury-Morrison type matrix identities and more expensive full Cholesky decompositions will be needed.

Apart from being easily extensible to multivariate and spatiotemporal settings with discretized time, the NNGP can fuel interest in process-based modeling over graphs. Examples include networks, where data arising from nodes are posited to be similar to neighboring nodes. It also offers new modeling avenues and alternatives to the highly pervasive Markov random field models for analyzing regionally aggregated spatial data. Also, there is scope for innovation when space and time are jointly modeled as processes using spatiotemporal covariance functions. One will need to construct neighbor sets both in space and time and effective strategies, in terms of scalability and inference, will need to be explored. Comparisons with alternate approaches \citep[see, e.g.,][]{katz12} will also need to be made. Finally, a more comprehensive study on the alternate algorithms, including direct methods for executing sparse Cholesky factorizations, in Section~\ref{Sec: alt} is being undertaken. More immediately, we plan to migrate our lower-level \texttt{C++} code to the existing \texttt{spBayes} package \citep[][]{fin13} in the \texttt{R} statistical environment (\url{http://cran.r-project.org/web/packages/spBayes}) to facilitate wider user accessibility to NNGP models.

\begin{description}


\item[Acknowledgments:] 
We express our gratitude to Professors Michael Stein and Noel Cressie for discussions which helped to enrich this work. The work of the first three authors was supported by federal grants NSF/DMS 1106609 and NIH/NIGMS RC1-GM092400-01.

\end{description}

\section*{Appendix}

\appendix

\section{Densities on directed acyclic graphs}\label{sec:dag}
We will show that if $\calG=(\calS, N_\calS)$ is acyclic then $\tp(\bw_\calS)$ defined in (\ref{eq:cl}) corresponds to a true density over $\calS$. For any directed acyclic graph, there exists a node with zero in-degree i.e. no directed edge pointing towards it. We denote this node by $\bs_{\pi(1)}$  This means $\bs_{\pi(1)}$ does not belong to the neighbor set of
any other location in $\calS$. The only term where it appears on the right hand side of (\ref{Eq: nngp_gen}) is
$p(\bw(\bs_{\pi(1)} \given \bw_{N(\bs_{\pi(1)})})$ which integrates out to one with respect to $d\bw(\bs_{\pi(1)})$.
We now have a new acyclic directed graph $\calG_1$ obtained by removing vertex $\bs_{\pi(1)}$ and its directed edges from
$\calG$. Now we can find a new vertex $\bs_{\pi(2)}$ with zero out-degree in $\calG_1$ and continue as before to get a
permutation $\pi(1),\pi(2),\ldots,\pi(k)$ of $1,2,\ldots,k$ such that
\[ \int \prod_{i=1}^k p(\bw(\bs_i)\given \bw_{N(\bs_i)}) d\bw(\bs_{\pi(1)}) d\bw(\bs_{\pi(2)}) \ldots d\bw(\bs_{\pi(k)}) = 1 \]
An easy application of Fubini's theorem now ensures that this is a proper joint density.

\section{Properties of $\tildebC_\calS^{-1}$}\label{sec:sparse}
If $p(\bw_\calS) = N(\bw_\calS \given \bzero, \bC_\calS)$, then $\bw(\bs_i) \given \bw_{N(\bs_i)} \sim N(\bB_{\bs_i}\bw_{N(\bs_i)},\bF_{\bs_i} )$, where $\bB_{\bs_i}$ and $\bF_{\bs_i}$ are defined in (\ref{eq:cl}). So, the likelihood in (\ref{Eq: nngp_gen}) is proportional to
\[
 \frac 1{\prod_{i=1}^k \sqrt{ \det(\bF_{\bs_i})}} \exp\left (-\frac 12 \sum_{i=1}^k (\bw(\bs_i) - \bB_{\bs_i}\bw_{N(\bs_i)})'\bF_{\bs_i}^{-1}(\bw(\bs_i) - \bB_{\bs_i}\bw_{N(\bs_i)}) \right) \]
For any matrix $\bA$, let $\bA[,j:j']$ denote the submatrix formed using columns $j$ to $j'$ where $j < j'$. For $j=1,2,\ldots,k$, we define $q \times q$ blocks $\bB_{\bs_i,j}$ as
\begin{align*}
\bB_{\bs_i,j} = \left \{
\begin{array}{l}
\bI_q \mbox{ if } j=i; \\
-\bB_{\bs_i}[,(l-1)q+1:lq] \mbox{ if } \bs_j = N(\bs_i)(l) \mbox{ for some } l; \\
\bO \mbox{ otherwise,}
\end{array} \right .
\end{align*}
where, for any location $\bs$, $N(\bs)(l)$ is the $l$-th neighbor of $\bs$. So, $\bw_{\bs_i}-\bB_{\bs_i}\bw_{N(\bs_i)} = \bB^*_{\bs_i}\bw_\calS$, where $\bB^*_{\bs_i}=[\bB_{\bs_i,1},\bB_{\bs_i,2},\ldots,\bB_{\bs_i,k}]$ is $q \times kq$ and sparse with at most $m+1$ non-zero blocks. Then,
\begin{align*}
\sum_{i=1}^k (\bw(\bs_i) - \bB_{\bs_i}\bw_{N(\bs_i)})'\bF_{\bs_i}^{-1}(\bw(\bs_i) - \bB_{\bs_i}\bw_{N(\bs_i)})
=\sum_{i=1}^k \bw_\calS'(\bB^*_{\bs_i})'\bF_{\bs_i}^{-1}\bB^*_{\bs_i}\bw_\calS
= \bw_\calS'\bB_\calS'\bF_\calS^{-1}\bB_\calS\bw_\calS\;,
\end{align*}
where $\bF=diag(\bF_{\bs_1},\bF_{\bs_2},\ldots,\bF_{\bs_k})$ and $\bB_\calS=((\bB^*_{\bs_1})',(\bB^*_{\bs_2})',\ldots,(\bB^*_{\bs_k})')'$. So, we have:
\begin{equation}\label{eq:dec}
(\tildebC_ \calS)^{-1} = \bB_ \calS'\bF_ \calS^{-1}\bB_ \calS
\end{equation}
From the form of $\bB_{\bs_i,j}$, it is clear that $\bB_\calS$ is sparse and lower triangular with ones on the diagonals. So, $\det(\bB_\calS)=1$, $\det((\bB_\calS'\bF_\calS^{-1}\bB_\calS)^{-1}) = \prod \det(\bF_{\bs_i})$ and (\ref{Eq: nngp_gen}) simplifies to $N(\bw_\calS \given \bzero, \tildebC_\calS)$. 


Let $\tildebC_\calS^{ij}$ denote the $(i,j)^{th}$ block of $\tildebC_\calS^{-1}$. Then from equation (\ref{eq:dec}) we see that for $i < j$, $\tildebC_\calS^{ij} = \sum_{l=j}^k (\bB^*_{\bs_l,i})'\bF_{\bs_l}^{-1}\bB^*_{\bs_l,j}$. So, $\tildebC_\calS^{ij}$ is non-zero only if there exists at least one location $\bs_l$ such that $\bs_i \in N(\bs_l)$ and $\bs_j$ is either equal to $\bs_l$ or is in $N(\bs_l)$. Since every neighbor set has at most $m$ elements, there are at most $km(m+1)/2$ such pairs $(i,j)$. This demonstrates the sparsity of $\tildebC_\calS^{-1}$ for $m \ll k$.

\clearpage
\section{Simulation Experiment: Robustness of NNGP to ordering of locations}\label{sec:order}
We conduct a simulation experiment demonstrating the robustness of NNGP to the ordering of the locations. We generate the data for $n=2500$ locations using the model in Section \ref{Sec: Simulation_Experiments}. However instead of a square domain we choose a long skinny domain (see Figure~\ref{uni2-w-obs}) which can bring out possible sensitivity to ordering due to scale disparity between the $x$ and $y$ axes. We use three different orderings for the locations: ordering by $x$-coordinates, by $y$-coordinates and by the function $f(x,y) = x+y$. 

Table~\ref{tab:uni2-params} demonstrates that the point estimates and the $95\%$ credible intervals for the process parameters from all three NNGP models are extremely consistent with the estimates from the full Gaussian process model. 

Posterior estimates of the spatial residual surface from the different models are shown in Figure~\ref{fig:uni2-w}. Again, the impact of the different ordering is negligible.    As one of the reviewers suggested, we also plotted the difference between the posterior estimates of the random effects of the true GP and NNGP for all 3 orderings in Figure~\ref{fig:diff}. It was seen that this difference was negligible compared to the difference between the true spatial random effects and full GP estimates. This shows the inference obtained from the NNGP (using any ordering) closely emulates the corresponding full GP inference. 

\begin{figure}[t!]
\begin{center}
\subfigure[True $\bw$]{\includegraphics[width=16cm,height=2.4cm]{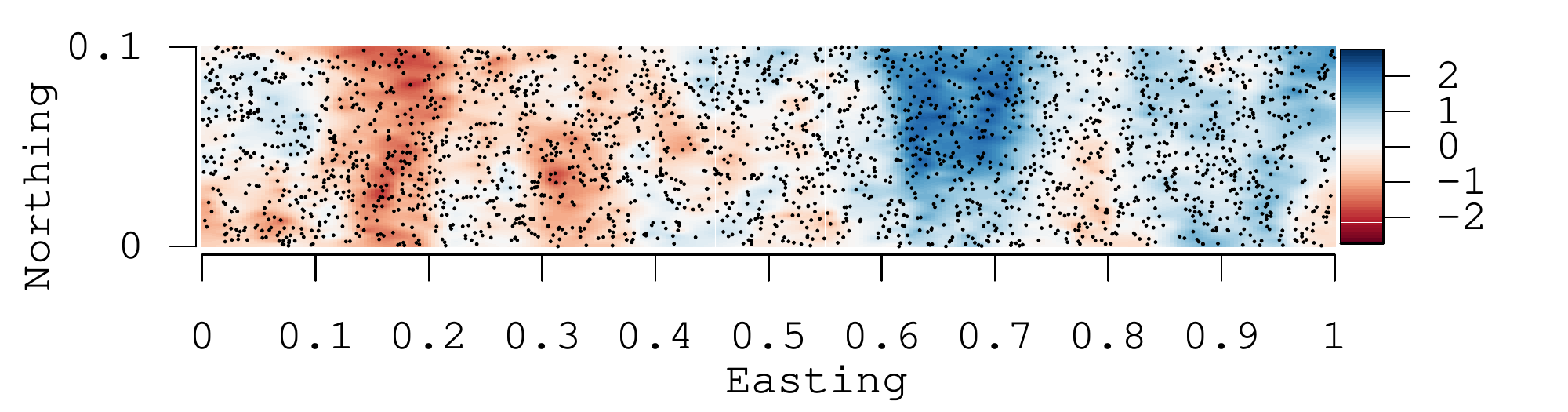}\label{uni2-w-obs}}
\subfigure[Full GP]{\includegraphics[width=16cm,height=2.4cm]{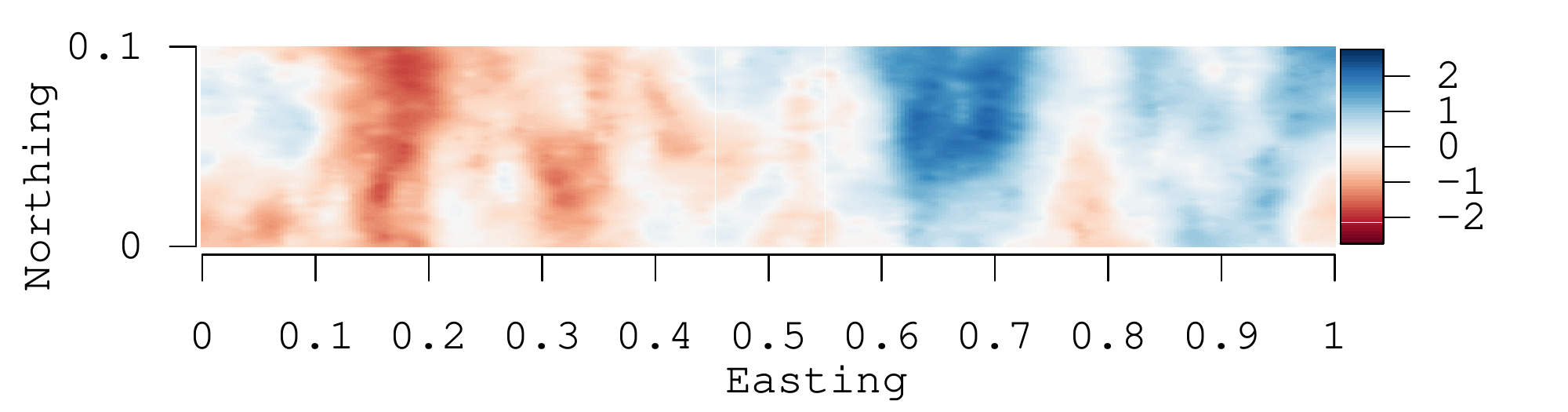}\label{uni2-w-gp}}
\subfigure[NNGP order by $y$-coordiantes]{\includegraphics[width=16cm,height=2.4cm]{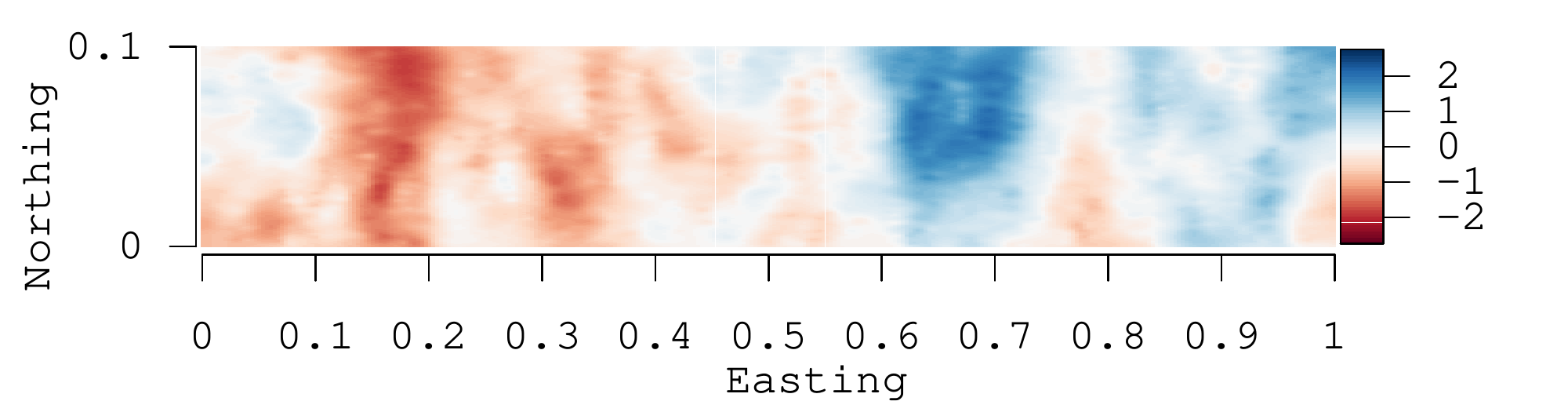}\label{uni2-y-order}}
\subfigure[NNGP order by $x$-coordiantes]{\includegraphics[width=16cm,height=2.4cm]{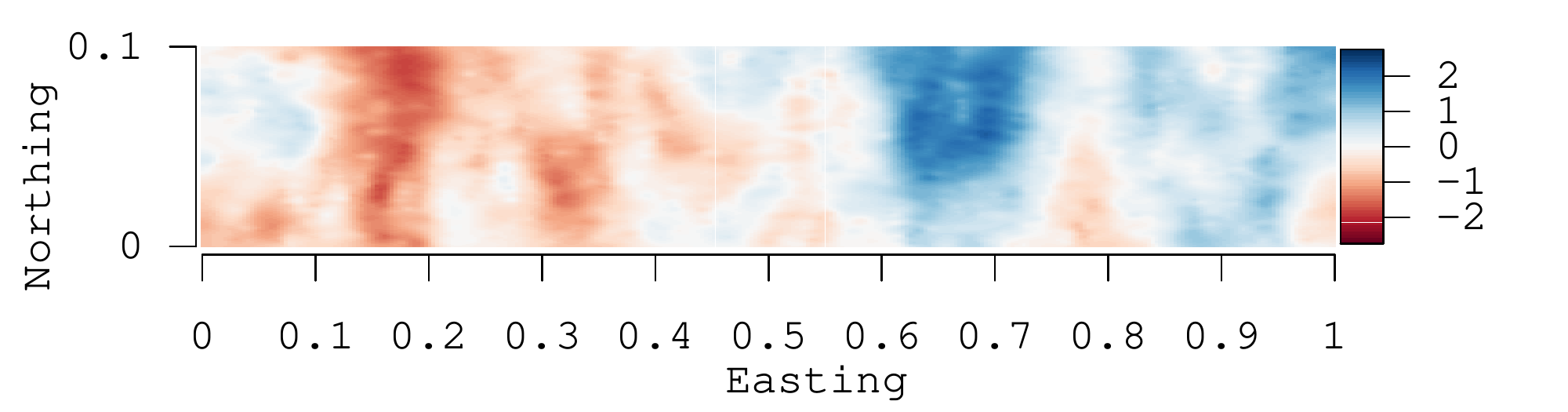}\label{uni2-x-order}}
\subfigure[NNGP order by $x+y$-coordiantes]{\includegraphics[width=16cm,height=2.4cm]{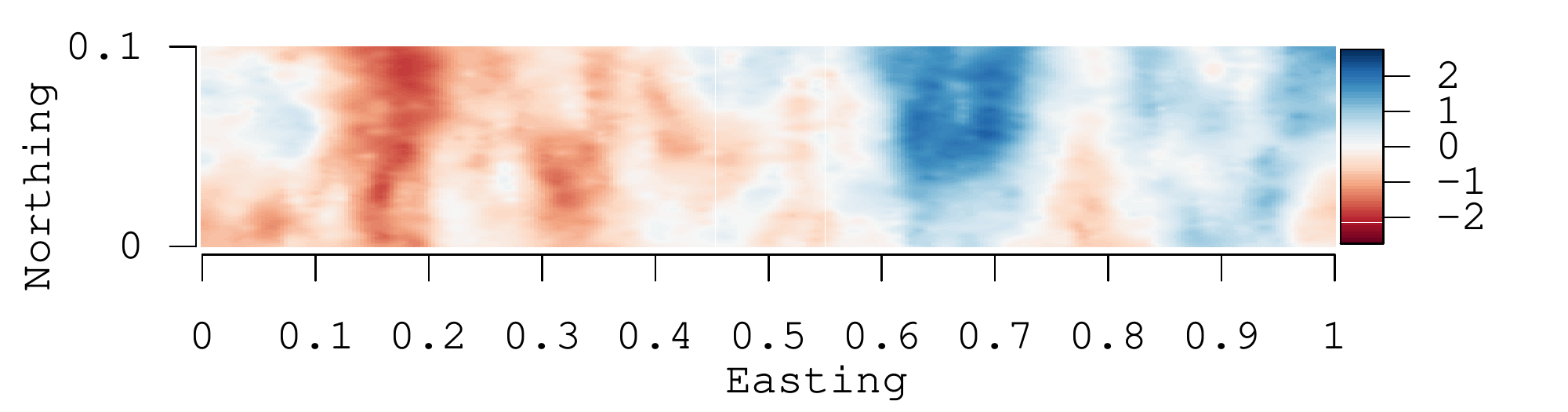}\label{uni2-xy-order}}
\end{center}
\caption{Robustness of NNGP to ordering: Figures~\subref{uni2-w-obs} and \subref{uni2-w-gp} show interpolated surfaces of the true spatial random effects and posterior median estimates for full geostatistical model respectively. Figures \subref{uni2-y-order}, \subref{uni2-x-order}, and \subref{uni2-xy-order} show interpolated surfaces of the posterior median estimates for NNGP model with $\calS=\calT$, $m=10$, and alternative coordinate ordering. Corresponding true and estimated process parameters are given in Table~\ref{tab:uni2-params}.}
\label{fig:uni2-w}
\end{figure}

\begin{figure}[t!]
\begin{center}
\subfigure[True $\bw -$ Full GP $\widehat \bw$]{\includegraphics[width=16cm]{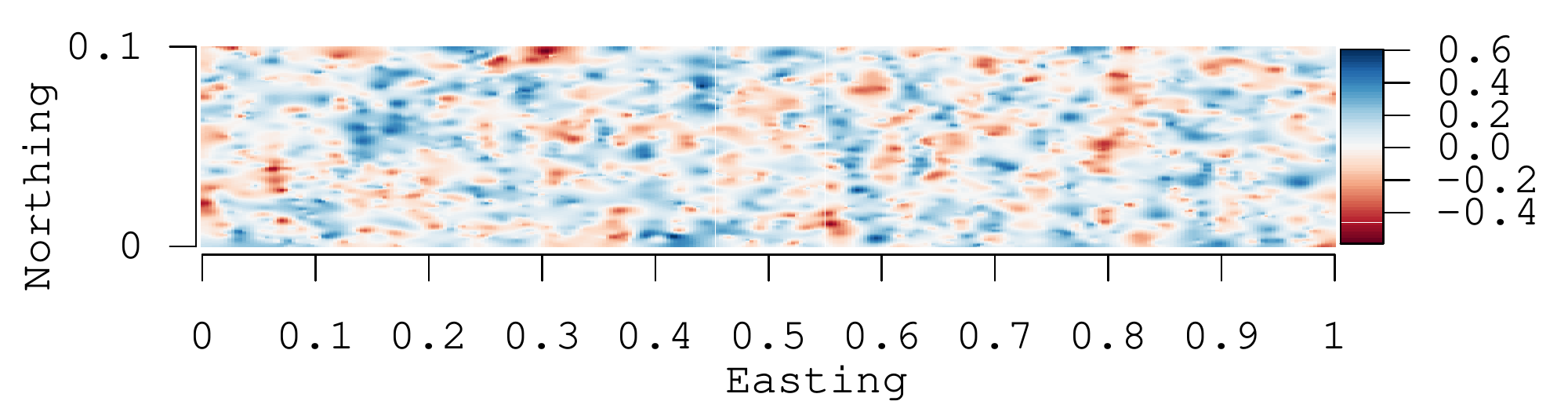}\label{truefull}}
\subfigure[Full GP $\widehat \bw - $ NNGP (order by $x$) $ \widehat \bw$]{\includegraphics[width=16cm]{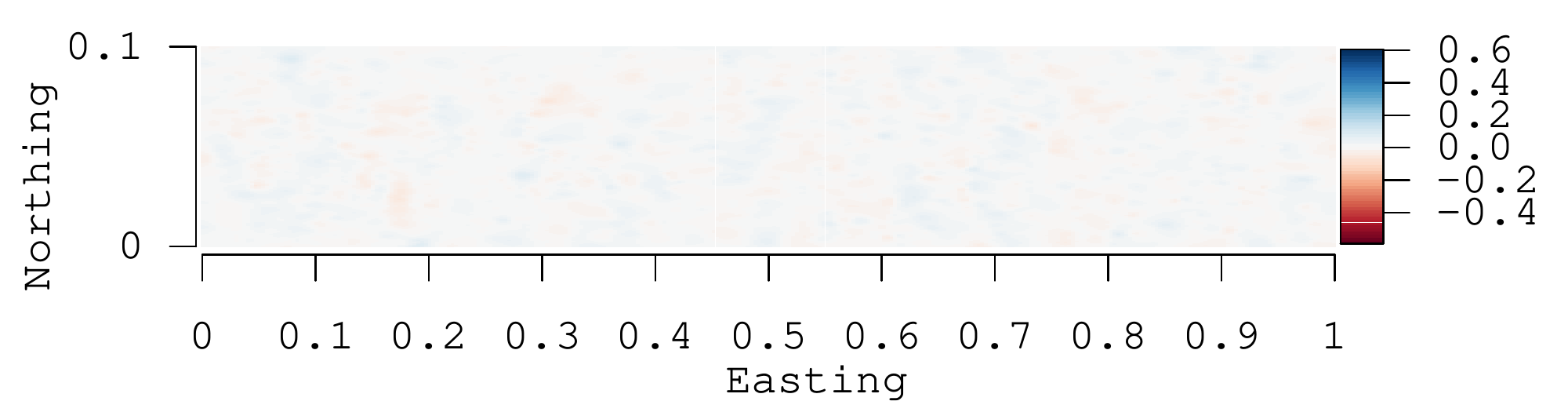}\label{fullx}}
\subfigure[Full GP $\widehat \bw - $ NNGP (order by $y$) $ \widehat \bw$]{\includegraphics[width=16cm]{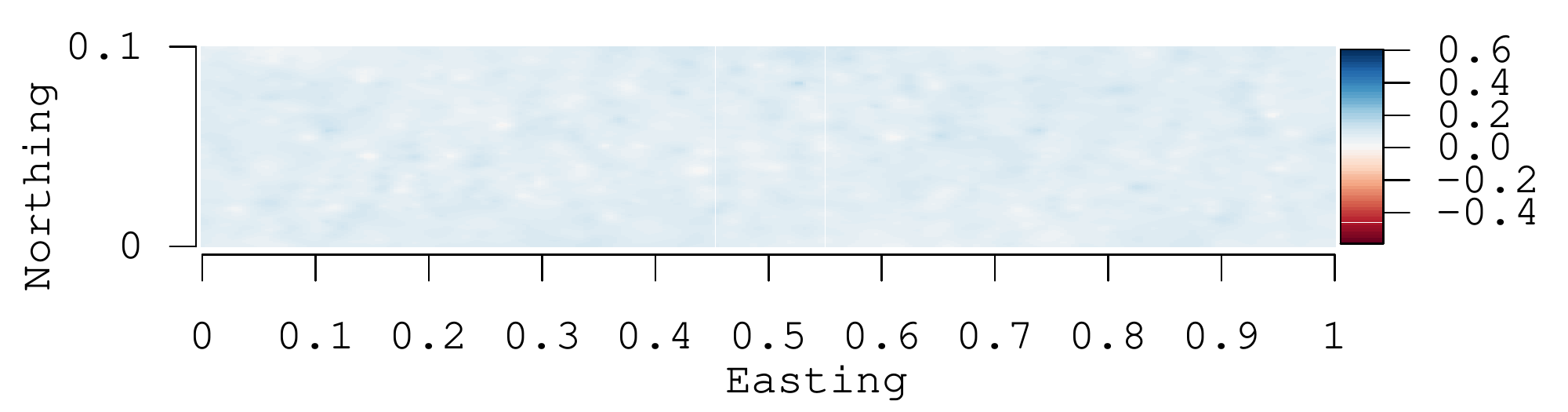}\label{fully}}
\subfigure[Full GP $\widehat \bw - $ NNGP (order by $x+y$) $ \widehat \bw$]{\includegraphics[width=16cm]{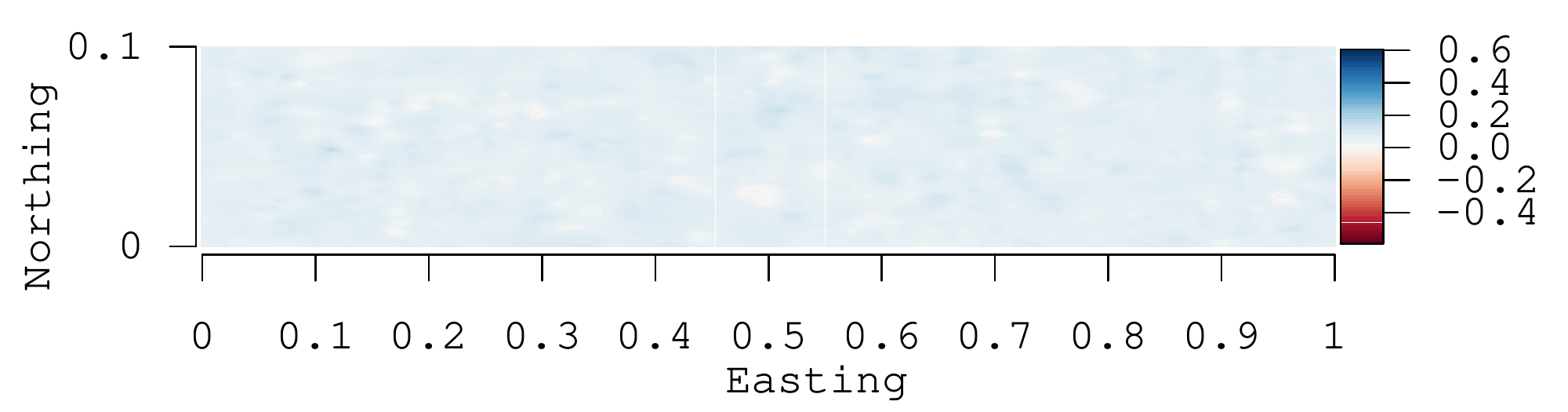}\label{fullplus}}
\end{center}
\caption{Difference between Full GP and NNGP estimates of spatial effects: Figure~\subref{truefull} shows the difference between the true spatial random effects and the full GP posterior median estimates. Figures \subref{fullx}, \subref{fully} and \subref{fullplus} plots the difference between posterior median estimates of full GP and NNGP ordered by $x$, $y$ and $x+y$ co-ordinates respectively. All the figures are in the same color scale.}
\label{fig:diff}
\end{figure}

\begin{table}[h!]
\centering
\caption{Univariate synthetic data analysis parameter estimates and computing time in minutes for NNGP $m$=10 and full GP models. Parameter posterior summary 50 (2.5, 97.5) percentiles. }
\begin{tabular}{cccccc}
  \hline
  &      & & \multicolumn{3}{c}{NNGP ($\calS=\calT$)}\\
&      & Full& Order by &Order by &Order by \\
   & True & Gaussian Process & $y$-coordinates&  $x$-coordinates&  $x+y$-coordinates\\
  \hline
  $\sigma^2$&1&0.640 (0.414, 1.297)&0.712 (0.449, 1.530)&0.757 (0.479, 1.501)&0.718 (0.464, 1.436)\\
  $\tau^2$&0.1&0.107 (0.098, 0.117)&0.106 (0.097, 0.114)&0.107 (0.099, 0.117)&0.107 (0.098, 0.115)\\
  $\phi$&6&8.257 (4.056, 13.408)&8.294 (3.564, 12.884)&7.130 (3.405, 11.273)&7.497 (3.600, 11.911)\\
  \hline
\end{tabular}
\label{tab:uni2-params}
\end{table}

\clearpage
\section{Kolmogorov Consistency for NNGP}\label{sec:kolcon}
Let $\{\bw(\bs) \given \bs \in \calD\}$ be a random process over some domain $\calD$ with density $p$ and let $\tp(\bw_\calS)$ be a probability density for observations over a fixed finite set $\cal S \subset \calD$. The conditional density $\tp(\bw_ \calU \given \bw_\calS)$ for any finite set $\calU \subset \calD$ outside of $\calS$ is defined in (\ref{Eq: nngp_t}). 

We will first show that for every finite set $\calV=\{\bv_1,\bv_2,\ldots,\bv_n\}$ in $\calD$, $n \in \{1,2,\ldots\}$ and for every permutation $\pi(1),\pi(2),\ldots,\pi(n)$ of $1,2,\ldots,n$ we have, \[ \tp\left(\bw(\bv_1),\bw(\bv_2),\ldots,\bw(\bv_n) \right) = \tp \left(\bw(\bv_{\pi(1)}),\bw(\bv_{\pi(2)}),\ldots,\bw(\bv_{\pi(n)}) \right)\;. \]. We begin by showing that for any finite set $\calV$, the expression given in (\ref{Eq: nngp_finite}) is a proper density. Let $\calU= \calV \setminus \calS$. Since $\calV \cup (\calS \setminus \calV) = \calS \cup \calU$, we obtain
\begin{align*}
 & \int \tp(\bw_\calV) \prod_{\bv_i \in \calV} d(\bw(\bv_i)) = \int \tp(\bw_{\calU} \given \bw_\calS)\tp(\bw_\calS) \prod_{\bv_i \in \calU } d(\bw(\bv_i)) \prod_{\bs_i \in \calS} d(\bw(\bs_i)) \\
 = & \int \tp(\bw_\calS) \left ( \int \tp(\bw_{\calU} \given \bw_\calS) \prod_{\bv_i \in \calU} d(\bw(\bv_i)) \right) \prod_{\bs_i \in \calS} d(\bw(\bs_i)) =   \int \tp(\bw_\calS) \prod_{\bs_i \in \calS} d(\bw(\bs_i)) = 1
\end{align*}
Note that $\calS$ is fixed. Therefore, the expression for the joint density of $\bw_\calV$ depends only on the the neighbor sets
$N(\bv_i)$ for $\bv_i \in \calU$. So the NNGP density for $\calV$ is invariant under any permutation of locations inside $\calV$.

We now prove that for every location $\bv_0 \in \calD$, we have, $\tp(\bw_\calV) = \int \tp(\bw_{\calV \cup  \{\bv_0\}})d(\bw(\bv_0))$. let $\calV_1=\calV \cup \{\bv_0\}$. We split the proof into two cases. If $\bv_0 \in \calS$, then using the
fact $\calV_1 \setminus \calS = \calV \setminus \calS = \calU$, we obtain
\begin{align*}
\int \tp(\bw_{\calV_1}) d(\bw(\bv_0)) = \int \tp(\bw_\calS)\tp(\bw_{\calV_1 \setminus \calS} \given \bw_\calS) \prod_{\bs_i \in \calS \setminus \calV_1} d(\bw(\bs_i)) d(\bw(\bv_0) \\
=\int \tp(\bw_\calS)\tp(\bw_{\calV \setminus \calS} \given \bw_\calS) \prod_{\bs_i \in \calS \setminus \calV} d(\bw(\bs_i)) = \tp(\bw_{\calU})\;.
\end{align*}
If $\bv_0 \notin \calS$, then $\bw(\bv_0)$ does not appear in the neighborhood set of any other term.  So, $p(\bw(\bv_0) \given
\bw_ \calS)$ integrates to one with respect to $d(\bw(\bv_0))$. The result now follows from $\int p(\bw_{\calV_1} \given
\bw_\calS) d(\bw(\bv_0)) = p(\bw_\calV \given \bw_\calS) $.

\section{Properties of NNGP}\label{sec:nngp}
Standard Gaussian conditional distribution facts reveal that the conditional distribution $\bw(\bu_i) \given \bw_{\calS} \sim N(\bB_{\bu_i}\bw_{N(\bu_i)},\bF_{\bu_i})$ where $\bB_{\bu_i}$ and $\bF_{\bu_i}$ be defined analogous to (\ref{eq:cl}) based on the neighbor sets $N(\bu_i)$. From (\ref{Eq: nngp_t}),
we see that
\[ \tp(\bw_\calU \given \bw_\calS)=\frac 1{\prod_{i=1}^r \sqrt{ \det(\bF_{\bu_i})}} \exp\left (-\frac 12 \sum_{i=1}^r (\bw(\bu_i) - \bB_{\bu_i}\bw_{N(\bu_i)})'\bF_{\bu_i}^{-1}(\bw(\bu_i) - \bB_{\bu_i}\bw_{N(\bu_i)}) \right) \]
It then follows that $\tp(\bw_\calU \given \bw_\calS) \sim N(\bB_\calU  \bw_ \calS, \bF_ \calU)$ where $\bB_\calU=(\bB_{\bu_1}',\bB_{\bu_2}',\ldots,\bB_{\bu_r}')'$ and $\bF_\calU=diag(\bF_{\bu_1},\bF_{\bu_2}, \ldots, \bF_{\bu_r})$. Since each row of $\bB_\calU$ has at most $m$ non-zero entries, $\bB_\calU$ is sparse for $m \ll k$.

As the nearest neighbor densities of $\bw_\calS$ and $\bw_\calU \given \bw_\calS$ for every finite $\calU$ outside $\calS$ are Gaussian, all finite dimensional realizations of an NNGP process will be Gaussian. Let $\bv_1$ and $\bv_2$ be any two
locations in $\calD$ and let $\widetilde E$ and $\widetilde {Cov}$ denote, respectively, the expectation and covariance operator for a NNGP. Then, if $\bv_1=\bs_i$ and $\bv_2=\bs_j$ are both in $\calS$ then we obviously have $\widetilde{ Cov}
(\bw(\bv_1),\bw(\bv_2) \given \btheta) = \tildebC_{\bs_i,\bs_j}$. If $\bv_1$ is outside $\calS$ and $\bv_2=\bs_j$, then \[ \begin{array}{rl}
\widetilde{ Cov} (\bw(\bv_1),\bw(\bv_2) \given \btheta) & = \widetilde E( \widetilde {Cov} (\bw(\bv_1),\bw(\bv_2) \given \bw_\calS , \btheta) ) + \widetilde {Cov}( \widetilde E (\bw(\bv_1)),\widetilde E (\bw(\bv_2)) \given \bw_\calS , \btheta)) \\
\therefore \tildebC(\bv_1 , \bv_2 \given \btheta) & = 0 + \widetilde {Cov} (\bB_{\bv_1} \bw_{N(\bv_1)}, \bw(\bs_j) \given \btheta) = \bB_{\bv_1} \tildebC_{N(\bv_1),\bs_j}
 \end{array}
 \]
If both $\bv_1$ and $\bv_2$ are outside $\calS$, then $\tildebC(\bv_1 , \bv_2 \given \btheta)=\delta(\bv_1=\bv_2)\bF_{\bv_1}+ \bB_{\bv_1} \tildebC_{N(\bv_1),N(\bv_2)}\bB_{\bv_2}'$, which yields (\ref{Eq: tildefunc}).

For any two set of locations $A$ and $B$, let $||A,B||$ denote the pairwise Euclidean distance matrix. Let $\calZ_1$ denote set of all points $\bv$ such that $\bv$ is equidistant from any two points in $\calS$. Since $\calS$ is finite, the set $\calZ_2=(\calZ_1 \times \calZ_1) \cup \{(\bv,\bv) \given \bv \in \calD \}$ has Lebesgue measure zero in the Euclidean domain $\Re^d \times \Re^d$. We will show that $\tildebC(\bv_1 , \bv_2 \given \btheta)$ is continuous for any pair $(\bv_1, \bv_2)$ in $\calD \times \calD \setminus \calZ_2$. Observe that for any pair of points $(\bv_1,\bv_2)$ in $\calD \times \calD \setminus \calZ_2$, it is easy to verify that $\displaystyle \lim_{\bh_i \rightarrow 0} ||(\bv_i+\bh_i, N(\bv_i+\bh_i) || \rightarrow ||\bv_i,N(\bv_i)||$, for $i=1,2$, and $\displaystyle \lim_{\bh_1 \rightarrow 0,\bh_2 \rightarrow 0} ||N(\bv_1+\bh_1), N(\bv_2+\bh_2) || \rightarrow  ||N(\bv_1),N(\bv_2)||$. We prove the continuity of $\tildebC(\bv_1,\bv_2 \given \btheta)$ for the case when $\bv_1$ is outside $\calS$ and $\bv_2=\bs_j$. The other cases are proved similarly. We assume that the covariance function for the original GP is isotropic and continuous. The two distance results yield $\bB_{\bv_1+\bh_1} = \bC_{\bv_1+\bh_1,N(\bv_1+\bh_1)} \bC_{N(\bv_1+\bh_1)}^{-1} \rightarrow \bC_{\bv_1,N(\bv_1)} \bC_{N(\bv_1)}^{-1}  = \bB_{\bv_1}$. Also, as $\bv_2+\bh_2 \rightarrow \bv_2 = \bs_j$, then $\bs_j \in N(\bv_2+\bh_2)$ for small enough $\bh_2$. Let $\bs_j = N(\bv_2+\bh_2)(1)$ and, hence, $\bC_{\bv_2+\bh_2,N(\bv_2+\bh_2)} \bC_{N(\bv_2+\bh_2)}^{-1} \rightarrow \be_1$ where $\be_1=(1,0,\ldots,0)_ {m\times 1}$. Therefore,
\begin{align*}
& \lim_{\bh_1 \rightarrow 0,\bh_2 \rightarrow 0} \tildebC(\bv_1+\bh_1,\bv_2+\bh_2 \given \btheta) =  \bB_{\bv_1} \; lim_{\bh_1 \rightarrow \bzero,\bh_2 \rightarrow \bzero} \; \widetilde {Cov} (\bw_{N(\bv_1+\bh_1)},\bw_{N(\bv_2+\bh_2)} \given \btheta)\be_1 \\
&\qquad\qquad = \bB_{\bv_1} \; lim_{\bh_1 \rightarrow \bzero} \; \widetilde {Cov} (\bw_{N(\bv_1+\bh_1)},\bw(\bs_j) \given \btheta) = \bB_{\bv_1} \tildebC_{N(\bv_1),\bs_j} = \tildebC(\bv_1,\bv_2 \given \btheta)\;.
\end{align*}

\clearpage
\section{Simulation experiment: NNGP credible intervals as function of $m$}\label{sec:ci}
From a classical viewpoint, NNGP can be regarded as a computationally convenient approximation to the full GP model. The accuracy of the approximation is expected to improve with increase in $m$ as NNGP model becomes identical to the full model when $m$ equals the sample size. However, we construct the NNGP as an independent model and found that inference from this model closely emulates that from the full GP model. Figure~\ref{Fig: uni-nn-pred} demonstrates how root mean square predictive error and parameter CI width vary with choice of $m$. We conduct another simulation experiment to investigate how the parameter estimation of the hierarchical NNGP model depends on $m$. 

\begin{figure}[b!]
\begin{center}
\includegraphics[width=17cm]{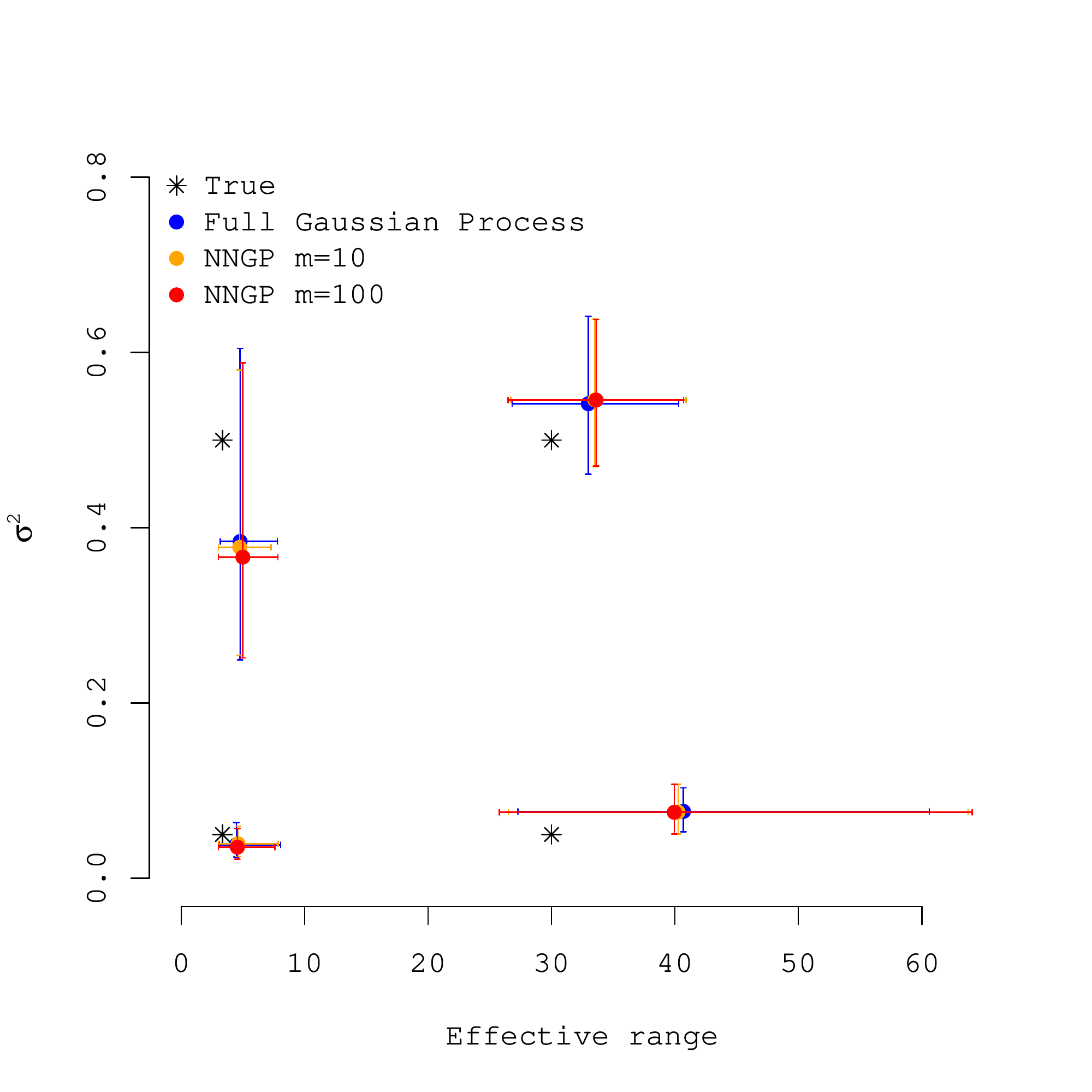}
\end{center}
\caption{NNGP credible intervals for small and large values of $m$}
\label{fig:ci}
\end{figure}

We generated a dataset of size $1000$ using the model described in Section 5.1 for 4 combination of values of $\phi$ and $\sigs$. Other parameters and prior choices were similar to those in section~\ref{sec:nn}. Figure~\ref{fig:ci} gives true values of $\sigs$ and effective range ($3/\phi$) alongwith the posterior medians and credible intervals for the full GP, NNGP with $m=10$ and $m=100$. We see that the CI's for NNGP $m=10$ and $m=100$ are almost identical and are very close to the CI for full GP. This suggests that even for small values of $m$ NNGP, parameter CI's closely resemble full GP parameter CI's. 

\clearpage
\section{Simulation experiment: Data with gaps}\label{sec:gaps}
One possible area of concern for NNGP is that if the data have large gaps and the NNGP is constructed using the data locations as the reference set $\calS$, then NNGP covariance function may be a poor approximation of the full GP covariance function. This arises from the fact that if the reference set has large gaps then two very close locations outside $\calS$ can have very different neighbor sets. Since, in a NNGP, locations outside $\calS$ are correlated through their neighbors sets this may lead to little correlation among very close points in certain regions of the domain. 

\begin{figure}[h!]
\begin{center}
\includegraphics[width=17cm,height=13cm]{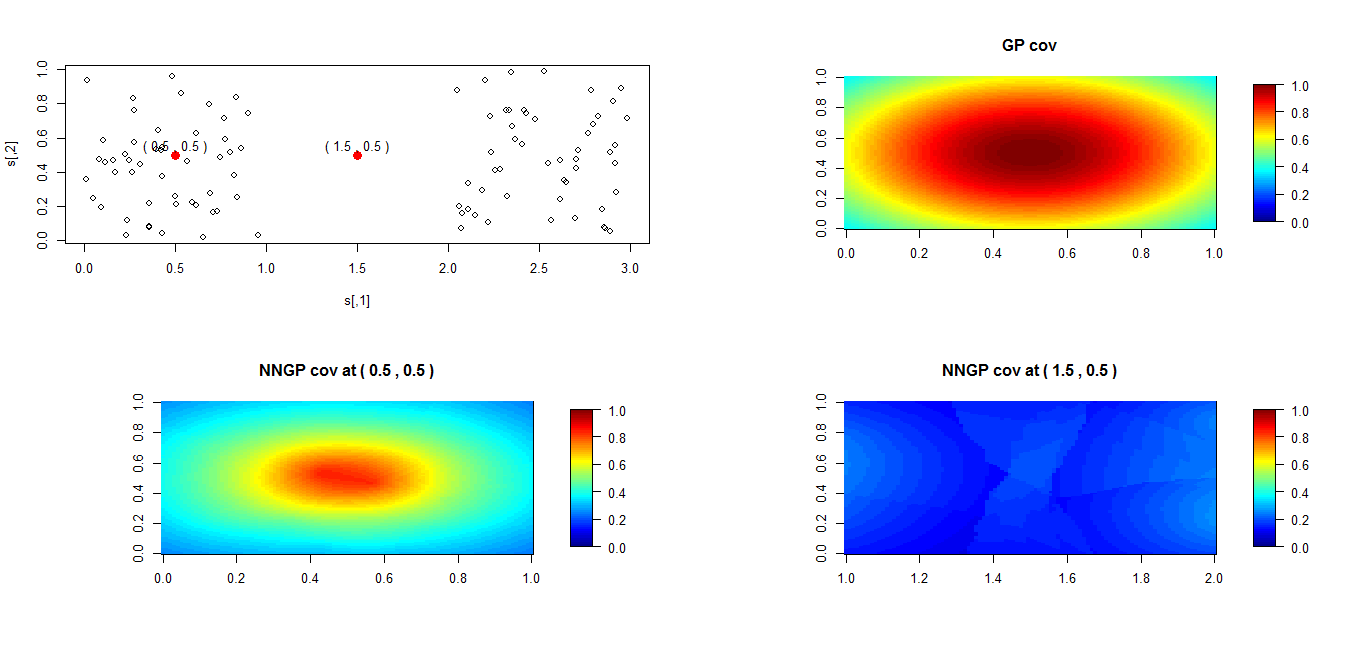}
\end{center}
\caption{Full GP and NNGP ($m=10$) covariance function for data with gaps}
\label{fig:gapscov}
\end{figure}

Figure~\ref{fig:gapscov} demonstrates this issue. We generate a set $\calT$ of $100$ locations (topleft) on the domain $[0,3] \times [0,1]$ with half the locations in $[0,1] \times [0,1]$ and the remaining half in $[2,3] \times [0,1]$. This creates a large gap in the middle where there are no datapoints. The topright panel shows the heatmap of the full GP covariance function with $\sigs$=1 and $\phi=2$ (so that the effective range is $1.5$). The NNGP is a non-stationary process and the covariance function depends on the locations. We evaluate this covariance at two points (red dots in the topleft figure) --- $(0.5,0.5)$ (which is surrounded by many points in $\calS$) and $(1.5,0.5)$ (which is at the middle of the gap and equidistant from the two sets of locations in $\calS$). 

The NNGP field at $(0.5,0.5)$ (bottomleft) closely resembles the GP field. This is because the neighbors of $(0.5,0.5)$ are close to the point and provides strong information about the true GP at that point. The NNGP field at $(1.5,0.5)$ (bottomright) is almost non-existent with near zero correlations even at very small distances. This is an expected consequence of the way NNGP is constructed. Any two points outside $\calS$ are correlated via their neigbhor sets only. The neighbors for $(1.5,0.5)$ are far away from the point it provides weak information about the point as it is in the middle of the gap. 

This suggests that a NNGP constructed using a reference set with large gaps is a poor approximation to the full GP as a process in certain regions of the domain. If the data locations do have large gaps, perhaps a NNGP with $\calS$ as a grid over the domain provides a much better approximation to the full GP. To observe this we used a $14\times 7 $ grid over the domain $[0,3]\times [0,1]$ as $\calS$. So the size of this new $\calS$ was similar to the original sample size of $100$. Figure~\ref{fig:gapsgrid} demonstrates the NNGP covariance function at the two points using this new $\calS$. We see that using the grid $\calS$, the NNGP covariance function at the two points are very similar and closely resemble the true GP covariance function. This suggests that in order for the NNGP to resemble full GP, the reference set needs to have points uniformly distributed over the domain.

\begin{figure}[h!]
\begin{center}
\includegraphics[width=\textwidth]{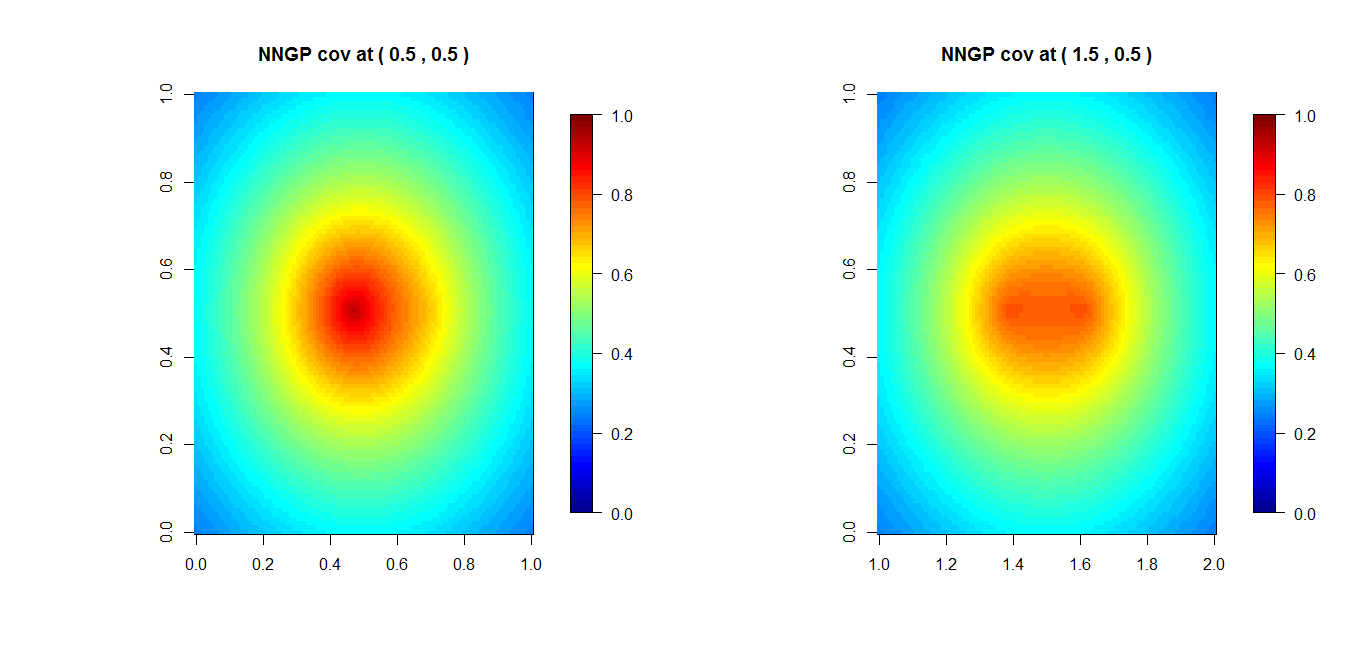}
\end{center}
\vskip -1cm \caption{NNGP covariance function using a grid $\calS$}
\label{fig:gapsgrid}
\end{figure}

However, from a kriging perspective, if the data have large gaps, inference from a NNGP with $\calS=\calT$ may not differ a lot from the full GP inference. Even when one uses the full GP, kriging is usually done one point at a time and thereby ignores the covariances between points outside the data locations and assumes conditional independence.
\begin{figure}[h!]
\begin{center}
\includegraphics[width=\textwidth]{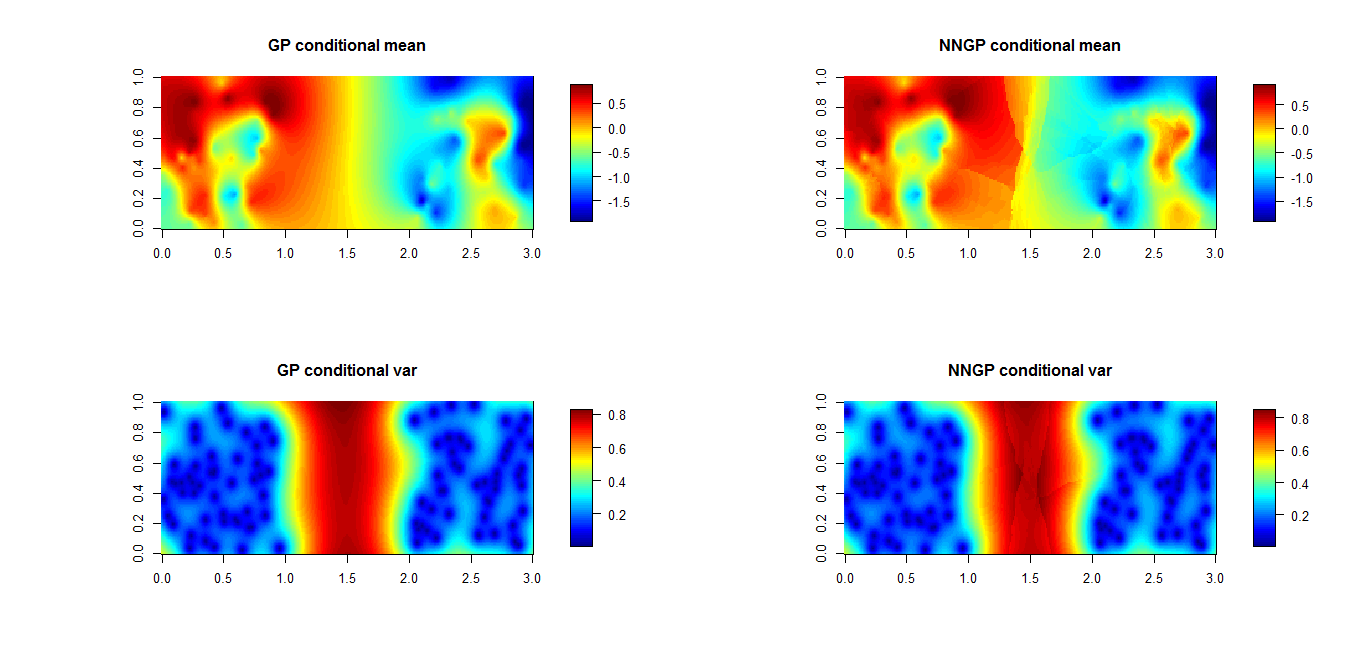}
\end{center}
\caption{Kriging means and variances for full GP and NNGP ($\calS$ = data locations)}
\label{fig:gapskrig}
\end{figure}
 Figure~\ref{fig:gapskrig} plots the kriging mean and variances over the entire domain for the full GP and the NNGP. They are very close. This suggests even for data with gaps the kriging performance of NNGP and GP are similar.

We also generated a dataset over $\calT$ and fitted the full GP and NNGP ($\calS=\calT$) model to compare parameter estimation and kriging performance. In addition to the conventional independent kriging, we also used the computationally expensive joint kriging for the full GP to see if it improves kriging quality at locations in the gap. Table~\ref{tab:gaps} provide the parameter estimates and model fitting metrics. Figures~\ref{fig:gapsfitm} and \ref{fig:gapsfitv} gives the posterior median and the variance surface over the domain. We see that the the NNGP and full GP produce very similar parameter estimates and kriging. Hence, for data with large gaps both the full GP and NNGP ($\calS =\calT$) doesn't provide enough information for locations inside the gaps. So even if NNGP ($\calS =\calT$) poorly approximates the full GP as a process, in terms of model fitting, their performances are very similar. 

\begin{table}
\begin{center}
\begin{tabular}{ccccc}
    & True & Full GP          & NNGP m=10         & NNGP m=20\\ \hline
$\beta$& 1    & 0.72 (0.00, 1.32)& 0.65 (-0.14, 1.30)& 0.69 (0.02, 1.16)\\
$\tau$ & 0.01 & 0.03 (0.01, 0.05)& 0.03 (0.01, 0.06) & 0.03 (0.01, 0.06)\\
$\sigs$ & 1    & 0.63 (0.38, 1.31)& 0.65 (0.39, 1.29) & 0.62 (0.38, 1.27)\\
$\phi$ & 2    & 2.94 (1.27, 5.19)& 2.76 (1.27, 5.25) & 2.91 (1.34, 5.20)\\
RMSPE    & -- & 0.58 (ind)             & 0.57              & 0.57\\
    & -- & 0.58  (joint)           & --                & --\\
$95\%$ CI cover   & -- & 94.00 (ind)            & 95.66             & 95.33\\
			    & -- & 95.33 (joint)           & --                & --\\					   
Mean $95\%$ CI width   & -- & 2.12 (ind)             &2.12               & 2.13\\
					    & -- & 2.11 (joint)             & --                & --\\
\end{tabular}
\end{center}
\caption{Data analysis for locations with gaps}\label{tab:gaps}
\end{table}

\begin{figure}[t!]
\begin{center}
\subfigure[Full GP (independent)]{\includegraphics[width=8cm]{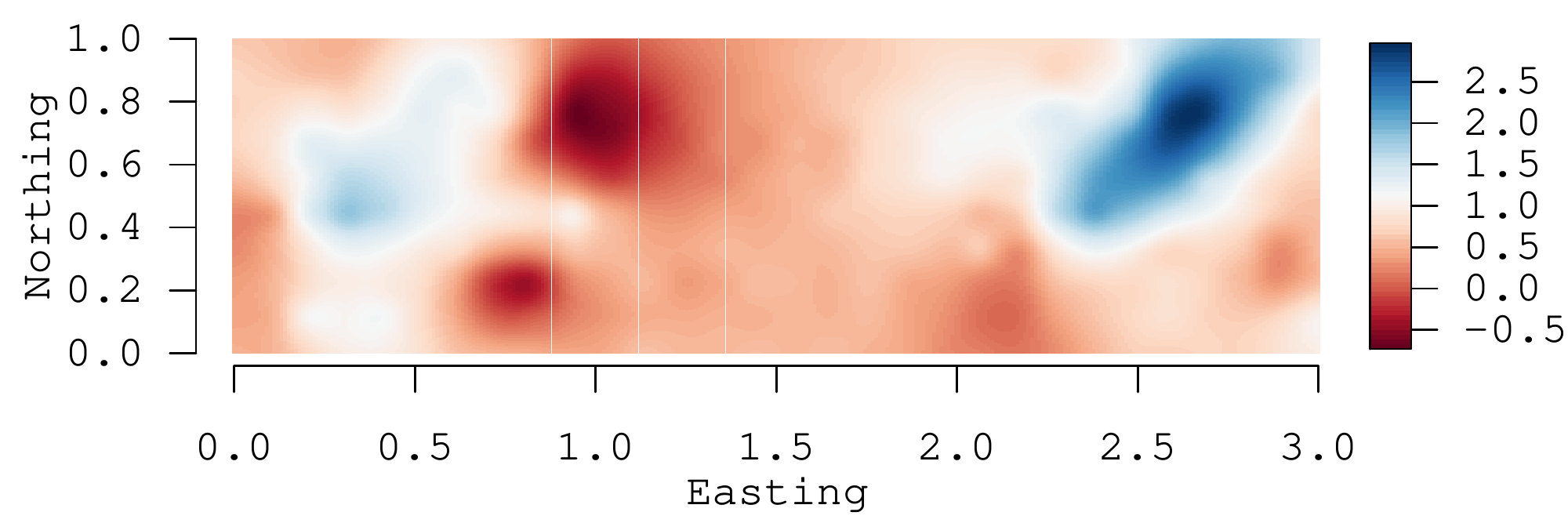}}
\subfigure[Full GP (joint)]{\includegraphics[width=8cm]{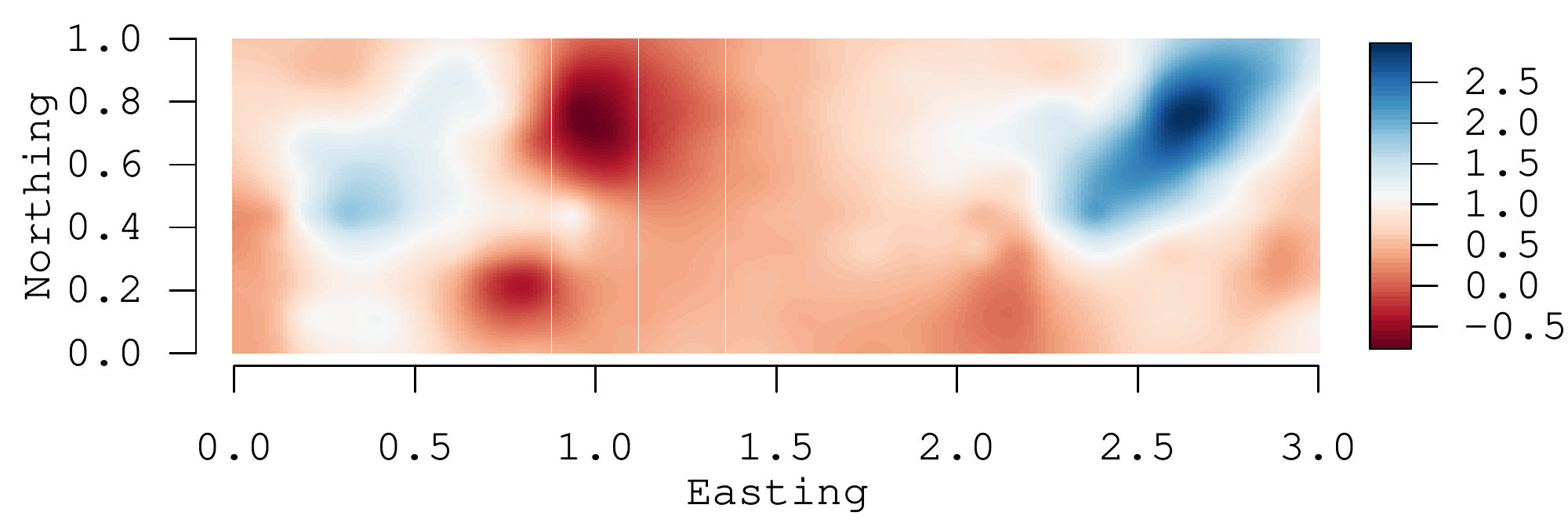}}\\
\subfigure[NNGP $m=10$]{\includegraphics[width=8cm]{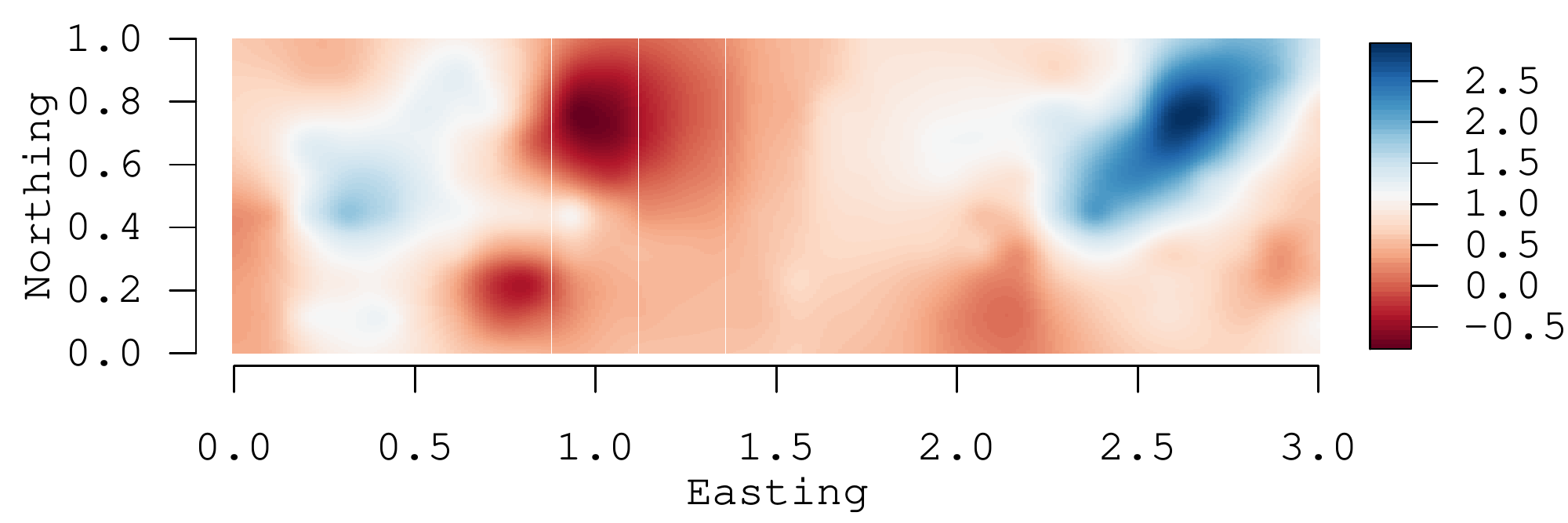}}
\subfigure[NNGP $m=20$]{\includegraphics[width=8cm]{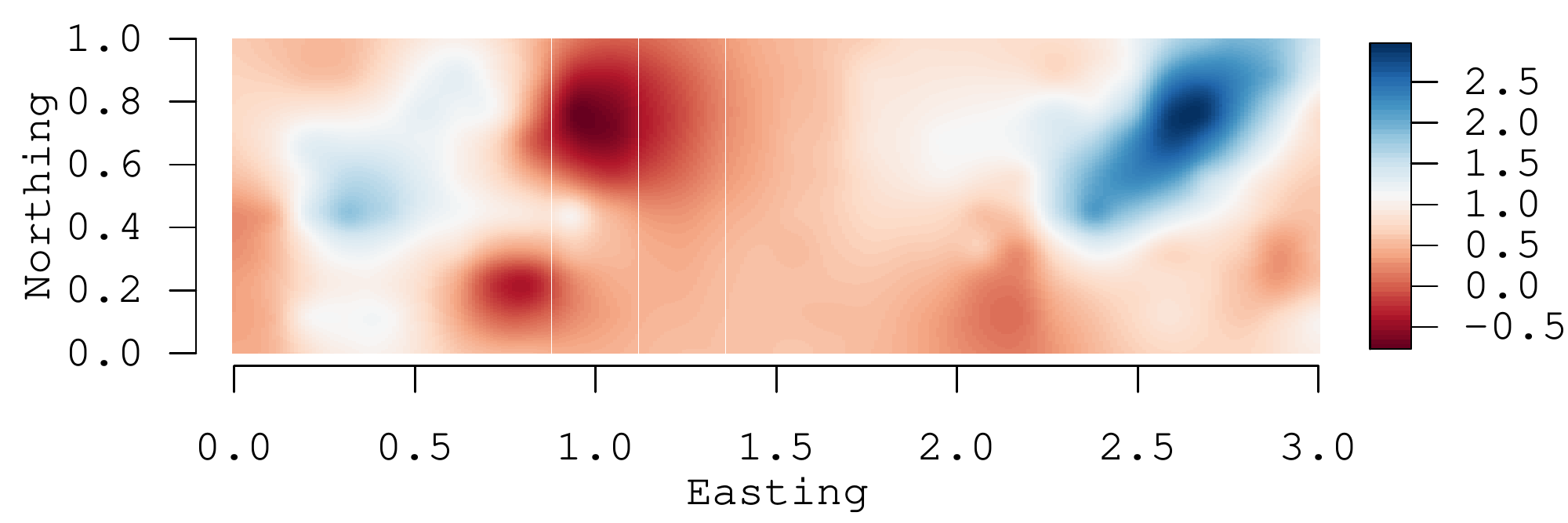}}
\end{center}
\caption{Posterior median surface for data with gaps}
\label{fig:gapsfitm}
\end{figure}

\begin{figure}
\begin{center}
\subfigure[Full GP (independent)]{\includegraphics[width=8cm]{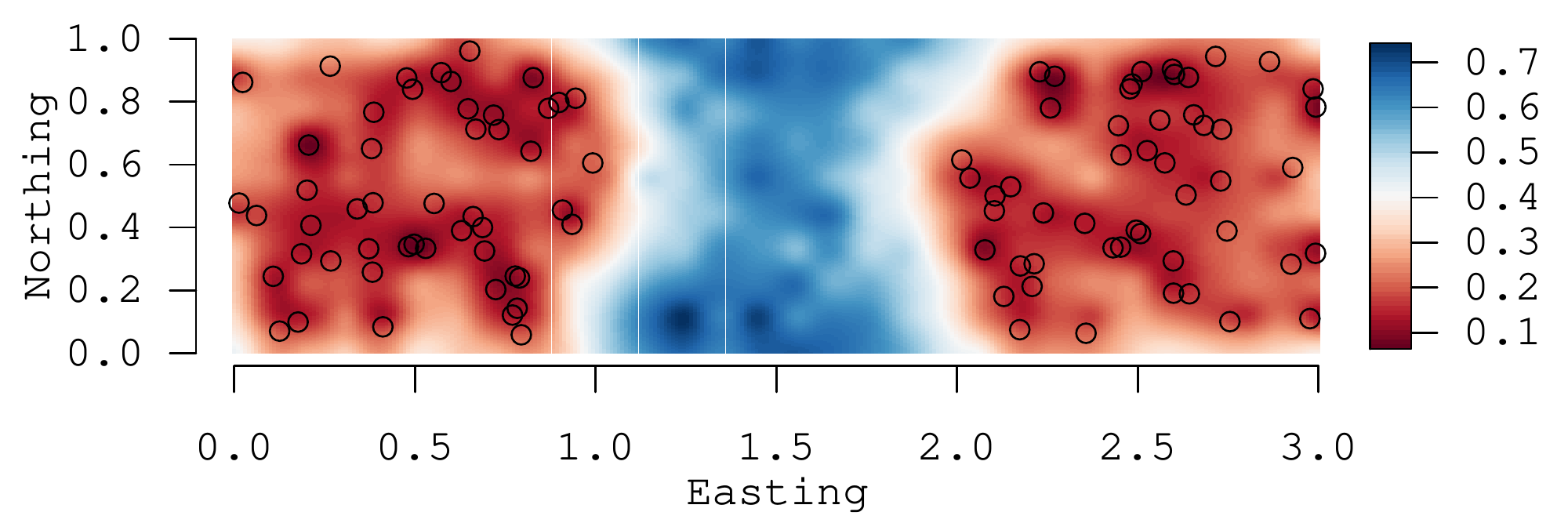}}
\subfigure[Full GP (joint)]{\includegraphics[width=8cm]{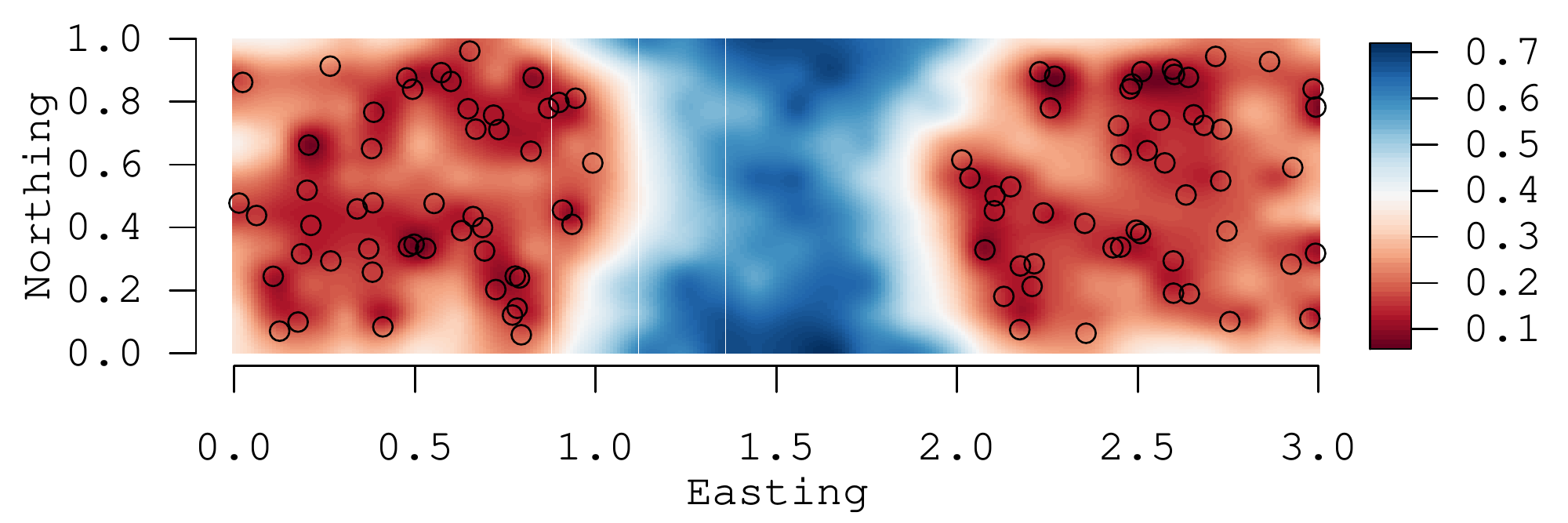}}\\
\subfigure[NNGP $m=10$]{\includegraphics[width=8cm]{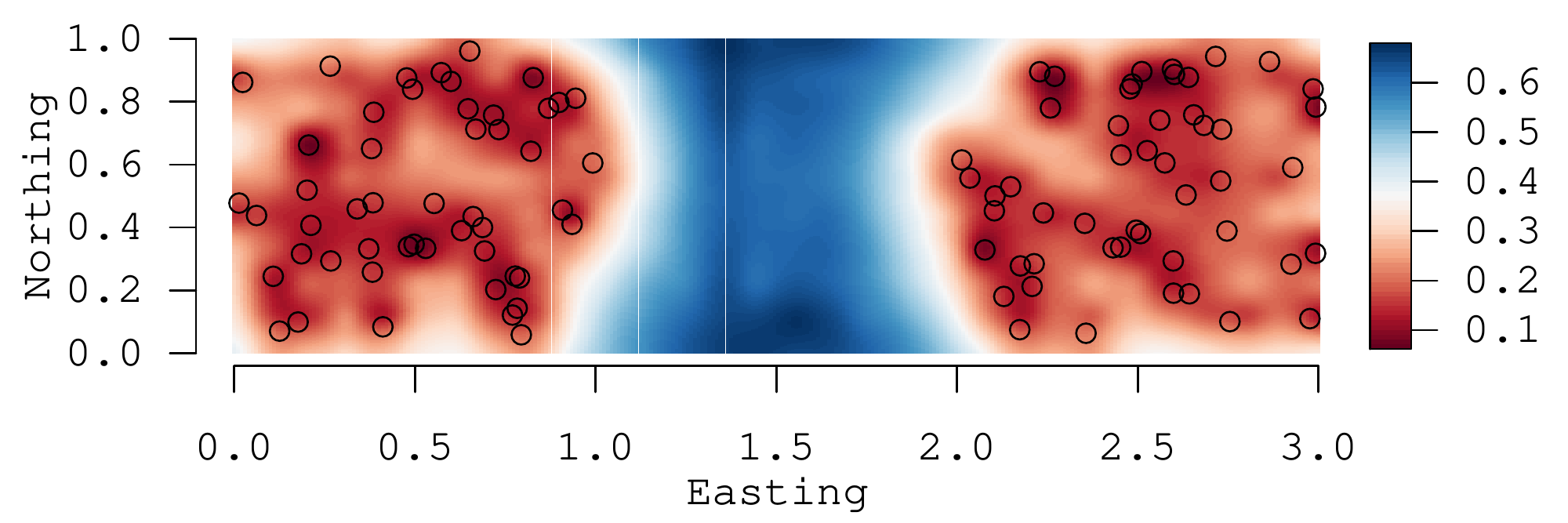}}
\subfigure[NNGP $m=20$]{\includegraphics[width=8cm]{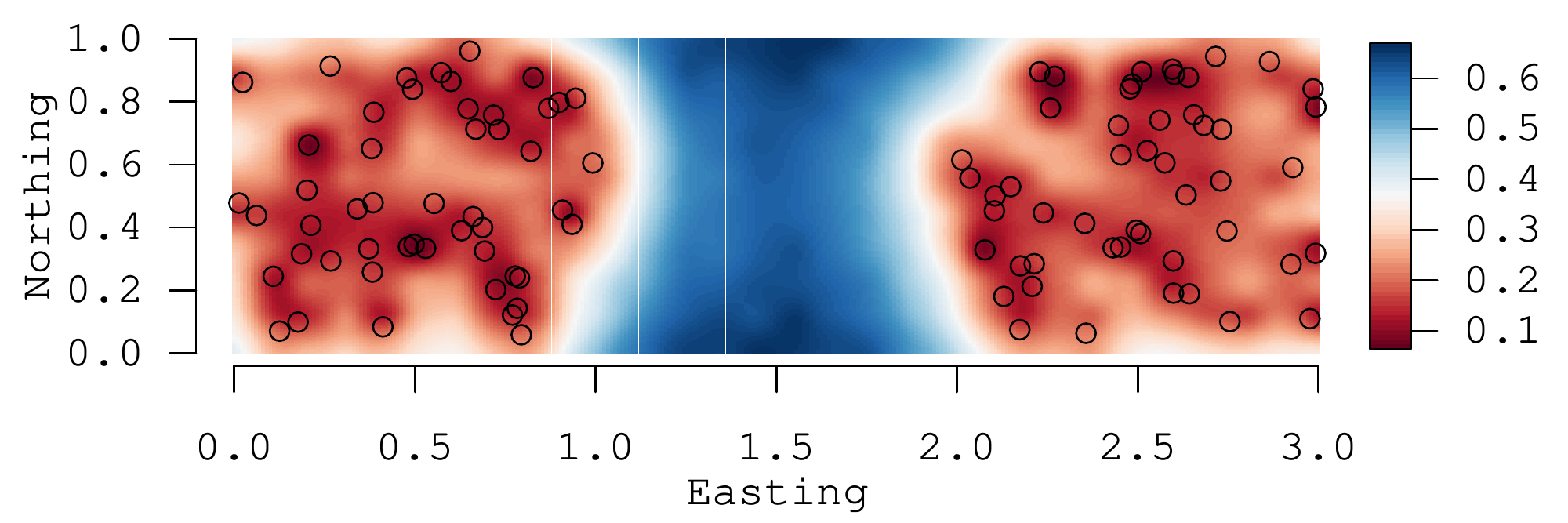}}
\end{center}
\caption{Posterior variance surface for data with gaps}
\label{fig:gapsfitv}
\end{figure}

\clearpage
\section{Simulation experiment: Slow decaying covariance functions}\label{sec:nn}
We note in Section~\ref{Sec: NNGD} that several valid choices of neighbor sets can be used to construct a NNGP. However, our choice of using $m$-nearest neighbors to construct neighbor sets performed extremely well for all the data analysis in Section \ref{Sec: Illustrations}. Since, our design of NNGP just includes $m$-nearest neighbors it is natural to be skeptical of the performance of NNGP when the data arises from a Gaussian process with very flat tailed covariance function. Such a covariance function implies that even distant observations are significantly correlated with the given observation and $m$-nearest neighbors may fail to capture all the information about the covariance parameters. 

We generate datasets of size $2500$ in a unit domain using the model described in Section 5.1 for a wide range of values for the parameters $\sigma^2$ and $\phi$. The marginal variance $\sigma^2$ was varied over $(0.05,0.1,0.2,0.5)$ and the `true effective range' $3/\phi$ phi was varied over $(0.1, 0.2, \ldots, 1)$. Larger values of the `true effective range' indicate higher correlation between points at large distances. The nugget variance $\tau^2$ was held constant at 0.1. The prior on $\phi$ was U(3,300) or 0.01 to 1 distance units. Also both $\tau^2$ and $\sigma^2$ were given Inverse Gamma$(2,0.1)$ priors in all cases.

Figure~\ref{fig:uni3-phi} gives the results for NNGP and full GP CIs. We see that for all choices of parameters, the posterior samples from the NNGP and full GP look identical. This strongly suggests that the NNGP model deliver inference similar to that of a full GP even for slow decaying covariance functions and justifies the choice of the neighbor sets.

\begin{figure}[t!]
\begin{center}
\subfigure[$\sigma^2=0.05$]{\includegraphics[width=8cm]{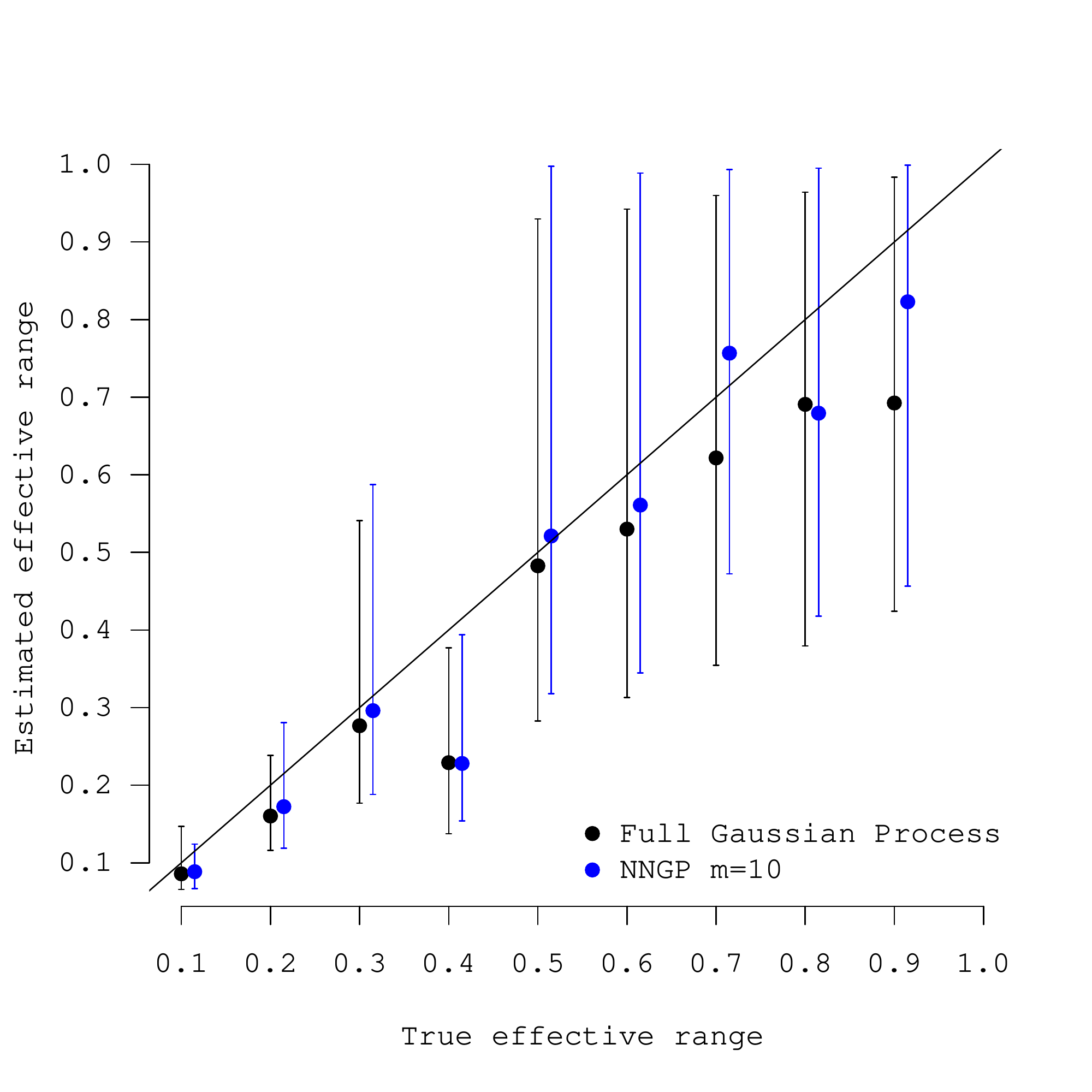}\label{uni3-phi-sig05}}
\subfigure[$\sigma^2=0.1$]{\includegraphics[width=8cm]{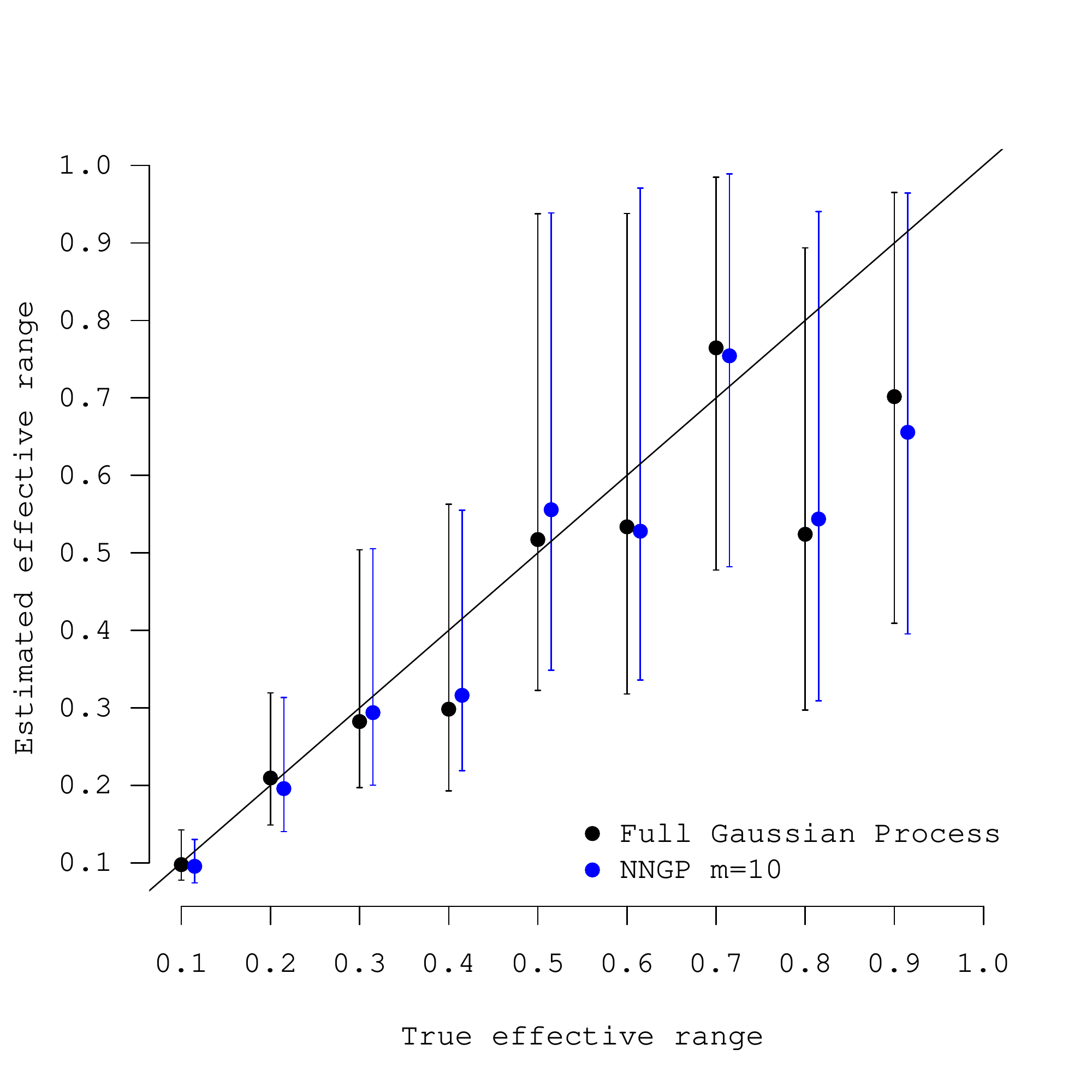}\label{uni3-phi-sig1}}\\
\subfigure[$\sigma^2=0.2$]{\includegraphics[width=8cm]{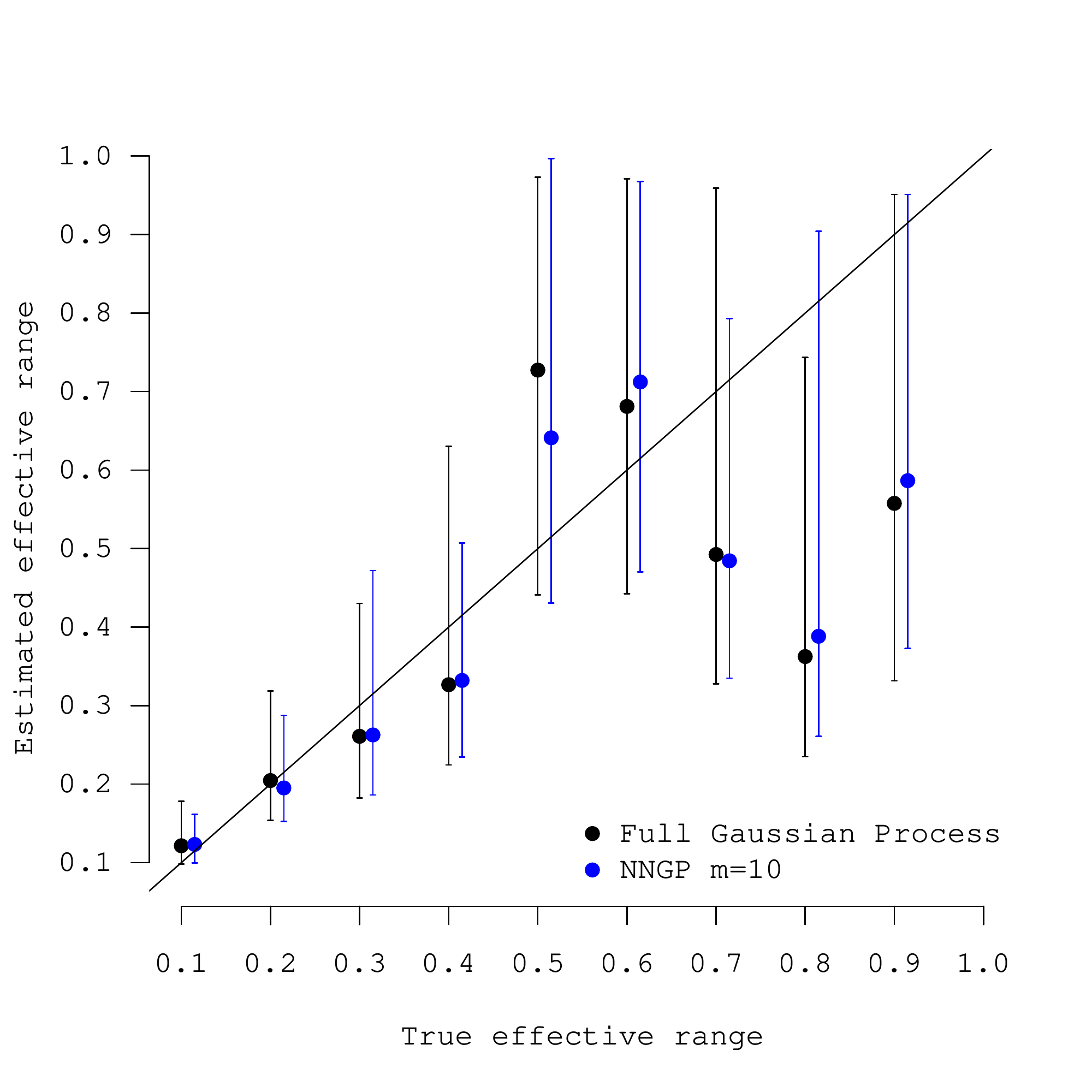}\label{uni3-phi-sig2}}
\subfigure[$\sigma^2=0.5$]{\includegraphics[width=8cm]{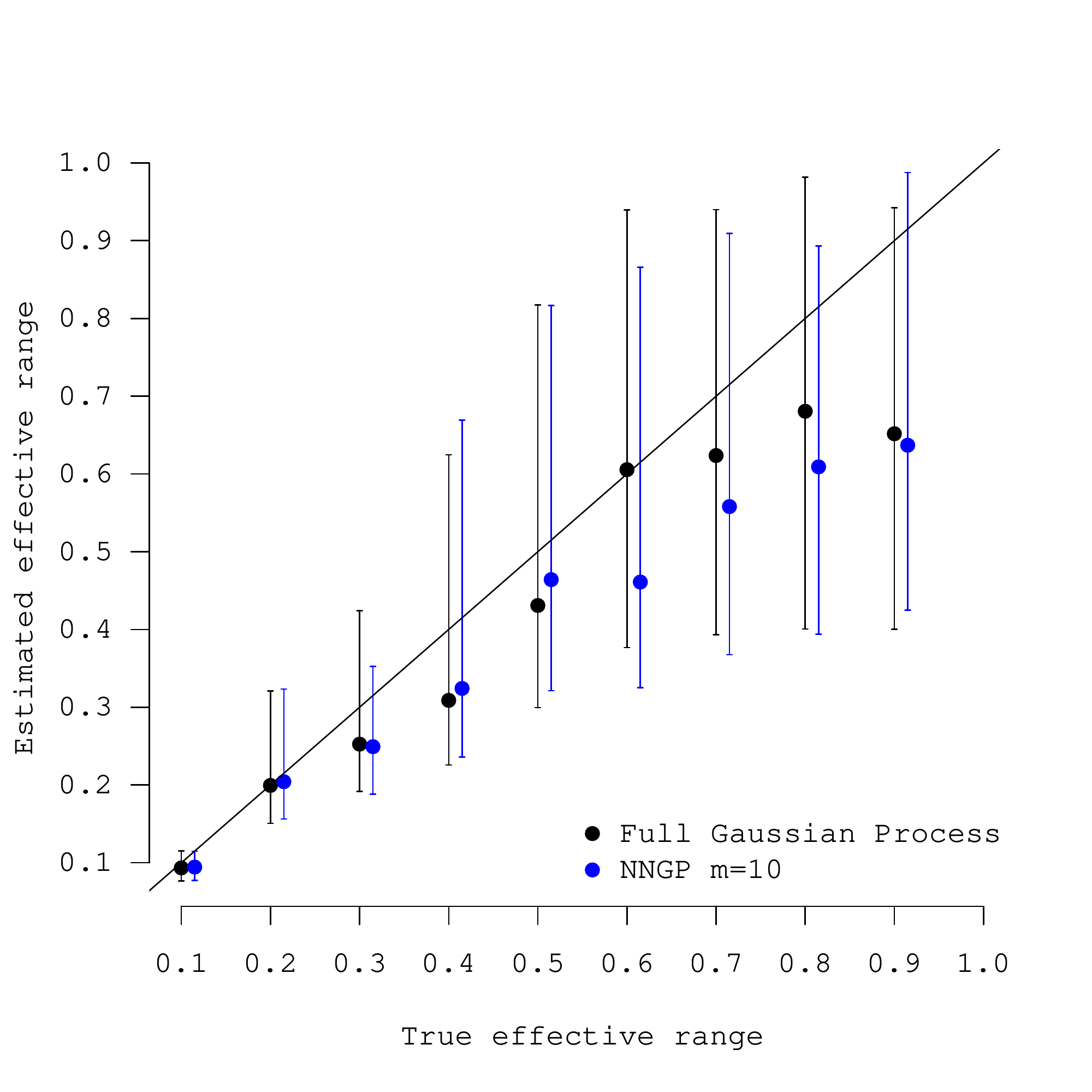}\label{uni3-phi-sig5}}
\end{center}
\caption{Univariate synthetic data analysis: true versus posterior $50\%$ ($2.5\%$, $97.5\%$) percentiles for the effective spatial range simulated for various values of $\sigma^2$ and $\tau^2=0.1$. NNGP model fit with $\calS=\calT$ and $m=10$.}
\label{fig:uni3-phi}
\end{figure}

\clearpage
\section{Simulation experiment: Wave covariance function}\label{sec:wave}
We have restricted most of our simulation experiments to Mat\'ern (or in particular exponential) covariance functions. Mat\'ern covariance functions like many other covariance functions decrease monotonically with distance and hence nearest neighbors of a location have highest correlation with that location. We wanted to investigate the performance of NNGP for covariance functions which do not monotonically decrease with distance. We use the two-dimensional damped cosine covariance function given by:
\begin{equation}\label{eq:dampcos}
C(d) = \exp(- d/a) \cos(\phi d) \; , \; a \leq 1/ \phi
\end{equation}

First, we generated the Kullback-Leibler (KL) divergence numbers for the NNGP model with respect to the full GP model using damped cosine covariance. In addition to the default neighbor selection scheme, we also used an alternate scheme described by \cite{stein04}. This scheme includes $m'=\lceil 0.75m \rceil$ nearest neighbors and $m-m'$ neighbors whose ranked distances from the $i^{th}$ location equal $m+\lfloor  l(i-m-1)/(m-m') \rfloor$ for $l=1,2,\ldots,m-m'$. \cite{stein04} suggested that this scheme choice often improves parameter estimation. The two schemes are referred to as NNGP and NNGP (alt) respectively. We used $\phi=10$, $a=.099$, sample sizes of $100, 200$ and $500$ and varied $m$ from $5$ to $50$ in increments of $5$.

\begin{figure}[!]
\centering
\includegraphics[width=\textwidth]{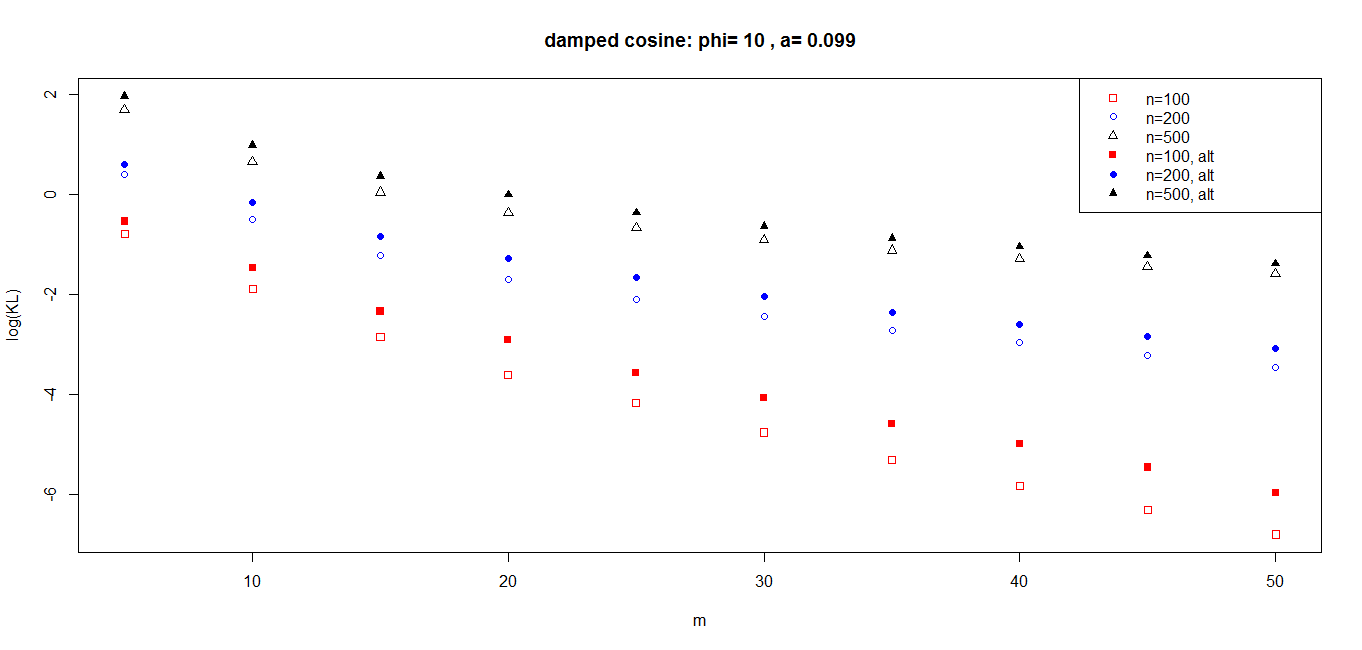}
\vskip -8mm \caption{NNGP KL divergence numbers (log scale) for damped cosine covariance}\label{fig:kl}
\end{figure}

Figure~\ref{fig:kl} plots the KL divergence numbers (in log-scale) for varying $m$, $n$ and neighbor selection schemes. We see that larger sample size implies higher KL divergence numbers which is expected as with increasing sample size the size of the neighbor set $m$ becomes smaller in proportion. Also, we see that KL numbers for the alternate neighbor selection scheme are always higher indicating that nearest neighbors perform better even for such wave covariance functions. In general we observed that the KL numbers are quite small for $m \geq 25$ for all $n$ and neighbor selection schemes indicating that the NNGP models closely approximate the true damped cosine GP. 

\begin{table}[b!]
\centering
\begin{tabular}{cccc}
&	True &  m=10 &      m=20 \\ \hline
$\beta_0$  &  1  &   1.03 (0.65, 1.34) &   1.06 (0.70, 1.32) \\
$\beta_1$  &  5   &  5.00 (4.95, 5.06)  &  5.00 (4.95, 5.06) \\
$\tau^2$  &  0.1 &  0.06 (0.02, 0.12)  &  0.05 (0.03, 0.11)\\
$\sigma^2$ & 1   &  1.13 (0.90, 1.57)  &  1.14 (0.90, 1.57)\\
$\phi$   &   10  &  7.41 (1.63, 11.59) &  6.31 (1.61, 10.50)\\
$a$    &    0.099  & 0.093 (0.067, 0.135) & 0.09 (0.07, 0.14)
\end{tabular}
\caption{Damped cosine GP data analysis using NNGP}\label{tab:wave}
\end{table}

Next, we conducted a data analysis using the wave covariance function. We choose $n=500$, $m=10, 20$. The two values of $m$ yielded around $3.4\%$ and $18.7\%$ nearest neighbors which were negatively correlated with the corresponding locations. Table~\ref{tab:wave} gives the parameter estimates for the NNGP model. Figure~\ref{fig:wavecov} demonstrates how the NNGP approximates the wave covariance function while figure~\ref{fig:wavesurface} plots the true and fitted random effect surface. We observe that NNGP provides an excellent approximation of the the true wave GP in terms of model parameter estimation and kriging. 

\begin{figure}
\centering
\includegraphics[width=8cm]{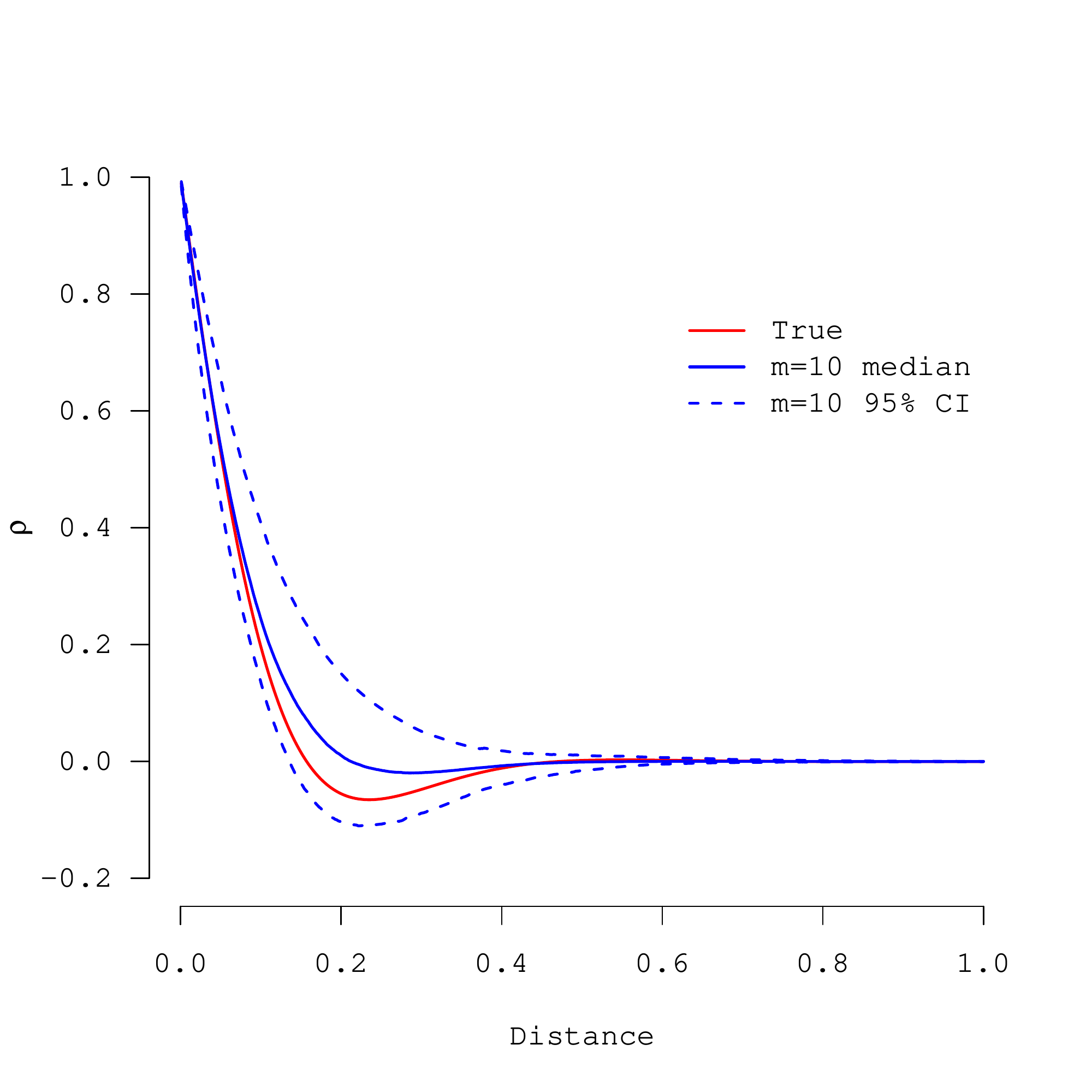}
\includegraphics[width=8cm]{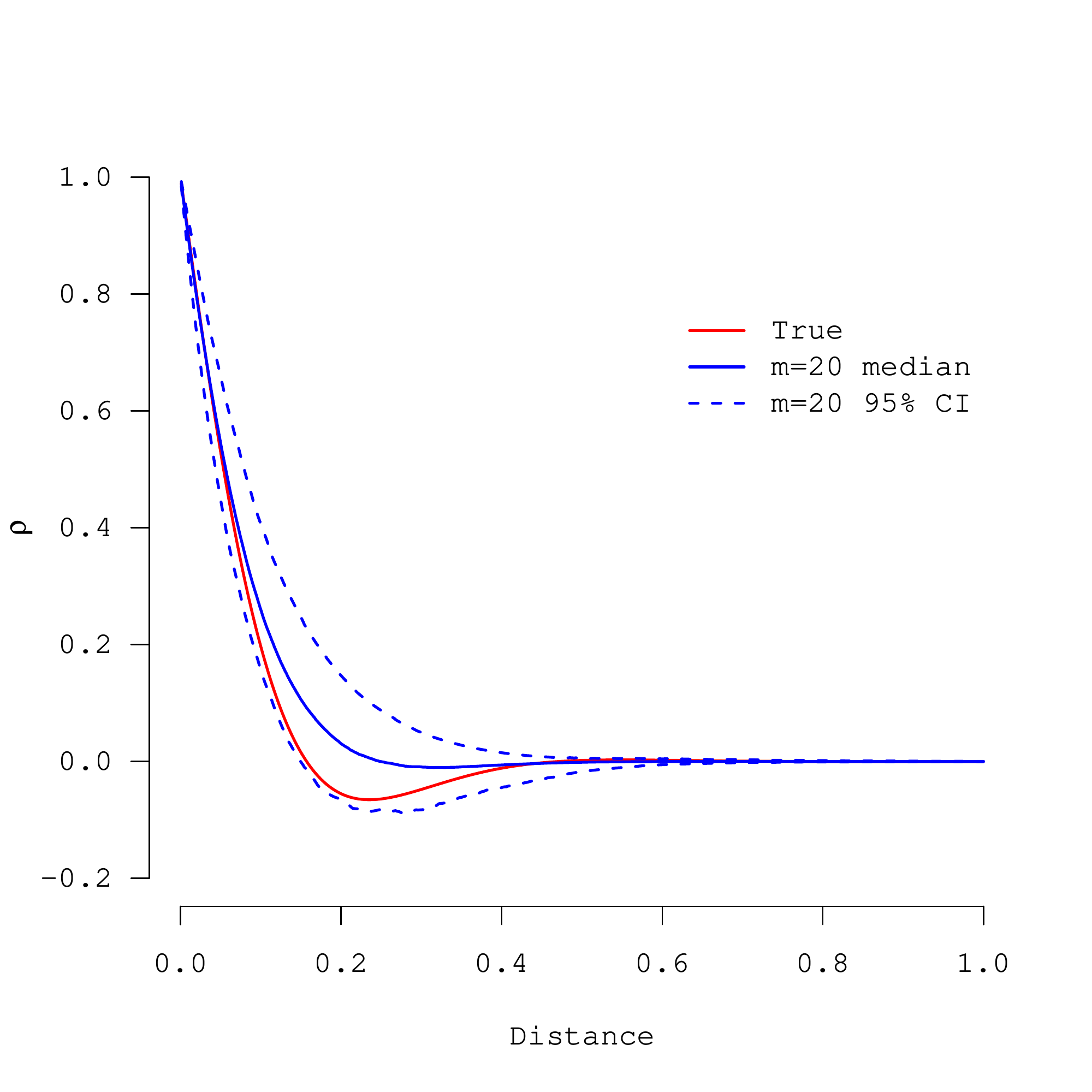}
\vskip -7mm \caption{Wave covariance function estimates using NNGP}\label{fig:wavecov}
\end{figure}

\begin{figure}
\centering
\subfigure[True]{\includegraphics[width=5cm]{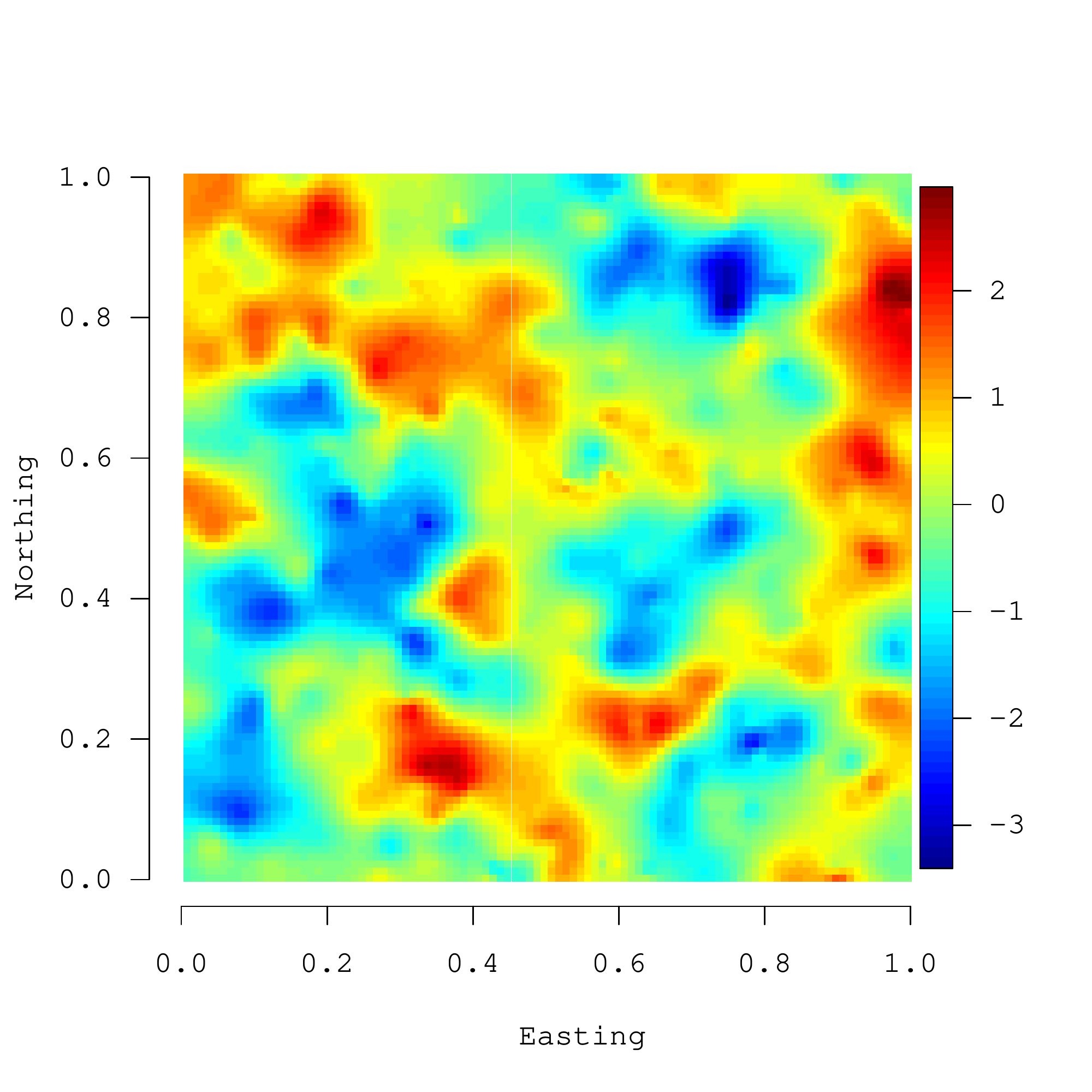}}
\subfigure[NNGP $m=10$]{\includegraphics[width=5cm]{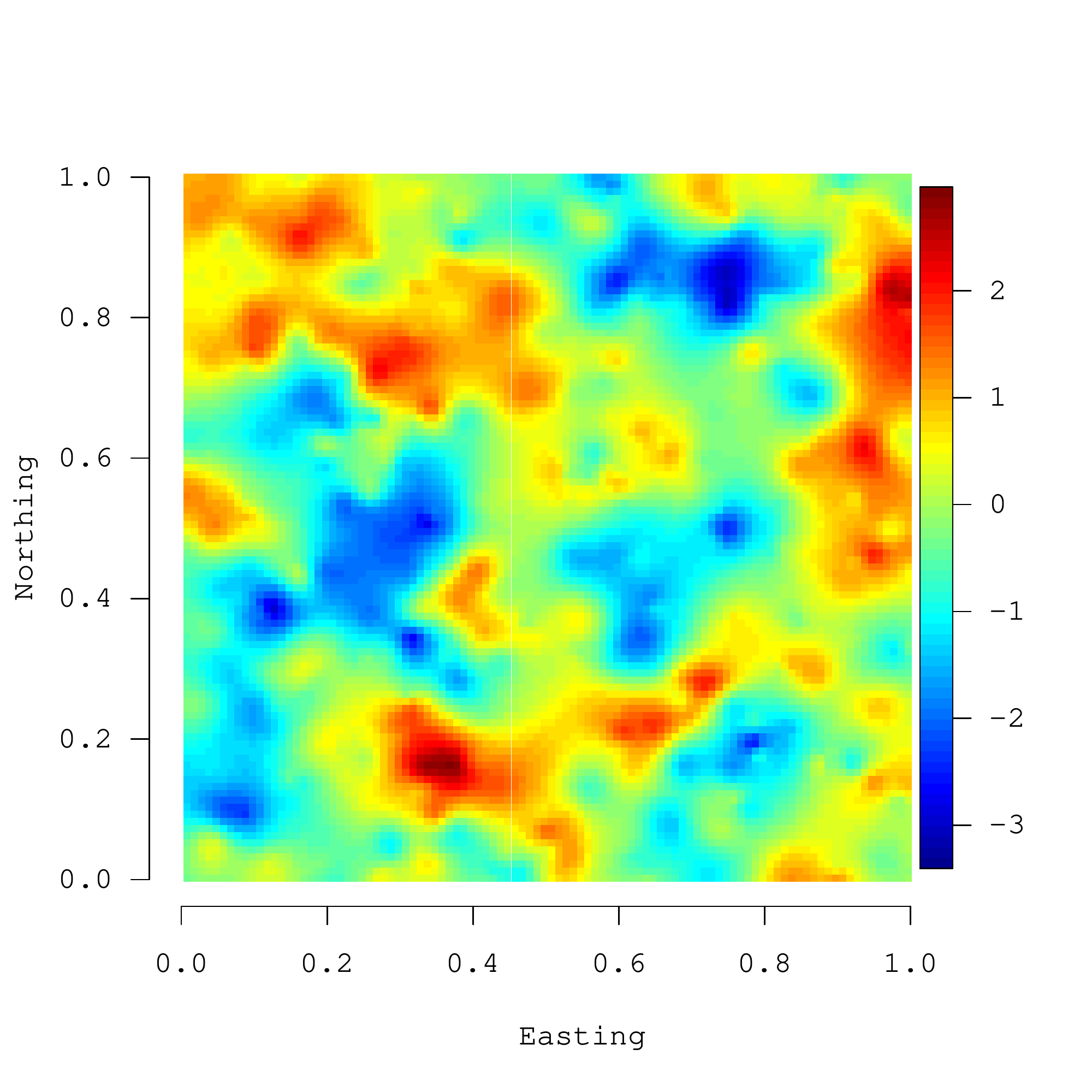}}
\subfigure[NNGP $m=20$]{\includegraphics[width=5cm]{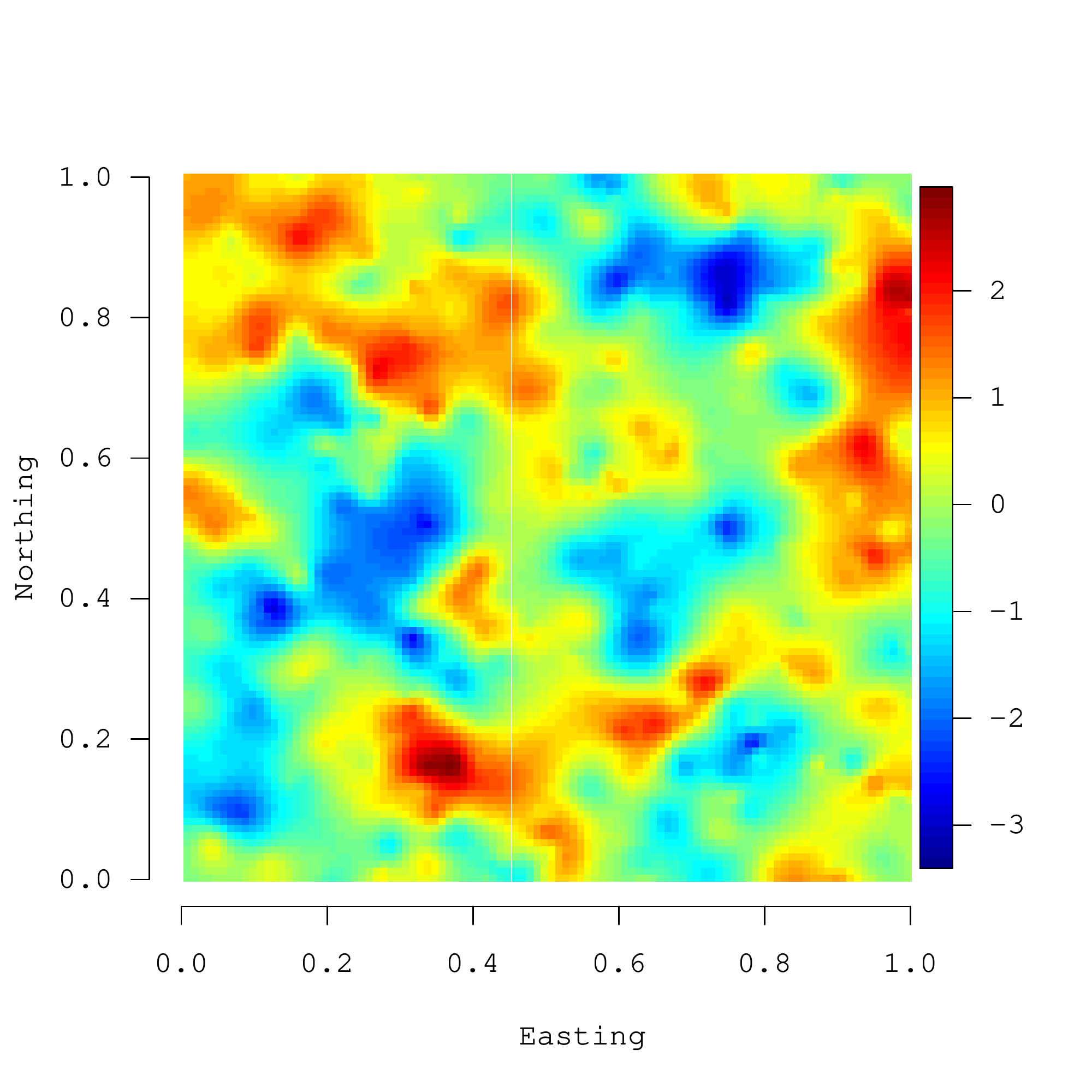}}
\caption{True and estimated (posterior median) random effect surface of the damped cosine GP}\label{fig:wavesurface}
\end{figure}

We could not fit the full GP model due to computation instability of the large wave covariance matrix. NNGP does not involve inverting large matrices and hence we could use it for model fitting. 

\bibliographystyle{asa}
\bibliography{NNGPjasanew}	
\label{lastpage}

\end{document}